\documentclass[10pt,a4paper]{article}
\usepackage{theorem}
\usepackage{amsmath}
\usepackage{amssymb}
\usepackage{graphics}
\usepackage{graphicx}
\usepackage{pstricks}
\usepackage{setspace}

\makeatletter
\newcommand{\sech}{\mathop{\operator@font sech}}
\makeatother

\makeatletter
\newcommand{\cn}{\mathop{\operator@font cn}}
\makeatother
 \makeatletter
\newcommand{\curl}{\mathop{\operator@font curl}}
\makeatother

\makeatletter
\newcommand{\ddiv}{\mathop{\operator@font div}}
\makeatother

\makeatletter
\newcommand{\grad}{\mathop{\operator@font grad}}
\makeatother

\makeatletter
\newcommand{\dist}{\mathop{\operator@font dist}}
\makeatother

\makeatletter
\newcommand{\dexp}{\mathop{\operator@font exp}}
\makeatother

\makeatletter
\newcommand{\ess}{\mathop{\operator@font ess}}
\makeatother

\renewcommand{\ge }{\geqslant }

\renewcommand{\exp}{ {\rm e}}

\newtheorem{lemma}{Lemma}[section]
\newtheorem{theorem}{Theorem}[section]
\newtheorem{proposition}{Proposition}[section]

\newenvironment{proof}
{\textbf{Proof:}} 
    {\hfill $\Box$}

\begin{document}
\doublespacing

\title{BOUSSINESQ SYSTEMS OF BONA-SMITH TYPE ON PLANE DOMAINS: THEORY AND
NUMERICAL ANALYSIS\thanks{This work was supported in part by a French-Greek scientific cooperation grant for the period 2006--08, funded jointly by EGIDE, France, and the General Secretariat of Research and Technology, Greece. D. Mitsotakis was also supported by Marie Curie Fellowship No. PIEF-GA-2008-219399 of the European Commission.}}
\author{V. A. Dougalis \footnote{Department of Mathematics, University of Athens, 15784 Zographou, Greece}
\footnote{Institute of Applied and Computational Mathematics
FO.R.T.H., 70013 Heraklion, Greece}, D. E.
Mitsotakis\footnotemark[4] and J.-C. Saut \footnote{UMR de Math\'{e}matiques, Universit\'{e} de Paris-Sud, B\^{a}timent 425, 91405 Orsay, France}}
\date{\today}

\maketitle

\begin{abstract}
We consider a class of Boussinesq systems of Bona-Smith type in two space dimensions approximating surface wave flows modelled by the three-dimensional Euler equations. We show that various initial-boundary-value problems for these systems, posed on a bounded plane domain are well posed locally in time. In the case of reflective boundary conditions, the systems are discretized by a modified Galerkin method which is proved to converge in $L^2$ at an optimal rate. Numerical experiments are presented with the aim of simulating two-dimensional surface waves in complex plane domains with a variety of initial and boundary conditions, and comparing  numerical solutions of Bona-Smith systems with analogous solutions of the BBM-BBM system. 
\end{abstract}

\section{Introduction}

In this paper we will study the Boussinesq system
\begin{equation}\label{E1.1}
\begin{array}{l}
\eta_t+\nabla\cdot {\bf v}+\nabla\cdot \eta{\bf v}- b\Delta \eta_t=0,\\
{\bf v}_t+\nabla\eta+\frac{1}{2}\nabla|{\bf v}|^2+c\nabla\Delta\eta-b\Delta{\bf v}_t=0,
\end{array}
\end{equation}
where $b>0$ and $c<0$ are constants. This system is the two-dimensional version of a system in one space variable originally derived and analyzed by Bona and Smith, \cite{BS}. It belongs to a family of Boussinesq systems, \cite{BCSI}, \cite{BCL}, that approximate the three-dimensional Euler equations for the irrotational free surface flow of an ideal fluid over a horizontal bottom. In (\ref{E1.1}) the independent variables ${\bf x}=(x,y)$ and $t$ represent spatial position and elapsed time, respectively. The dependent variables $\eta=\eta({\bf x},t)$ and ${\bf v}={\bf v}({\bf x},t)=(u({\bf x},t), v({\bf x},t))$, represent quantities proportional, respectively, to the deviation of the free surface from its level of rest, and to the horizontal velocity of the fluid particles at some height above the bottom. In (\ref{E1.1}) the variables are nondimensional and unscaled and the horizontal velocity ${\bf v}$ is evaluated at a height $z=-1+\theta(1+\eta({\bf x},t))$, for some $\theta\in [\sqrt{2/3},1]$. (In these variables the bottom lies at height $z=-1$.) In terms of the parameter $\theta$, the constants in (\ref{E1.1}) are given by the formulas $b=(3\theta^2-1)/6$ and $c=(2-3\theta^2)/3$. (The system originally analyzed by Bona and Smith in \cite{BS} corresponds to $\theta^2=1$.) The value $\theta^2=2/3$ (i.e. $c=0$) yields the BBM-BBM system, \cite{BC}, \cite{DMSII}, \cite{Ch}.

The system (\ref{E1.1}) is derived from the Euler equations, \cite{BCSI}, \cite{BCL}, under a long wavelength, small amplitude assumption. Specifically, one assumes that $\varepsilon:=A/h_0\ll 1$, $\lambda/h_0\gg 1$ with the Stokes number $S:=A\lambda^2/h_0^3$ being of $O(1)$. (Here $A$ is the maximum amplitude of the surface waves measured over the level of rest, $z=-h_0$ is the (constant) depth of the bottom, and $\lambda$ is a typical wavelength.) If one takes $S=1$, then, in nondimensional, scaled variables, appropriate asymptotic expansions in the Euler equations yield equations of the form 
\begin{equation}\label{E1.2}
\begin{array}{l}
\eta_t+\nabla\cdot {\bf v}+\varepsilon(\nabla\cdot \eta{\bf v}- b\Delta \eta_t)=O(\varepsilon^2),\\
{\bf v}_t+\nabla\eta+\varepsilon\left(\frac{1}{2}\nabla|{\bf v}|^2+c\nabla\Delta\eta-b\Delta{\bf v}_t\right)=O(\varepsilon^2),
\end{array}
\end{equation}
from which (\ref{E1.1}) follows by unscaling to remove $\varepsilon$, and replacing the right-hand side by zero.

The Cauchy problem for (\ref{E1.1}) in the case of one spatial variable has been proved to be globally well posed for $2/3<\theta^2\leq 1$ in appropriate classical and Sobolev space pairs, \cite{BS}, \cite{BCSI}. The analogous problem for $\theta^2=2/3$ is locally well posed, \cite{BCSI}. In \cite{DMSI} we considered a more general class of systems in the two-dimensional case and proved that the corresponding Cauchy problem is locally well posed in appropriate pairs of Sobolev spaces. Initial-boundary-value problems (ibvp's) for (\ref{E1.1}) for $2/3\leq\theta^2\leq 1$ on a finite interval in one space variable were analyzed in \cite{ADMI}, and in \cite{BC} for $\theta^2=2/3$. It was proved in \cite{ADMI} that the ibvp with Dirichlet boundary conditions, wherein $\eta$ and $u$ are given functions of $t$ at the endpoints of the interval, is locally well posed. The corresponding ibvp with {\em reflection} boundary conditions ($\eta_x=u=0$ at both endpoints of the interval) was shown in \cite{ADMI} to be globally well posed; so is also the {\em periodic} ivp. In \cite{DMSII} we analyzed three ibvp's for the BBM-BBM system on a smooth plane domain $\Omega$, corresponding to homogeneous Dirichlet boundary conditions for $\eta$ and ${\bf v}$ on $\partial\Omega$, to homogeneous Neumann boundary conditions for $\eta$ and ${\bf v}$ on $\partial\Omega$, and to the (normal) {\em reflective} boundary conditions $\frac{\partial \eta}{\partial n}=0$, and ${\bf v}=0$ on $\partial\Omega$, where $n$ is the normal direction to the boundary. We showed that these ibvp's are well posed locally in time in the appropriate sense.

In Section 2 of the paper at hand we consider the Bona-Smith system (\ref{E1.1}) and pose it as an ibvp on a plane domain $\Omega$ under a variety of homogeneous boundary conditions on $\partial\Omega$, including e.g. homogeneous Dirichlet b.c.'s for $\eta$ and ${\bf v}$ and reflective b.c.'s. We prove that the corresponding ibvp's are well posed, locally in time. 

Turning now to the numerical solution of ibvp's for systems of the type (\ref{E1.1}) by Galerkin-finite element methods, we note first that it is quite straightforward to construct and analyze such schemes for the BBM-BBM system. For example, in \cite{DMSI} we proved optimal-order $L^2$-error estimates for the standared Galerkin semidiscretization of the BBM-BBM system with homogeneous boundary conditions on a smooth domain with a general triangulation. When $c>0$, i.e. in the case of the proper Bona-Smith systems, the presence of the term $\nabla\Delta\eta$ complicates issues. In \cite{DMSI} we analyzed the standard Galerkin semidiscretization with bicubic splines for this class of systems posed on rectangles with homogeneous Dirichlet boundary conditions, and proved optimal-order $H^2$-error estimates for the approximation of $\eta$. (Experimental evidence indicates that the $L^2$-errors for the approximation of $\eta$, $u$ and $v$ with this scheme have suboptimal -- $O(h^3)$ -- rate of convergence. In the one-dimensional case one may derive optimal-order estimates for the approximations of $\eta,u,v$ in $W^{1,\infty}\times L{^\infty}\times L^{\infty}$, cf. \cite{ADMII}.)

In Section 3 of the present paper we consider the Bona-Smith systems with $c\ge 0$ posed on convex smooth planar domains with reflective boundary conditions. The systems are discretized on an arbitrary triangulation of the domain using a modified Galrkin method, wherein the Laplacian in the $\nabla\Delta \eta$ terms in (\ref{E1.1}) is discretized weakly by an appropriate discrete Laplacian operator. This enables us to prove optimal-order $L^2$- and $H^1$- error estimates on finite element subspaces of continuous, piecewise polynomial functions that include the case of piecewise linear functions utilized in most applications. The systems of ordinary differential equations representing the semidiscretizations of the Bona-Smith systems are shown to be non-stiff. One may thus use any explicit method for their temporal discretization.

We close the paper by showing the results of a series of numerical experiments of simulations of surface waves in complex domains, aimed at comparing the numerical solution of a Bona-Smith system with analogous results obtained by solving the BBM-BBM system.

\section{Well-posedness of ibvp's for the Bona-Smith system}
\setcounter{equation}{0}

Let $\Omega$ be a bounded plane domain with smooth boundary (or a convex polygon). We consider the Bona-Smith system 
\begin{subequations}\label{E2.1}
\begin{gather}
\eta_t+\nabla\cdot \left({\bf v}+\eta{\bf v}\right)- b\Delta \eta_t=0,\label{E2.1a}\\
{\bf v}_t+\nabla\left(\eta+\frac{1}{2}|{\bf v}|^2+c\Delta\eta\right)-b\Delta{\bf v}_t=0,\label{E2.1b}
\end{gather}
\end{subequations}
for $({\bf x},t)\in\Omega\times \mathbb{R}_{+}$, with $b>0$, $c<0$, with initial conditions 
\begin{equation}\label{E2.2}
\eta({\bf x},0)=\eta_0({\bf x}),\qquad {\bf v}({\bf x},0)={\bf v}_0({\bf x})\qquad {\bf x}\in\Omega.
\end{equation}

\begin{figure}[h]
\begin{picture}(150,160)(0,0)
\put(100,0){\includegraphics[scale=0.3]{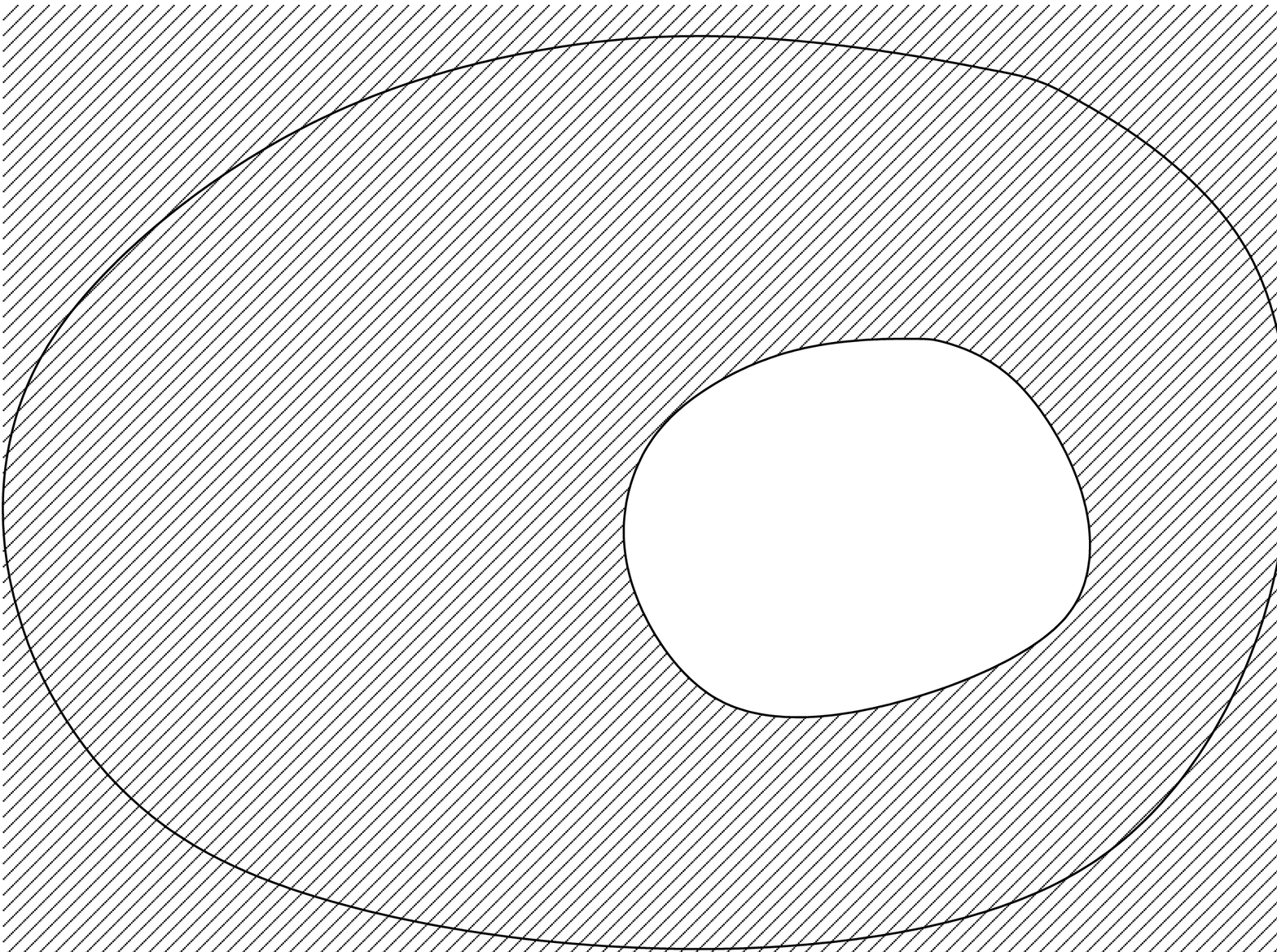}}
\put(135,90){$\Omega$}
\put(200,70){$\Gamma_1$}\put(95,25){$\Gamma_2$}
\end{picture}
\caption{Plane domain $\Omega$ with $\left.\frac{\partial\eta}{\partial n}\right|_{\Gamma_1}=0$, $\eta|_{\Gamma_2}=0$.}\label{F1}
\end{figure}

We now describe the class of boundary conditions on $\partial\Omega$ under which the problem (2.1)--(\ref{E2.2}) will be solved. Let $H^s=H^s(\Omega)$, $s\in \mathbb{R}$, denote the $L^2$-based, real-valued, Sobolev classes on $\Omega$ and $H_0^1$ the subspace of $H^1$ whose elements have zero trace on $\partial\Omega$. In the sequel, we shall denote by $\|\cdot\|$ and $(\cdot,\cdot)$ the norm and inner product, respectively, of $L^2=L^2(\Omega)$, by $\|\cdot\|_s$ the norm of $H^s$, and by $\|\cdot\|_{\infty}$ the norm of $L^{\infty}=L^{\infty}(\Omega)$. The boundary conditions on $\eta$ will be of the form 
\begin{subequations}
\begin{equation}
{\mathcal B}\eta=0,\quad {\bf x}\in\partial\Omega,\quad t\in\mathbb{R}_{+},\label{E2.3a}
\end{equation}
where the linear operator ${\mathcal B}$ and the domain $\Omega$ will be assumed to be such that the boundary-value problem 

$$\left\{\begin{array}{l}
-b\Delta w+w=f,\quad \mbox{in}\,\,\Omega,\\
{\mathcal B}w=0,\quad \mbox{in}\,\,\partial\Omega,
\end{array}
\right.$$

has for each $f$ in $L^2$ a unique solution $w\in H^2$ for which ${\mathcal B}w|_{\partial\Omega}$ is well defined.
(We will also assume that an $H^1$ solution $w$ of the problem is defined whenever $f\in H^{-1}$). The boundary conditions on ${\bf v}$ will be of homogeneous Dirichlet type, i.e. 
\begin{equation}
{\bf v}=0,\quad {\bf x}\in\partial\Omega,\quad t\in\mathbb{R}_{+}. \label{E2.3b}
\end{equation}
\end{subequations}
We note that examples of suitable boundary conditions of the type (\ref{E2.3a}) include, among other, homogeneous Dirichlet ($\eta=0$), Neumann ($\frac{\partial\eta}{\partial n}=0$) or Robin ($\alpha\frac{\partial \eta}{\partial n}+\beta\eta=0$) boundary conditions on the boundary $\partial\Omega$ if $\Omega$ is a bounded plane domain or a convex polygon, boundary conditions of the form $\frac{\partial\eta}{\partial n}|_{\Gamma_1}=0$, $\eta|_{\Gamma_2}=0$ for a multiply connected domain, for example such as the one shown in Figure \ref{F1}, {\em et al.}
%
%

In the sequel we let $X:= H^2(\Omega)\cap\{w:{\mathcal B} w=0\,\,\,\mbox{on}\,\,\partial\Omega\}$ and ${\bf H}_0^1=(H_0^1)^2$ etc.
The main result of this section is

\begin{theorem}\label{T2.1}
Given $\eta_0\in X$, ${\bf v}_0\in H_0^1$, there exists $T>0$ and a unique solution $(\eta,{\bf v})\in C([0,T]; X)\cap C([0,T]; {\bf H}_0^1)$ of the ibvp (2.1), (\ref{E2.2}), (2.3). Moreover, for each integer $k\ge 0$ $\left(\frac{\partial^k\eta}{\partial t^k},
\frac{\partial^k{\bf v}}{\partial t^k}\right)\in C([0,T];X)\cap C([0,t];{\bf H}_0^1)$.
\end{theorem}
\begin{proof}
Write (\ref{E2.1a}) and (\ref{E2.1b}) as
\begin{subequations}\label{E2.4}
\begin{gather}
\eta_t+(I-b\Delta)^{-1}\nabla\cdot \left({\bf v}+\eta{\bf v}\right)=0,\label{E2.4a}\\
{\bf v}_t+(I-b\Delta)^{-1}\nabla\left(\eta+\frac{1}{2}|{\bf v}|^2+c\Delta\eta\right)=0,\label{E2.4b}
\end{gather}
\end{subequations}
for ${\bf x}\in\Omega$, $t>0$. In (\ref{E2.4a}) $(I-b\Delta)^{-1}$ denotes the inverse of the operator $I-b\Delta$ with domain $X$, while in (\ref{E2.4b}) $(I-b\Delta)^{-1}$ represents the inverse of $I-b\Delta$ with domain ${\bf H}^2\cap {\bf H}^1_0$. Let $F$ be the vector field on $X\times {\bf H}_0^1$ defined by 
$$F(\eta,{\bf v}):=\left((I-b\Delta)^{-1}\nabla\cdot ({\bf v}+\eta{\bf v}), (I-b\Delta)^{-1}\nabla (\eta+\frac{1}{2}|{\bf v}|^2+c\Delta\eta) \right).$$
${\bf F}$ is well defined on $X\times{\bf H}_0^1$, since, by the Sobolev imbedding theorem, $\eta{\bf v}\in {\bf H}^1$ and $|{\bf v}|^2\in {\bf L}^2$. Hence, $(I-b\Delta)^{-1}\nabla\cdot ({\bf v}+\eta{\bf v})\in X$ and $(I-b\Delta)^{-1} \nabla (\eta+\frac{1}{2} |{\bf v}|^2+c\Delta\eta)\in {\bf H}^1_0$. Moreover, $F$ is $C^1$ on $X\times {\bf H}_0^1$, with derivative $F'(\eta^{\ast}, {\bf v}^{\ast})$ given by 
$$F'(\eta^{\ast},{\bf v}^{\ast})(\eta,{\bf v})=\left((I-b\Delta)^{-1}\nabla\cdot ({\bf v}+\eta{\bf v}^{\ast}+\eta^{\ast}{\bf v}), (I-b\Delta)^{-1}\nabla (\eta+{\bf v}^{\ast}\cdot {\bf v}+c \Delta\eta) \right). $$
The continuity of $F'$ follows from the Sobolev imbedding theorem and the regularity properties of the operators $(I-b\Delta)^{-1}$, considered as inverses of $(I-b\Delta)$ on $X$ for the first component, and of $I-b\Delta$ on ${\bf H}^2\cap {\bf H}_0^1$ for the second.

By the standard theory of ordinary differential equations in Banach spaces, we conclude therefore that there exists a unique maximal solution $(\eta,{\bf v})\in C^1([0,T];X)\times C^1([0,T];{\bf H}_0^1)$ of (\ref{E2.4}) for some $T>0$, with $\eta|_{t=0}=\eta_0$ and ${\bf v}|_{t=0}={\bf v}_0$. The first conclusion of the theorem follows. The assertion on $\frac{\partial^k\eta}{\partial t^k}$, $\frac{\partial^k{\bf v}}{\partial t^k}$ follows by differentiating (\ref{E2.4}) with respect to $t$ $k-1$ times.
\end{proof}

{\bf Remark 2.1} It is not hard to see, using the energy method on the nondimensional but scaled system (\ref{E1.2}) (with right-hand side replaced by zero), that at least in the cases of homogeneous Dirichlet boundary conditions for $\eta$ and ${\bf v}$ and reflective boundary conditions $\frac{\partial\eta}{\partial n}=0$, ${\bf v}=0$ on $\partial\Omega$, the maximum existence time $T_{\varepsilon}$ is independent of $\varepsilon$. 

{\bf Remark 2.2} If the domain $\Omega$ is smooth, one gets smooth solutions from smooth data. Namely, let $k\in \mathbb{N}, \;k\geq 3$, and assume that  $\eta_0\in X\cap H^k$, ${\bf v}_0\in {\bf H}_0^1\cap {\bf H}^{k-1}$, then  $(\eta,{\bf v})\in C([0,T]; X\cap H^k)\cap C([0,T]; {\bf H}_0^1\cap {\bf H}^{k-1})$. This follows directly from the elliptic regularity estimates on $I-b\Delta$ (with the {\em ad hoc} boundary conditions). One is thus reduced to an ODE in $X\cap H^k\times {\bf H}_0^1\cap {\bf H}^{k-1}.$

{\bf Remark 2.3} Consider the ibvp (\ref{E2.1})--(\ref{E2.2}) for the Bona-Smith system with (normal) reflective boundary conditions 
\begin{equation}\label{E2.5}
\frac{\partial \eta}{\partial n}=0,\quad {\bf v}=0,\qquad \mbox{for}\quad {\bf x}\in\partial\Omega,\,t\in\mathbb{R}_{+}.
\end{equation}
It is known that in one dimension, cf. \cite{ADMI}, the Hamiltonian of the ibvp (\ref{E2.1}), (\ref{E2.2}), (\ref{E2.5}),
\begin{equation}\label{E2.6}
H=H(\eta,{\bf v})=\frac{1}{2}\int_{\Omega} \left[\eta^2+(1+\eta)|{\bf v}|^2-c|\nabla\eta|^2\right]d{\bf x},
\end{equation}
is conserved. (Indeed this implies global existence-uniqueness of the solution of this ibvp in 1D provided that $\eta_0(x)+1>0$ and $H(\eta_0,u_0)$ is suitably restricted.) In the two-dimensional case $H$ is not conserved in general. To see this, write the system (\ref{E2.1}) for $({\bf x},t)\in \Omega\times \mathbb{R}_{+}$ as
\begin{subequations}\label{E2.7}
\begin{gather}
\eta_t+\nabla\cdot {\bf P}=0,\label{E2.7a}\\
{\bf v}_t+\nabla Q+b\nabla\times \boldsymbol{\omega}_t=0,\label{E2.7b} 
\end{gather}
\end{subequations}
where ${\bf P}:={\bf v}+\eta{\bf v}-b\nabla\eta_t$, $Q:=\eta+\frac{1}{2}|{\bf v}|^2+c\Delta\eta-b\nabla\cdot {\bf v}_t$, and $\boldsymbol{\omega}$ is the vorticity of the flow given by $\boldsymbol{\omega}=\nabla\times {\bf v}=(0,0,\omega)$, $\omega:=v_x-u_y$. From (\ref{E2.7}) we obtain 
\begin{equation}\label{E2.8}
\int_{\Omega}\left[\eta_t Q+{\bf v}_t\cdot {\bf P}+\nabla\cdot (Q{\bf P})+b{\bf P}\cdot (\nabla\times \boldsymbol{\omega}_t) \right]d{\bf x}=0
\end{equation}
Using now the reflective boundary conditions in (\ref{E2.8}) and integrating by parts we see that in the maximal temporal interval of existence of a solution of the ibvp (\ref{E2.1}), (\ref{E2.2}), (\ref{E2.5}),
\begin{equation}\label{E2.9}
\frac{dH}{dt}+b\int_{\Omega}{\bf P}\cdot (\nabla\times \boldsymbol{\omega}_t) d{\bf x}=0.
\end{equation}
Hence, a simple sufficient condition for the conservation of the Hamiltonian is irrotationality of the flow. In one space dimension, the flow is trivially irrotational. In 2D taking the $\curl$ of (\ref{E2.1a}) we see that 
\begin{equation}\label{E2.10}
\partial_t (\omega-b\Delta \omega)=0,\quad t>0.
\end{equation}
Now, if the flow is, for example, irrotational at $t=0$, the reflective boundary conditions do not allow us to conclude from (\ref{E2.10}) that $\omega=0$ for $t>0$. However, in the case of the Cauchy problem or the ibvp with {\em periodic} boundary conditions on $\eta$ and ${\bf v}$ on a rectangle (wherein (\ref{E2.8}) holds as well), $\omega(0)=0$ implies by (\ref{E2.10}) that $\omega=0$ for $t>0$. (This was also noticed for the BBM-BBM system in \cite{CI}.) In these cases, it follows by (\ref{E2.9}) that the Hamiltonian is invariant. Note however that conservation of $H$ does not imply global well-posedness in 2D.

\section{A modified Galerkin method}
\setcounter{equation}{0}

We turn now to the numerical solution of the ibvp (\ref{E2.1}), (\ref{E2.2}), (\ref{E2.5}) by Galerkin-finite element methods.
The usual Galerkin method for Bona-Smith systems that was analyzed in \cite{DMSI} requires finite element subspaces consisting of $C^2$ functions. Hence, it is not very useful in practice, as it cannot be applied to arbitrary plane domains and triangulations. In this section a {\em modified Galerkin method} for the numerical solution of the ibvp  with (normal) reflection boundary conditions is analyzed. The method is more versatile, being applicable to general triangulations and domains (even with piecewise linear continuous functions) and is shown to have error estimates with optimal convergence rates in $L^2$ and $H^1$.

For ease of reference we rewrite here the ibvp that we will approximate. We seek $\eta$ and ${\bf v}$, defined for $({\bf x},t)\in \Omega\times [0,T]$ and satisfying for constants $b>0$ and $c\leq 0$
\begin{equation}\label{ER}
\begin{array}{l}
\begin{array}{l}
\eta_t+\nabla\cdot {\bf v}+\nabla\cdot \eta{\bf v}-b\Delta \eta_t=0,\\
{\bf v}_t+\nabla (\eta+\frac{1}{2}|{\bf v}|^2+c\Delta\eta)-b\Delta {\bf v}_t=0,
\end{array} ({\bf x},t)\in \Omega\times [0,T],\\
\eta({\bf x},0)=\eta_0({\bf x}),\,\,{\bf v}({\bf x},0)={\bf v}_0({\bf x}),\,\,\, {\bf x}\in \Omega,\\
\frac{\partial\eta}{\partial n}({\bf x},t)=0,\,\, {\bf v}({\bf x},t)=0,\,\, ({\bf x},t)\in \partial\Omega\times [0,T].
\end{array}\tag{${\mathcal R}$}
\end{equation}
We assume that $\Omega$ is a convex domain with smooth enough boundary $\partial\Omega$ and that the ibvp (\ref{ER}) has a unique solution $(\eta,{\bf v})$ which is smooth enough for the purposes of its numerical approximation. In the sequel, we put ${\bf x}=(x,y)$.

We suppose that ${\cal T}_h$ is a regular triangulation of $\Omega$ with triangles $\tau$ of maximum sidelength $h$ and let $\widetilde{S}_h$ be a finite-dimensional subspace of $C(\Bar{\Omega})\cap H^1$, which, for small enough $h$ and integer $r\geq 2$, satisfies the approximation property
\begin{equation}\label{E3.1}
\underset{\chi\in \widetilde{S}_h}{\operatorname{\inf}} \left\{\|w-\chi\|+h\|w-\chi\|_1 \right\}\leq Ch^s\|w\|_s,\quad 1\leq s\leq r,
\end{equation}
when $w\in H^s$. (In this section $C$ will denote generic constants independent of $h$.) On the same triangulation we denote by $S_h$ the subspace of $\widetilde{S}_h$ consisting of the elements of $\widetilde{S}_h$ that vanish on the boundary i.e. $S_h=\widetilde{S}_h\cap H^1_0$. Hence, on $S_h$ we have 
\begin{equation}\label{E3.2}
\underset{\chi\in S_h}{\operatorname{\inf}} \left\{\|w-\chi\|+h\|w-\chi\|_1 \right\}\leq Ch^s\|w\|_s,\quad 1\leq s\leq r,
\end{equation}
for $w\in H^s\cap H^1_0$. We will assume that the elements of $S_h$ and  $\widetilde{S}_h$ are piecewise polynomial functions defined on ${\cal T}_h$, of degree at most $r-1$ on each $\tau\in {\cal T}_h$.

We consider the symmetric bilinear form $a_D:H_0^1\times H_0^1\rightarrow\mathbb{R}$ defined by 
\begin{equation}\label{E3.3}
a_D(u,v):=(u,v)+b(\nabla u,\nabla v),\quad u,v\in H^1_0,
\end{equation}
which is coercive on $H^1_0\times H^1_0$. In addition, we consider the symmetric bilinear form $a_N:H^1\times H^1\rightarrow \mathbb{R}$ that is defined by
\begin{equation}\label{E3.4}
a_N(u,v):=(u,v)+b(\nabla u,\nabla v),\quad u,v\in H^1,
\end{equation}
and is coercive on $H^1\times H^1$. With the aid of $a_D$, $a_N$ we define the {\em elliptic projection operators} $R_h:H^1_0\rightarrow S_h$, $\widetilde{R}_h:H^1\rightarrow \widetilde{S}_h$ as follows:
\begin{subequations}\label{E3.5}
\begin{gather}
a_D(R_hw,\chi)=a_D(w,\chi),\qquad \forall w\in H^1_0,\quad \chi\in S_h, \label{E3.5a}\\
a_N(\widetilde{R}_hw,\chi)=a_N(w,\chi),\qquad \forall w\in H^1,\quad \chi\in \widetilde{S}_h. \label{E3.5b}
\end{gather}
\end{subequations}
As a consequence of (\ref{E3.1}), (\ref{E3.2}) and elliptic regularity we have then
\begin{subequations}\label{E3.6}
\begin{equation}
\|w-R_hw\|_k\leq Ch^{s-k}\|w\|_{s},\quad w\in H^s\cap H^1_0,\,2\leq s\leq r,\,\, k=0,1, \label{E3.6a}
\end{equation}
and
\begin{equation} \label{E3.6b}
\|w-\widetilde{R}_hw\|_k\leq Ch^{s-k}\|w\|_{s},\quad w\in H^s,\,2\leq s\leq r,\,\, k=0,1. \end{equation}
\end{subequations}
We assume that the triangulation ${\cal T}_h$ is quasiuniform. Then, the following inverse assumptions hold on $S_h$, \cite{Ci}
\begin{gather}
\|\chi\|_1\leq Ch^{-1}\|\chi\|,\quad \forall \chi\in S_h,\label{E3.7}\\
\|\chi\|_{\infty} \leq Ch^{-1}\|\chi\|,\quad \forall \chi\in S_h,\label{E3.8}
\end{gather}

The same inverse assumptions hold on $\widetilde{S}_h$ as well.

We will also assume that for the elliptic projections we have the following approximation properties in the $L^{\infty}$ norm:
\begin{subequations}\label{E3.10}
\begin{gather}
\|w-R_h w\|_{\infty}\leq C\gamma (h)\|w\|_{r,\infty},\quad \forall w\in W^r_{\infty}\cap H^1_0,\label{E3.10a}\\
\|w-\widetilde{R}_h w\|_{\infty}\leq C\gamma (h)\|w\|_{r,\infty},\quad \forall w\in W^r_{\infty},\label{E3.10b}
\end{gather}
\end{subequations}
where $\gamma(h)=h^r|\log h|^{\bar{r}}$ with $\bar{r}=0$ if $r>2$ and $\bar{r}=1$ when $r=2$, cf. \cite{S}. Here, $W^r_{\infty}$ (with norm $\|\cdot\|_{r,\infty}$) denotes the $L^{\infty}$-based Sobolev space on $\Omega$ of order $r$.

In the sequel we will use the {\em discrete Laplacian operator} $\widetilde{\Delta_h}:H^1\rightarrow \widetilde{S}_h$, defined for $w\in H^1$ by
\begin{equation}\label{E3.12}
(\widetilde{\Delta}_hw,\chi)=-(\nabla w,\nabla \chi),\quad \forall \chi\in \widetilde{S}_h.
\end{equation}
By the divergence theorem, it is easy to check that for $w\in H^2$ there holds
\begin{equation}\label{E3.13}
(\widetilde{\Delta}_h\widetilde{R}_h w,\chi)=(\Delta w,\chi)-\int_{\partial\Omega}\frac{\partial w}{\partial n}\chi{\rm ds}-\frac{1}{b}(w,\chi)+\frac{1}{b}(\widetilde{R}_h w,\chi),\quad \chi\in\widetilde{S}_h.
\end{equation}
Hence, if $w\in H^2$ with $\frac{\partial w}{\partial n}|_{\partial\Omega}=0$, then
\begin{equation}\label{E3.14}
\widetilde{\Delta}_h\widetilde{R}_h w=\widetilde{P}_h\Delta w+\frac{1}{b}(\widetilde{R}_h-\widetilde{P}_h)w,
\end{equation}
where $\widetilde{P}_h$ is the $L^2$-projection onto $\widetilde{S}_h$. The identity (\ref{E3.13}) is proved as follows: For $w\in H^2$, $\chi\in \widetilde{S}_h$, using the definition of $\widetilde{R}_h$ we have 
\begin{eqnarray*}
(\widetilde{\Delta}_h\widetilde{R}_h w,\chi) &=& -(\nabla \widetilde{R}_h w,\nabla\chi)-\frac{1}{b}(\widetilde{R}_h w,\chi)+\frac{1}{b}(\widetilde{R}_h w,\chi)\\
& =& -(\nabla w,\nabla \chi)-\frac{1}{b}(w,\chi)+\frac{1}{b}(\widetilde{R}_h w,\chi),\\
&=& -\int_{\partial\Omega}\frac{\partial w}{\partial n}\chi ds+(\Delta w,\chi)-\frac{1}{b}(w,\chi)+\frac{1}{b}(\widetilde{R}_hw,\chi).
\end{eqnarray*}
Denoting the components of ${\bf v}$ as $(u,v)$, we define the semidiscrete modified Galerkin method as follows. We seek $\eta_h: [0,T]\rightarrow \widetilde{S}_h$, $u_h,v_h:[0,T]\rightarrow S_h$, approximations to $\eta,u,v$, respectively, such that
\begin{equation}\label{E3.15}
\begin{array}{c}
a_N({\eta_{h}}_t,\phi)+({u_h}_x,\phi)+({v_h}_y,\phi)+((\eta_h u_h)_x,\phi)+((\eta_h v_h)_y,\phi)=0,\quad \phi\in \widetilde{S}_h,\\
a_D({u_h}_t,\chi)+({\eta_h}_x,\chi)+(u_h{u_h}_x,\chi)+(v_h{v_h}_x,\chi)+c((\widetilde\Delta_h\eta_h)_x,\chi)=0,\quad \chi\in S_h,\\
a_D({v_h}_t,\psi)+({\eta_h}_y,\psi)+(u_h{u_h}_y,\psi)+(v_h{v_h}_y,\psi)+c((\widetilde{\Delta}_h\eta_h)_y,\psi)=0,\quad \psi\in S_h,\\
\eta_h(\cdot,0)=\widetilde{R}_h\eta_0,\,\, u_h(\cdot,0)=R_h u_0,\,\, v_h(\cdot,0)=R_h v_0.
\end{array}\,\, 0\leq t\leq T,
\end{equation}
These relations are discrete analogs to the corresponding variational forms of the first p.d.e. of (\ref{ER}) in $H^1$ and of the second in $(H^1_0)^2$. Note that since $\eta_hu_h\in C(\bar{\Omega})$, $\eta_hu_h|_{\tau}\in C^{\infty}(\tau)$ for each $\tau\in{\cal T}_h$, and $\eta_h u_h|_{\partial\Omega}=0$, it follows that $\eta_hu_h\in H^1_0$. Similarly, $\eta_h v_h\in H^1_0$, $u_h^2\in H^1_0$, $v_h^2\in H^1_0$. Hence, all terms in the inner products of (\ref{E3.15}) are well defined. We now consider the mappings $\hat{f}_x,\hat{f}_y:L^2\rightarrow S_h$ defined for $w\in L^2$ by:
\begin{gather*}
a_D(\hat{f}_x(w),\phi)=(w,\phi_x),\quad \phi\in S_h,\\
a_D(\hat{f}_y(w),\phi)=(w,\phi_y),\quad \phi\in S_h,
\end{gather*}
and the mappings and $\hat{g}_x.\hat{g}_y:L^2\rightarrow \widetilde{S}_h$ defined for $w\in L^2$ by 
\begin{gather*}
a_N(\hat{g}_x(w),\chi)=(w,\chi_x),\quad \chi\in \widetilde{S}_h,\\
a_N(\hat{g}_y(w),\chi)=(w,\chi_y),\quad \chi\in \widetilde{S}_h.
\end{gather*}
Then (\ref{E3.15}) can be written as 
\begin{equation}\label{E3.16}
\begin{array}{l}
{\eta_h}_t=F(u_h,v_h,\eta_h),\\
{u_h}_t=G(u_h,v_h,\eta_h),\qquad \quad 0\leq t\leq T\\
{v_h}_t=Z(u_h,v_h,\eta_h),\\
\eta_h(0)=\widetilde{R}_h\eta_0,\,\, u_h(0)=R_h u_0,\,\, v_h(0)=R_h v_0,
\end{array}
\end{equation}
where 
\begin{gather*}
F(u_h,v_h,\eta_h):=\hat{g}_x(u_h)+\hat{g}_y(v_h)+\hat{g}_x(\eta_hu_h)+\hat{g}_y(\eta_h v_h),\\
G(u_h,v_h,\eta_h):=\hat{f}_x(\eta_h)+\frac{1}{2}(\hat{f}_x(u_h^2)+\hat{f}_x(v_h^2))+c\hat{f}_x(\widetilde{\Delta}_h\eta_h),\\
Z(u_h,v_h,\eta_h):=\hat{f}_y(\eta_h)+\frac{1}{2}(\hat{f}_y(u_h^2)+\hat{f}_y(v_h^2))+c\hat{f}_y(\widetilde{\Delta}_h\eta_h).
\end{gather*}
For the mappings $\hat{f}_x,\hat{f}_y,\hat{g}_x$ and $\hat{g}_y$ we have the following stability estimates:
\begin{lemma}\label{L3.1}
There exists a constant $C$ such that 
\begin{itemize}
\item[{\rm (i)}] $\|\hat{f}_x(w)\|_1\leq C\|w\|$, and $\|\hat{f}_y(w)\|_1\leq C\|w\|$, $w\in L^2$.
\item[{\rm (ii)}] $\|\hat{g}_x(w)\|_1\leq C\|w\|$, and $\|\hat{g}_y(w)\|_1\leq C\|w\|$, $w\in L^2$.
\end{itemize}
\end{lemma}
\begin{proof}
The proof follows immediately from the coercivity of $a_D$ on $H^1_0\times H^1_0$ and of $a_N$ on $H^1\times H^1$ and the definitions of $\hat{f}_x$, $\hat{f}_y$, $\hat{g}_x$, $\hat{g}_y$. 
\end{proof}

We define now the negative norms $\|\cdot\|_{-1}$ and $\|\cdot\|_{-2}$ for functions $w\in L^2$ as
$$\|w\|_{-1}:=\sup_{\substack{z\in H^1\\ z\not= 0}}\frac{(w,z)}{\|z\|_1}\,\,\,\mbox{and}\,\,\, \|w\|_{-2}:=\sup_{\substack{z\in H^2\cap H^1_0\\ z\not=0}}\frac{(w,z)}{\|z\|_2}.$$

\begin{lemma}\label{l3.2}
There exists a constant $C>0$ such that 
\begin{equation}\label{E3.17}
\|\hat{f}_x(\chi)\|\leq C\|\chi\|_{-1},\quad \|\hat{f}_y(\chi)\|\leq C\|\chi\|_{-1},\quad\forall\chi\in\widetilde{S}_h.
\end{equation}
\end{lemma}
\begin{proof}
Let $\chi\in\widetilde{S}_h$. Consider the problem $Lw=\chi_x$ with $w=0$ on $\partial\Omega$, where $L:=I-b\Delta$ with domain $H^2\cap H_0^1$. Then, by elliptic regularity and denoting by $L^{-1}$ the inverse of $L$, we have
\begin{eqnarray}
\|\chi_x\|_{-2} & = & \sup_{0\not= z\in H^2\cap H_0^1}\frac{(\chi_x,z)}{\|z\|_2}=\sup_{0\not= z\in H^2\cap H_0^1}\frac{(Lw,z)}{\|z\|_2}\nonumber\\
& \geq & \frac{(Lw,L^{-1} w)}{\|L^{-1}w\|_2}=\frac{\|w\|^2}{\|L^{-1}w\|_2}\geq C\frac{\|w\|^2}{\|w\|}\nonumber \\
&=& C\|w\|. \label{E3.18}
\end{eqnarray}

Let now $0\not= z\in H^2\cap H_0^1$. Then 
$$\frac{(\chi_x,z)}{\|z\|_2}=\frac{-(\chi,z_x)}{\|z\|_2}\leq\frac{\|\chi\|_{-1}\|z_x\|_1}{\|z\|_2}\leq \|\chi\|_{-1}.$$
Hence 
$$\sup_{0\not=z\in H^2\cap H_0^1}\frac{(\chi_x,z)}{\|z\|_2}\leq \|\chi\|_{-1}.$$
Therefore
$$\|\chi_x\|_{-2}\leq \|\chi\|_{-1},$$
and by (\ref{E3.17}) 
\begin{equation}\label{E3.19}
\|w\|\leq C\|\chi\|_{-1}.
\end{equation}

Consider $R_h w$, the elliptic projection of $w$ onto $S_h$. Note that $R_h w=\hat{f}_x(\chi).$ In addition, by (\ref{E3.6a})
$$\|R_h w-w\|\leq Ch^2\|w\|_2\leq Ch^2\|\chi_x\|\leq Ch^2 h^{-2}\|\chi\|_{-1}=C\|\chi\|_{-1}.$$
(In the last inequality we used the inverse inequality $\|\varphi\|_1\leq Ch^{-2}\|\varphi\|_{-1}$, $\forall\varphi\in \widetilde{S}_h$, which is valid since by (\ref{E3.7}): 
$$\|\varphi\|_{-1}=\sup_{0\not= z\in H^1}\frac{(\varphi,z)}{\|z\|_1}\geq \frac{\|\varphi\|^2}{\|\varphi\|_1}\geq Ch^2\frac{\|\varphi\|_1^2}{\|\varphi\|_1}=C h^2 \|\varphi\|_1.)$$
We conclude that 
$$\|R_h w\|-\|w\|\leq\|R_h w-w\|\leq C\|\chi\|_{-1},$$
and by (\ref{E3.19})
$$\|\hat{f}_x(\chi)\|=\|R_h w\|\leq C\|\chi\|_{-1},$$
which is the required conclusion. The proof for $\hat{f}_y$ is entirely analogous.
\end{proof}
\begin{lemma}\label{L3.3}
There exists a constant $C$ such that 
\begin{equation}\label{E3.20}
\|\widetilde{\Delta}_h u\|_{-1}\leq C\|\nabla u\|,\quad \forall u\in H^1.
\end{equation}
\end{lemma}
\begin{proof}
Let $u\in H^1$. Then, for each $w\not=0\in H^1$ we have
$$\frac{(\widetilde{\Delta}_h u,w)}{\|w\|_1}=\frac{(\widetilde{\Delta}_h u,\widetilde{P}_h w)}{\|w\|_1}\leq \frac{\|\nabla u\|\|\nabla \widetilde{P}_h w\|}{\|w\|_1}\leq C\|\nabla u\|,$$
from which, taking the supremum over $w\in H^1$ we obtain (\ref{E3.20}). We used the stability of $\widetilde{P}_h$ on $H^1$, i.e. the inequality 
$$\|\widetilde{P}_h w\|_1\leq C\|w\|_1,\quad \mbox{for}\,w\in H^1,$$
which follows by the argument in \cite{CT}.
\end{proof}

We now state and prove the main result of this section.

\begin{theorem}\label{T3.1}
For $h$ sufficiently small, the semidiscrete problem (\ref{E3.16}) has a unique solution $(\eta_h, u_h, v_h)$ in the interval $[0,T]$ of maximal existence of the solution $(\eta,u,v)$ of the ibvp (\ref{ER}).
Moreover, for some constant $C=C(\eta,u,v,T)$ independent of $h$ we have 
$$\|\eta-\eta_h\|+\|u-u_h\|+\|v-v_h\|\leq Ch^{r},$$
and 
$$\|\eta-\eta_h\|_1+\|u-u_h\|_1+\|v-v_h\|_1\leq Ch^{r-1},$$
for each $t\in [0,T]$.
\end{theorem}
\begin{proof}
We suppose that for some constant $M$ there holds that $\|\eta\|_{\infty}\leq M$, $\|u\|_{\infty}\leq M$ and 
$\|v\|_{\infty}\leq M$ for $0\leq t\leq T$. Then, from (\ref{E3.10b}) for $h$ sufficiently small 
$$\|\eta_h^0\|_{\infty}\leq\|\eta_h^0-\eta_0\|_{\infty}+\|\eta_0\|_{\infty}=\|\widetilde{R}_h\eta_0-\eta_0\|_{\infty}+\|\eta_0\|_{\infty}\leq C\gamma(h)\|\eta_0\|_{r,\infty}+\|\eta_0\|_{\infty}<2 M.$$
Similar estimates hold for $u_h^0$ and $v_h^0$. The o.d.e. system (\ref{E3.16}) has a unique solution locally in $t$. By continuity, we may assume that there exists $t_h\in (0,T]$ such that $\|u_h\|_{\infty}\leq 2M$, $\|\eta_h\|_{\infty}\leq 2M$ and $\|v_h\|_{\infty}\leq 2M$ for all $t\leq t_h$.

We let now 
$$\rho=\eta-\widetilde{R}_h\eta,\,\,\theta=\widetilde{R}_h\eta-\eta_h,\,\, \tau=v-R_h v,\,\,\zeta=R_hv-v_h,\,\,\sigma=u-R_hu,\,\,\xi= R_hu-u_h,$$
so that $\theta\in\widetilde{S}_h$, $\zeta,\xi\in S_h$ and $\eta-\eta_h=\rho+\theta$, $u-u_h=\sigma+\xi$, $v-v_h=\tau+\zeta$. By (\ref{ER}) and (\ref{E3.16}) we have 
\begin{eqnarray}
\theta_t & = & \hat{g}_x(\sigma+\xi)+\hat{g}_y(\tau+\zeta)+ \hat{g}_x(u\eta-u_h\eta_h)+\hat{g}_y(v\eta-v_h\eta_h),\label{E3.21}\\
\xi_t & = & \hat{f}_x(\theta+\rho)+ \frac{1}{2}\left\{ \hat{f}_x(u^2)-\hat{f}_x(u_h^2)+\hat{f}_x(v^2)-\hat{f}_x(v_h^2)\right\} +c\hat{f}_x(\Delta\eta-\widetilde{\Delta}_h\eta_h),\label{E3.22}\\
\zeta_t & = & \hat{f}_y(\tau+\zeta)+\frac{1}{2}\left\{\hat{f}_y(u^2)-\hat{f}_y(u_h^2)+\hat{f}_y(v^2)-\hat{f}_y(v_h^2)\right\}+c\hat{f}_y(\Delta\eta-\widetilde{\Delta}_h\eta_h).\label{E3.23}
\end{eqnarray}
The equation (\ref{E3.22}) may be written as
$$\xi_t=\hat{f}_x(\theta+\rho)+\frac{1}{2}\left\{ \hat{f}_x(u(\sigma+\xi))+\hat{f}_x((\sigma+\xi)u_h)+\hat{f}_x(v(\tau+\zeta))+\hat{f}_x((\tau+\zeta)v_h)\right\}+c\hat{f}_x(\Delta\eta-\widetilde{\Delta}_h\eta_h).$$
Taking $L^2$-norms and using Lemma \ref{L3.1}, (\ref{E3.17}), (\ref{E3.14}), (\ref{E3.6a}), (\ref{E3.6b}) and (\ref{E3.20}) we obtain, for $0\leq t\leq t_h$,
\begin{eqnarray*}
\|\xi_t\| &\leq & \|\hat{f}_x(\theta+\rho)\|+\frac{1}{2}\left(\|\hat{f}_x(u(\sigma+\xi))\|+\right.\\
& &\left. \|\hat{f}_x((\sigma+\xi)u_h)\|+\|\hat{f}_x(v(\tau+\zeta))\|+\|\hat{f}_x((\tau+\zeta)v_h)\|\right)
 +\|c\hat{f}_x(\Delta\eta-\widetilde{\Delta}_h\eta_h)\|\\
&\leq & C(\|\theta+\rho\|+\|u(\sigma+\xi)\|+\|(\sigma+\xi)u_h\|+\|v(\tau+\zeta)\|+\|(\tau+\zeta)v_h\|+\\
& &\|\hat{f}_x(\Delta\eta-\widetilde{\Delta}_h \widetilde{R}_h\eta)\|+\|\hat{f}_x(\widetilde{\Delta}_h(\widetilde{R}_h\eta-\eta_h)\|)\\
&\leq & C[\|\theta\|+\|\rho\|+\|u\|_{L^{\infty}}(\|\sigma\|+\|\xi\|)+\|v\|_{L^{\infty}}(\|\tau\|+\|\zeta\|)+\\
& &\|v_h\|_{L^{\infty}}(\|\tau\|+\|\zeta\|)+\|u_h\|_{L^{\infty}}(\|\sigma\|+\|\xi\|)+\|\Delta\eta-
\widetilde{\Delta}_h\widetilde{R}_h\eta\|+\|\widetilde{\Delta}_h(\widetilde{R}_h\eta-\eta_h)\|_{-1}]\\
&\leq & C\left[ h^r+\|\theta\|+\|\xi\|+\|\zeta\|+\|\Delta\eta-\widetilde{P}_h\Delta\eta\|+\|\widetilde{R}_h\eta-\widetilde{P}_h\eta\|+\|\widetilde{R}_h\eta-\eta_h\|_1 \right],
\end{eqnarray*}
and so
\begin{equation}\label{E3.24}
\|\xi_t\|\leq C(h^r+\|\theta\|_1+\|\xi\|+\|\zeta\|).
\end{equation}
Similarly, we have for $0\leq t\leq t_h$
\begin{equation}\label{E3.25}
\|\zeta_t\|\leq C(h^r+\|\theta\|_1+\|\xi\|+\|\zeta\|).
\end{equation}
The equation (\ref{E3.21}) may be written as
$$\theta_t=\hat{g}_x(\sigma+\xi)+\hat{g}_y(\tau+\zeta)+\hat{g}_x(u(\rho+\theta))+\hat{g}_x((\sigma+\xi)\eta_h)+\hat{g}_y(v(\rho+\theta))+\hat{g}_y((\tau+\zeta)\eta_h).$$
Hence, taking $H^1$-norms and using Lemma \ref{L3.1}, and (\ref{E3.6}a,b) we have for $0\leq t\leq t_h$ 
\begin{eqnarray*}
\|\theta_t\|_1 & \leq & C(\|\sigma+\xi\|+\|\tau+\zeta\|\\
& & +\|u(\rho+\theta)\|+\|(\sigma+\xi)\eta_h\|+\|v(\rho+\theta)\|+\|(\tau+\zeta)\eta_h\|)\\
&\leq & C(\|\sigma\|+\|\xi\|+\|\tau\|+\|\zeta\|+\|u\|_{L^{\infty}}(\|\rho\|+\|\theta\|)+(\|\sigma\|+\|\xi\|)\|\eta_h\|_{L^{\infty}}\\
& &+\|v\|_{L^{\infty}}(\|\rho\|+\|\theta\|)+(\|\tau\|+\|\zeta\|)\|\eta_h\|_{L^{\infty}})\\
& \leq & C(\|\sigma\|+\|\xi\|+\|\tau\|+\|\zeta\|+\|\rho\|+\|\theta\|),
\end{eqnarray*}
and therefore that 
\begin{equation}\label{E3.26}
\|\theta_t\|_1\leq C(h^r+\|\theta\|+\|\xi\|+\|\zeta\|),\quad 0\leq t\leq t_h.
\end{equation}
From (\ref{E3.24})--(\ref{E3.26}) we see that for $0\leq t\leq t_h$
$$\frac{1}{2}\frac{d}{dt}(\|\theta\|_1^2+\|\xi\|^2+\|\zeta\|^2)\leq Ch^{2r}+C(\|\theta\|_1^2+\|\xi\|^2+\|\zeta\|^2),$$
from which, using Gronwall's inequality, we have for $0\leq t\leq t_h$
\begin{equation}\label{E3.27}
\|\theta\|_1+\|\xi\|+\|\zeta\|\leq Ch^r.
\end{equation}
Hence, for $t\leq t_h$ there holds
\begin{equation}\label{E3.28}
\|\eta-\eta_h\|+\|u-u_h\|+\|v-v_h\|\leq Ch^r.
\end{equation}
Furthermore for $t\leq t_h$, we have, by (\ref{E3.8}), (\ref{E3.10b}) and (\ref{E3.27})
\begin{eqnarray*}
\|\eta_h-\eta\|_{L^{\infty}} & \leq & \|\eta_h-\widetilde{R}_h\eta\|_{L^{\infty}}+\|\widetilde{R}_h\eta-\eta\|_{L^{\infty}}\\
&\leq & Ch^{-1}\|\eta_h-\widetilde{R}_h\eta\|+C\gamma(h)\\
& =& Ch^{-1}\|\theta\|_1+C\gamma(h)\\
&\leq & Ch^{r-1}
\end{eqnarray*}
Therefore, $\|\eta_h\|_{L^{\infty}}\leq \|\eta\|_{L^{\infty}}+\|\eta_h-\eta\|_{L^{\infty}}\leq Ch^{r-1}+M< 2M$ for $h$ sufficiently small. Similar estimates hold for $u_h$, $v_h$. These contradict the maximal property of $t_h$ and we conclude that we may take $t_h=T$. Hence (\ref{E3.28}) holds up to $t=T$, giving the desired optimal-rate $L^2$-estimate. The $O(h^{r-1})$ $H^1$ estimate follows easily by (\ref{E3.6}a,b) and (\ref{E3.7}).
\end{proof}

It is worthwhile to note that temporal derivatives of arbitrary order of the semidiscrete solution $(\eta_h,u_h,v_h)$ are bounded on $[0,T]$ by constants independent of $h$, as the following proposition shows.

\begin{proposition}\label{P3.1}
For $h$ sufficiently small, let $(\eta_h, u_h, v_h)$ be the solution of the semidiscrete problem (\ref{E3.16}) for $t\in [0,T]$. Then, for $j=0,1,2,3,\ldots$, there exist constants $C_j$ independent of $h$, such that 
\begin{equation}\label{E3.29}
\max_{0\leq t\leq T}\left( \|\partial_t^j\eta_h\|_1+\|\partial_t^j u_h\|+\|\partial_t^j v_h\|\right)\leq C_j.
\end{equation}
\end{proposition}

\begin{proof}
From Theorem \ref{T3.1} we have, for some constant $C_0$ independent of $h$,
\begin{equation}\label{E3.30}
\max_{0\leq t\leq T}\left(\|\eta_h\|_1+\|u_h\|+\|v_h\| \right)\leq C_0.
\end{equation}
Now, from (\ref{E3.16}), (ii) of Lemma \ref{L3.1}, and (\ref{E3.30}), there follows for $0\leq t\leq T$ 
\begin{align}
\|{\eta_h}_t\|_1 & \leq \|\hat{g}_x(u_h)\|_1+\|\hat{g}_y(v_h)\|_1+\|\hat{g}_x(\eta_h u_h)\|_1+\|\hat{g}_y(\eta_h v_h)\|_1\nonumber\\
& \leq C(\|u_h\|+\|v_h\|+\|\eta_h u_h\|+\|\eta_h v_h\|)\nonumber\\
& \leq C(1+\|\eta_h\|_{\infty}).\label{E3.31}
\end{align}
In addition, from (\ref{E3.16}), (i) of Lemma \ref{L3.1}, (\ref{E3.17}), and (\ref{E3.30}) we have for $0\leq t\leq T$
\begin{align}
\|{u_h}_t\| & \leq \|\hat{f}_x(\eta_h)\|+\frac{1}{2}\|\hat{f}_x(u_h^2)\|+\frac{1}{2}\|\hat{f}_x(v_h^2)\|+|c|\|\hat{f}_x(\widetilde{\Delta}_h\eta_h)\| \nonumber\\
& \leq C(\|\eta_h\|+\|u_h^2\|+\|v_h^2\|+\|\widetilde{\Delta}_h\eta_h\|_{-1}) \nonumber\\
& \leq C(1+\|u_h\|_{\infty}\|u_h\|+\|v_h\|_{\infty}\|v_h\|+\|\eta_h\|_1) \nonumber\\
& \leq C(1+\|u_h\|_{\infty}+\|v_h\|_{\infty}).\label{E3.32}
\end{align}
A similar inequality holds for $\|{v_h}_t\|$. Now, from the closing argument of the proof of Theorem 3.1 we may infer that
\begin{equation}\label{E3.33}
\max_{0\leq t\leq T} (\|\eta_h\|_{\infty}+\|u_h\|_{\infty}+\|v_h\|_{\infty})\leq C,
\end{equation}
and, consequently, in view of (\ref{E3.31}) and (\ref{E3.32}), the validity of (\ref{E3.29}) for $j=1$. Differentiating now the equations in (\ref{E3.16}) with respect to $t$ and using again Lemma \ref{L3.1}, (\ref{E3.29}) for $j=1$, and (\ref{E3.33}) we see that (\ref{E3.29}) holds for $j=2$ as well. 

If we take the second temporal derivative of both sides of the first o.d.e. in (\ref{E3.16}), we see that in order to obtain a bound for $\|\partial_t^3\eta_h\|_1$, we need, in addition to already established estimates, an $h$-independent bound for $\|{\eta_h}_t\|_{\infty}$. This is obtained as follows: For $0\leq t\leq T$ we have, for $\theta=\widetilde{R}_h\eta-\eta_h$, from (\ref{E3.8}), (\ref{E3.10b}), and (\ref{E3.26}), that 
$
\|{\eta_h}_t\|_{\infty}  \leq \|\theta_t\|_{\infty}+\|\widetilde{R}_h\eta_t\|_{\infty}
 \leq Ch^{-1}\|\theta_t\|+C\gamma(h)+C
 \leq C. $
We similarly get $h$-independent bounds for $\|{u_h}_t\|_{\infty}$ and $\|{v_h}_t\|_{\infty}$ that yield in turn similar bounds for $\|\partial_t^3 u_h\|$ and $\|\partial_t^3 v_h\|$. Hence (\ref{E3.29}) holds for $j=3$ too. The case $j=4$ follows immediately, as it does not need any $L^{\infty}$ bounds on temporal derivatives of the semidiscrete approximations of order higher than one. 
To obtain (\ref{E3.29}) for $j=5$ one needs, in addition to already established bounds, $h$-independent bounds for
$\|\partial_t^2\eta_h\|_{\infty}$, $\|\partial_t^2 u_h\|_{\infty}$, and $\|\partial_t^2 v_h\|_{\infty}$ on $[0,T]$. These may be derived as follows: Differentiate with respect to $t$ the expression for $\theta_t$ after (\ref{E3.25}) and use the uniform bound on $\|{\eta_h}_t\|_{\infty}$ and (\ref{E3.24})--(\ref{E3.27}) to obtain that $\|\theta_{tt}\|_1\leq Ch^r$. The required bound for $\|\partial_t^2\eta_h\|_{\infty}$ follows then from (\ref{E3.8}) and (\ref{E3.10b}). Similarly, differentiating e.g. the expression for $\xi_t$ after (\ref{E3.23}) and using the bounds on $\|{u_h}_t\|_{\infty}$, $\|{v_h}_t\|_{\infty}$ and (\ref{E3.24})--(\ref{E3.27}) we obtain that $\|\xi_{tt}\|\leq Ch^r$, from which $\|\partial_t^2 u_h\|_{\infty}\leq C$ follows. The case $j=6$ requires no additional $L^{\infty}$ bounds.

We continue by induction. If $j=2k+1$, $L^{\infty}$-bounds for $\partial_t^k\eta_h$, $\partial_t^ku_h$, $\partial_t^kv_h$ are found by differentiating the expressions for $\theta_t$, $\xi_t$ and $\zeta_t$ and using previously established bounds. The even case $j=2k+2$ requires no additional $L^{\infty}$ bounds. (As a corrolary from the above proof it follows also that 
\begin{equation*}
\max_{0\leq t\leq T} (\|\partial_t^j\eta_h\|_{\infty}+\|\partial_t^j u_h\|_{\infty}+\|\partial_t^j v_h\|_{\infty})\leq C_j',
\end{equation*}
holds for $j=0,1,2,\ldots$, where $C_j'$ are constants independent of $h$.)
\end{proof}

From the result of this proposition we see that the system of o.d.e.'s (\ref{E3.16}) is {\em not stiff}. Therefore, one may use {\em explicit} time-stepping schemes to discretize (\ref{E3.16}) in the temporal variable without imposing stability mesh conditions on the time step $\Delta t$ in terms of $h$. Error estimates of optimal order in space and time in the case of explicit Runge-Kutta full discretizations may be established along the lines of the proof of Proposition 10 of \cite{ADMII}.

{\bf Remark 3.1} The $H^1$ error estimate in Theorem \ref{T3.1} may be strengthened as follows. For $w\in H^1$ define the discrete norm $\|\cdot\|_{2,h}$ as 
$$\|w\|_{2,h}:=(\|w\|_1^2+\|\widetilde{\Delta}_h w\|^2)^{1/2},\quad \forall w\in H^1.$$
Then, by letting $w\in H^1_0$ and considering the boundary-value problem $f-b\Delta f=-w_x$ in $\Omega$ with $\frac{\partial f}{\partial n}=0$ on $\partial\Omega$, we easily see that $\hat{g}_x(w)=\widetilde{R}_h f$. A straightforward computation using (\ref{E3.14}) yields then that $\|\hat{g}_x(w)\|_{2,h}^2\leq C\|w_x\|^2$, $\|\hat{g}_y(w)\|_{2,h}^2\leq C\|w_y\|^2$. We may take now the $\|\cdot\|_{2,h}$ norm in (\ref{E3.21}) and obtain 
$$\|\theta_t\|_{2,h}\leq C(h^r+\|\theta\|+\|\xi\|+\|\zeta\|),\quad 0\leq t\leq t_h,$$
instead of (\ref{E3.26}). We conclude, along the lines of the proof of Theorem \ref{T3.1}, that
$$\|\eta-\eta_h\|_{2,h}+\|u-u_h\|_1+\|v-v_h\|_1\leq Ch^{r-1}.$$ 

\section{Numerical experiments}

In this section we present the results of numerical experiments that we performed in the case of the Bona-Smith systems using the modified Galerkin method as a base spatial discretization scheme. For the temporal discretization of the system of o.d.e.'s (\ref{E3.16}) we used an explicit, second-order Runge-Kutta method, the so-called ``improved Euler'' scheme. For the solution of the resulting linear systems at each time step we used the Jacobi-Conjugate Gradient method of ITPACK, taking the relative residuals equal to $10^{-7}$ for terminating the iterations at each time step. In the computation of the discrete Laplacian $\widetilde{\Delta_h}$ we used lumping of the mass matrix. In what follows we present numerical results that confirm the expected rates of convergence of the fully discrete scheme and three numerical experiments illustrating the use of the method in various surface flows of interest.
\subsection{Experimental rates of convergence}

In order to check the spatial convergence rates of the fully discrete modified Galerkin method we applied the scheme to the  ibvp (\ref{ER}) for the Bona-Smith system with $\theta^2=9/11$ taking as exact solution
\begin{align*}
& \eta(x,y,t)=\cos (\pi x)\cos(\pi y) \exp^t,\\
& u(x,y,t)=x\cos((\pi x)/2)\sin(\pi y)\exp^t,\\
& v(x,y,t)=y\cos((\pi y)/2)\sin(\pi x)\exp^t.
\end{align*}
defined on $[0,1]\times [0,1]$, which was covered by a uniform mesh consisting of isosceles right-angle triangles with perpendicular sides of length $h=\sqrt{2/N}$, where $N$ is the number of the triangles. The solution was approximated in space by continuous, piecewise linear functions on this triangulation. We took $\Delta t=0.01$ and computed up to $T=1$. In Tables \ref{T1} and \ref{T2} we show the resulting $L^2$ and $H^1$ errors and the corresponding experimental convergence rates that confirm the result of Theorem \ref{T3.1}. 

\begin{table}[h]
\caption{$L^2$ errors and convergence rates for the modified Galerkin method for the Bona-Smith system with $\theta^2=9/11$. Linear elements on triangular mesh and second-order RK time-stepping.}\label{T1} \centerline{
\begin{tabular}{|c|c|c|c|c|c|c|}
\hline $N$ & $\|\eta-\eta_h\|$ & rate$(\eta)$ & $ \|u-u_h\|$ & rate$(u)$ & $\|v-v_h\|$ & rate$(v)$ \\
\hline\hline
 512 & 1.3016E-2  & -- &  9.0816E-3 & -- & 9.0429E-2 & -- \\\hline
 2048 &3.2571E-3 & 1.9986 & 2.3603E-3 & 1.9440 & 2.3604E-2 & 1.9378  \\\hline
 8192 & 8.1485E-4 & 1.9990 & 5.9882E-4 & 1.9788  & 5.9882E-4 & 1.9788 \\\hline
 32768 &  2.0523E-4 & 1.9893 & 1.5417E-4 & 1.9576 & 1.5417E-4 & 1.9576\\
\hline
\end{tabular}}
\end{table}
\begin{table}[h]
\caption{$H^1$ errors and convergence rates for the modified Galerkin method for the Bona-Smith system with $\theta^2=9/11$. Linear elements on triangular mesh and second-order RK time-stepping.}\label{T2} \centerline{
\begin{tabular}{|c|c|c|c|c|c|c|}
\hline $N$ & $\|\eta-\eta_h\|_1$ & rate$(\eta)$ & $ \|u-u_h\|_1$ & rate$(u)$ & $\|v-v_h\|_1$ & rate$(v)$ \\
\hline\hline
 512 &  4.3739E-1 & -- & 1.8597E-1 & -- & 1.8488E-1 &  --    \\\hline
 2048 & 2.2030E-1 & 0.9894 & 8.7012E-2 & 1.0958 & 8.7008E-2 & 1.0874   \\\hline
 8192 & 1.1038E-1 & 0.9970 & 4.2013E-2 & 1.0504 & 4.2013E-2 & 1.0503 \\\hline
 32768 & 5.5221E-2 & 0.9992 & 2.0680E-2 & 1.0226 & 2.0680E-2 &1.0226\\
\hline
\end{tabular}}
\end{table}

\subsection{Solitary-wave-like pulse hitting a cylindrical obstacle}

In our first experiment the domain that we consider is the rectangular channel $[-15,15]\times [-30,50]$ in which is placed a vertical impenetrable cylinder centered at $(0,10)$ with radius equal to $1.5$. We pose normal reflective boundary conditions on the boundary of the cylinder and along the lines $y=-30$ and $y=50$, and homogeneous Neumann boundary conditions for $\eta$ and ${\bf v}$ on the lateral boundaries $x=\pm 15$. As initial conditions we use the functions
\begin{equation}\label{E4.1}
\begin{array}{l}
\eta_0(x,y)=A{\rm sech}^2\left(\frac{1}{2}\sqrt{\frac{3A}{c_s}}(y+10)
\right),\\
u_0(x,y)=0,\\
v_0(x,y)=\eta_0(x,y)-\frac{1}{4}\eta_0^2(x,y),
\end{array}
\end{equation}
where $A=0.2$ and $c_s=1+\frac{A}{2}$, which represent a good approximation to a line solitary wave, \cite{DMSI}, of the BBM-BBM system. We integrate under these initial and boundary conditions the Bona-Smith system (\ref{E1.1}) with $\theta^2=9/11$ and compare its evolving solution with that of the BBM-BBM system, i.e. (\ref{E1.1}) with $\theta^2=2/3$. The Bona-Smith system was discretized in space by the modified Galerkin method (amended in a straightforward way to handle the Neumann conditions on ${\bf v}$ on the lateral boundaries) with continuous, piecewise linear elements on a triangulation consisting of $N=290112$ trangles; this mesh was fine enough for the `convergence' of the numerical solution. The same truangulation was used to discretize in space the BBM-BBM system with the standard Galerkin method, \cite{DMSI}. Both systems were discretized in time with the improved Euler method with $\Delta t=0.01$.

The line solitary-wave-like pulse is centered at $y=-10$ at $t=0$. It propagates, mainly in the positive $y$ direction, travelling with a speed $c_s$ of about $1.05$, cf. Figure 4.2. The bulk of the solitary wave travels past the cylinder, while smaller amplitude scattered waves are produced by the interaction of the wave with the obstacle. Figure 4.3 shows, for both systems, contours of the elevation of the wave, superimposed on velocity vector plots, near the obstacle during the passage of the solitary wave. Figure 4.4 shows the free surface elevation for both systems as a function of $y$ at $x=0$ at several temporal instances. It is worthwhile to note that the maximum run-up at the extreme upstream point $(x,y)=(0,8.5)$ of the cylinder was equal to $z=0.335107$ (achieved at $t=16.72$) for the BBM-BBM system; the corresponding run-up for the Bona-Smith system was $z=0.319960$ at $t=16.74$. The analogous values for the run-up on the downstream point $(0,11.5)$ of the cylinder were $z=0.234310$, $t=22.71$ for the BBM-BBM system, and $z=0.230590$, $t=22.68$ for the Bona-Smith system. Figure 4.5 shows the history of the free surface elevation for both systems at the extreme points $y=8.5$, $y=11.5$ along the $x=0$ diameter of the cylinder, perpendicular to the impinging wave. In general, we did not observe great differences in the behaviour of the solutions of the two systems.

\begin{figure}[p] 
\begin{center}
\begin{tabular}{cc}
\includegraphics[width=2.8in]{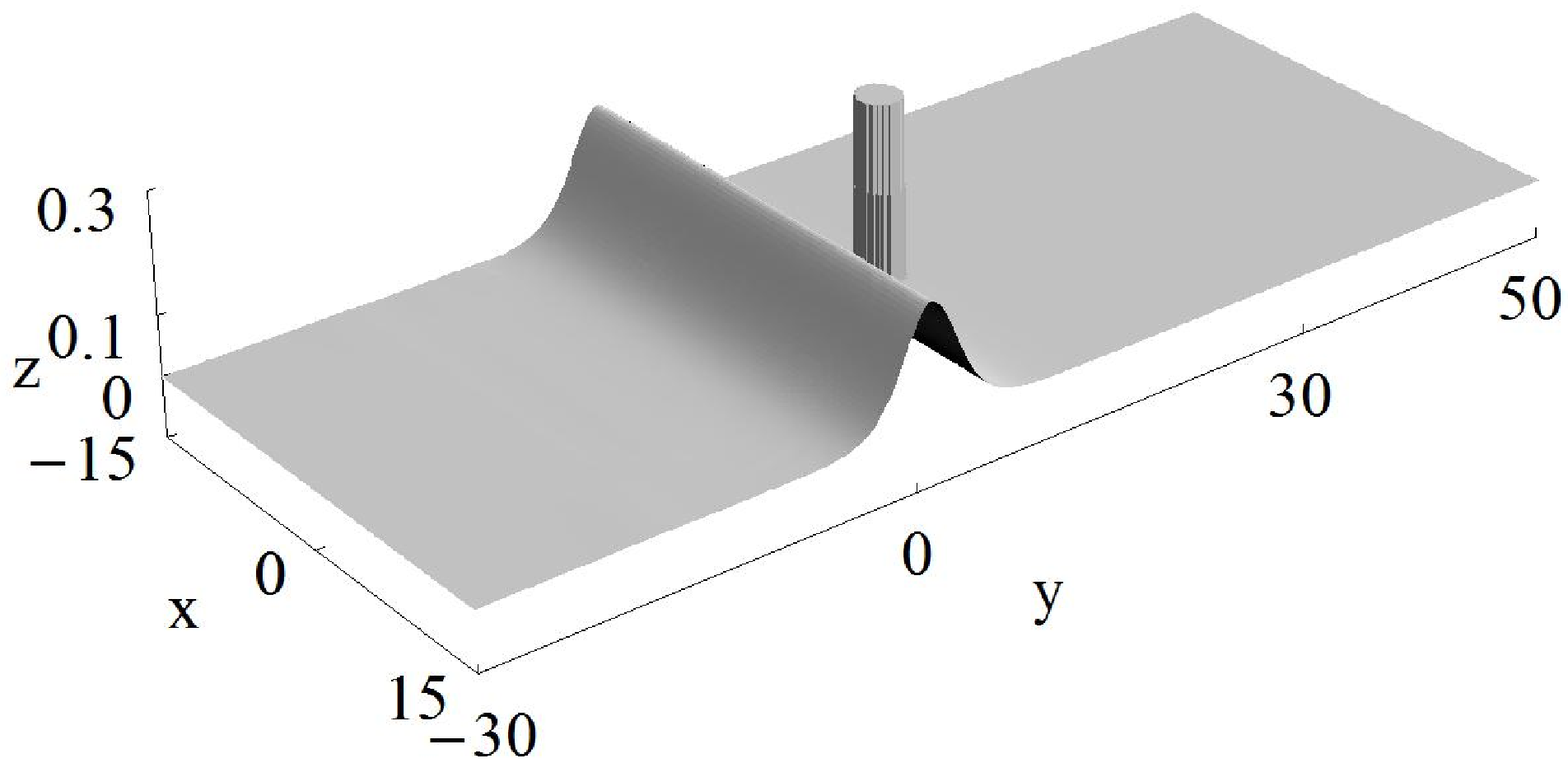} &\includegraphics[width=2.8in]{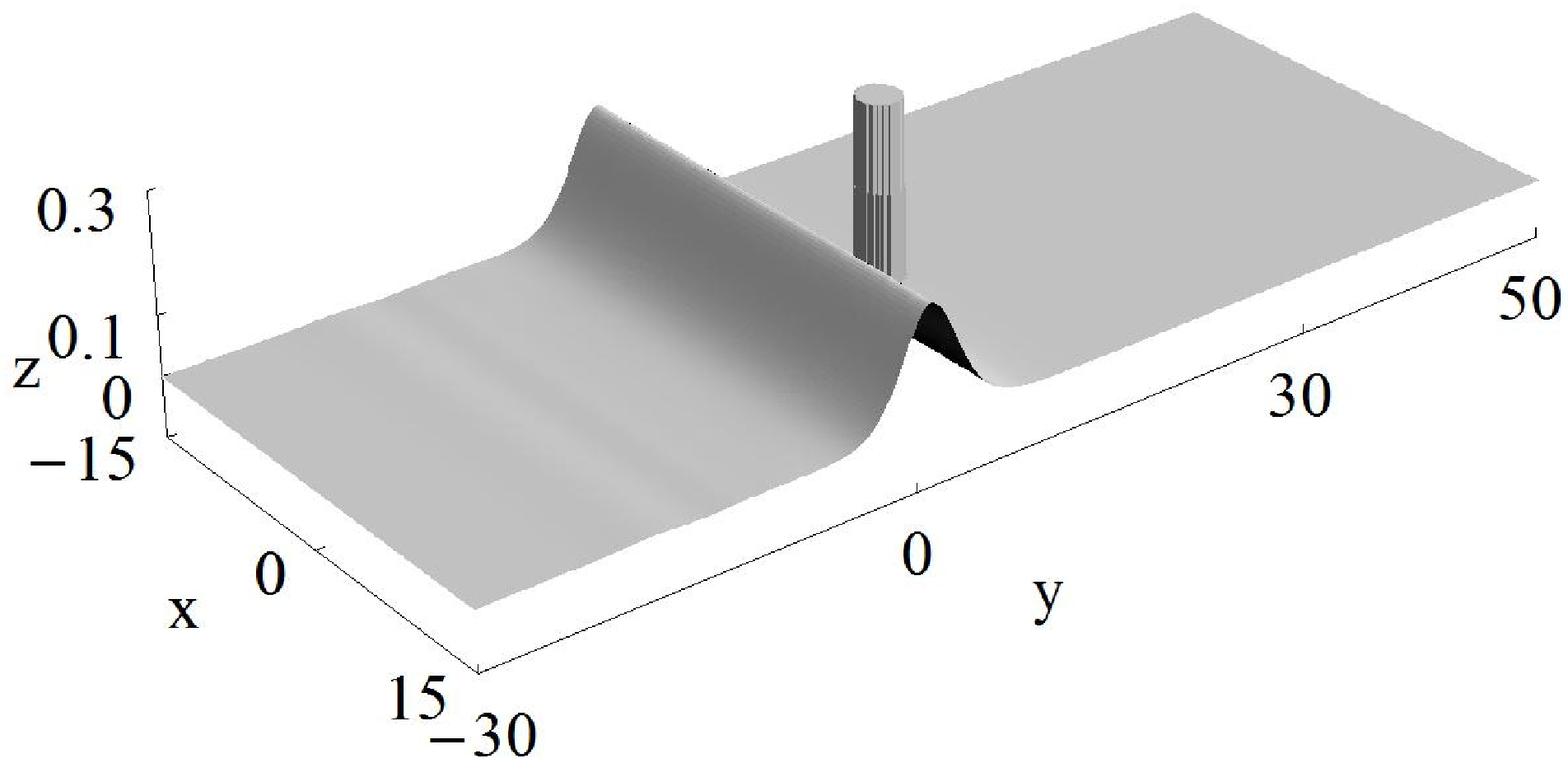}\\
BBM-BBM $t=10$ & Bona-Smith $t=10$\\
\includegraphics[width=2.8in]{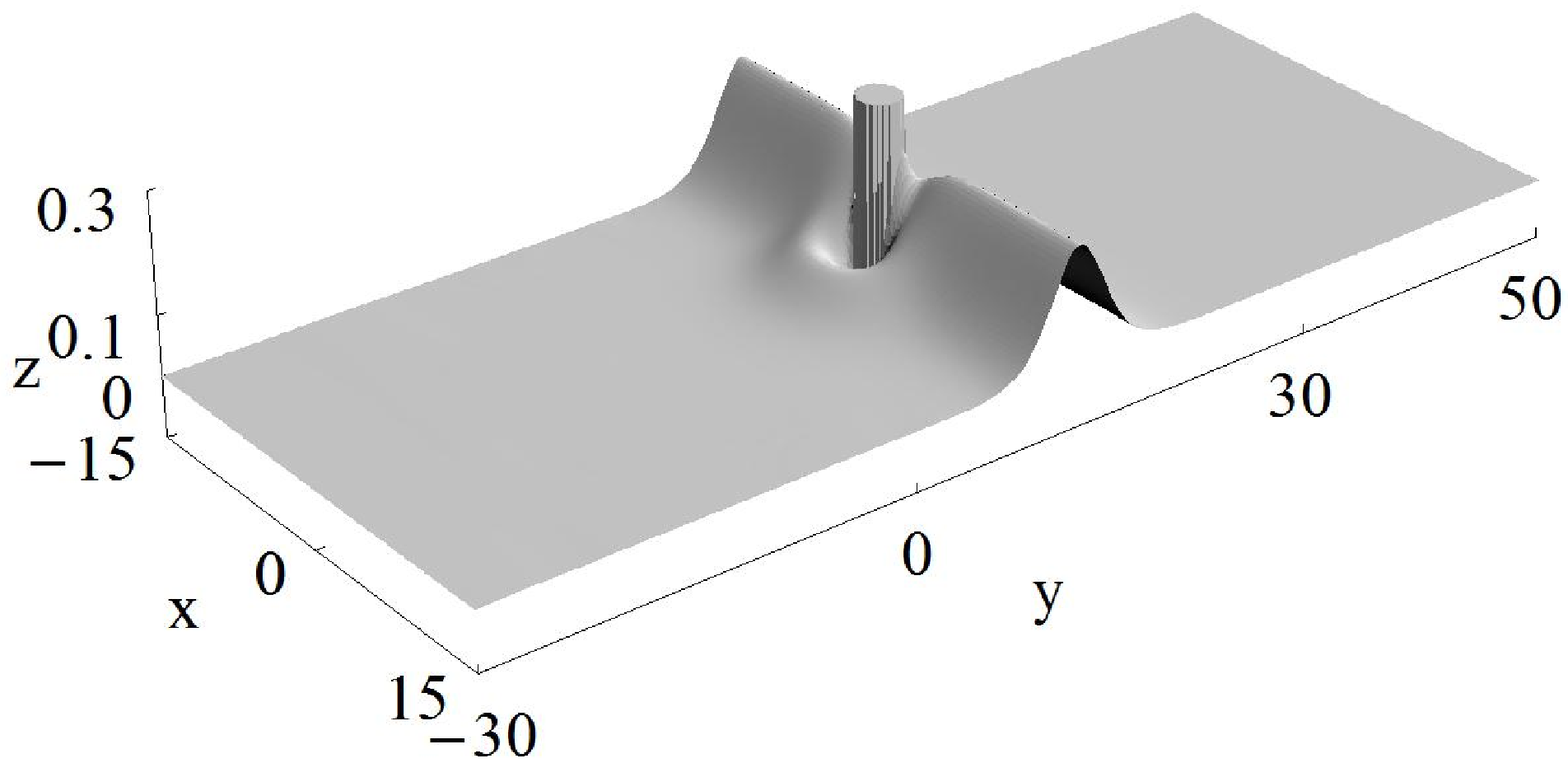} &\includegraphics[width=2.8in]{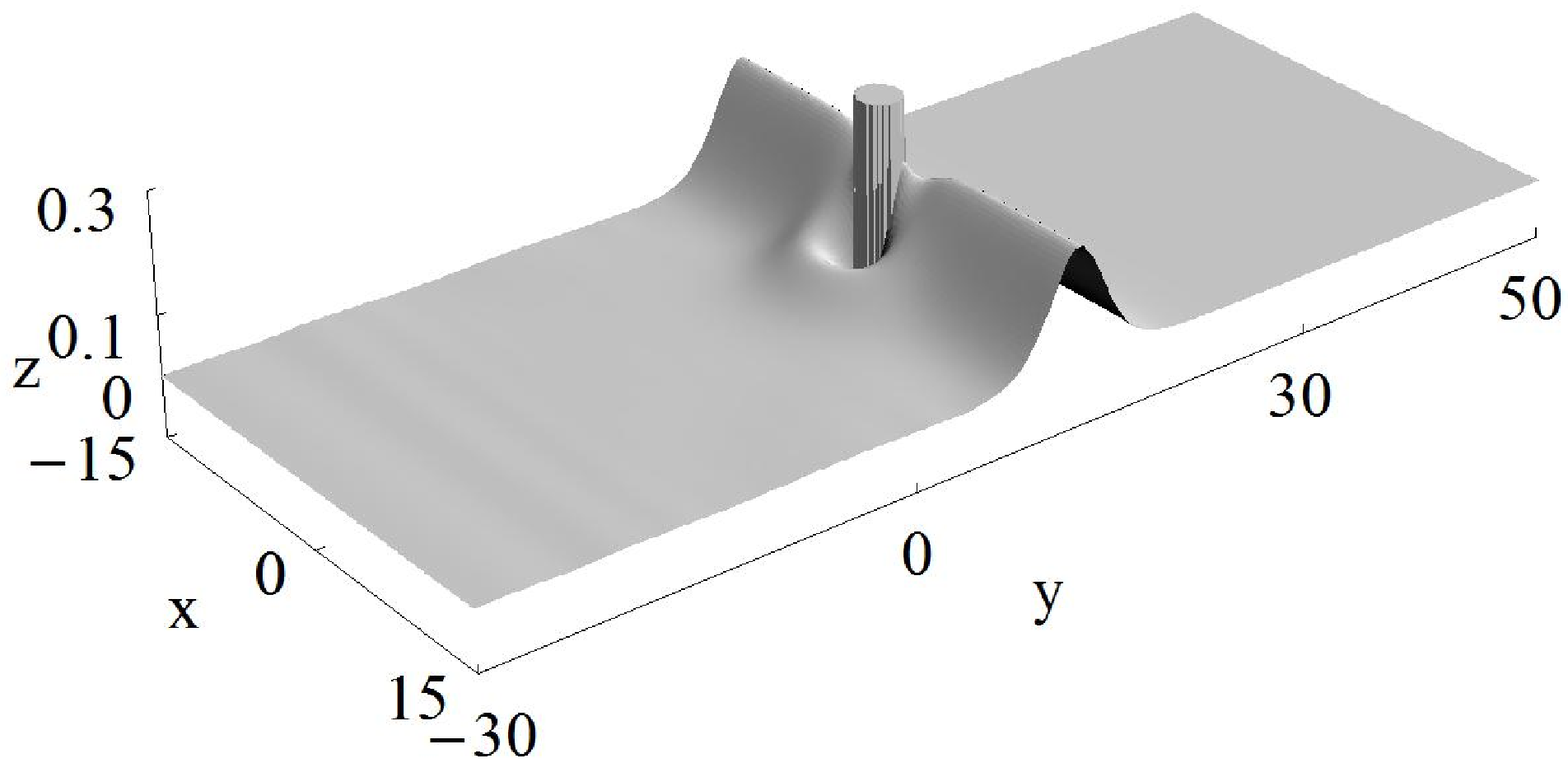}\\
BBM-BBM $t=20$ & Bona-Smith $t=20$\\
\includegraphics[width=2.8in]{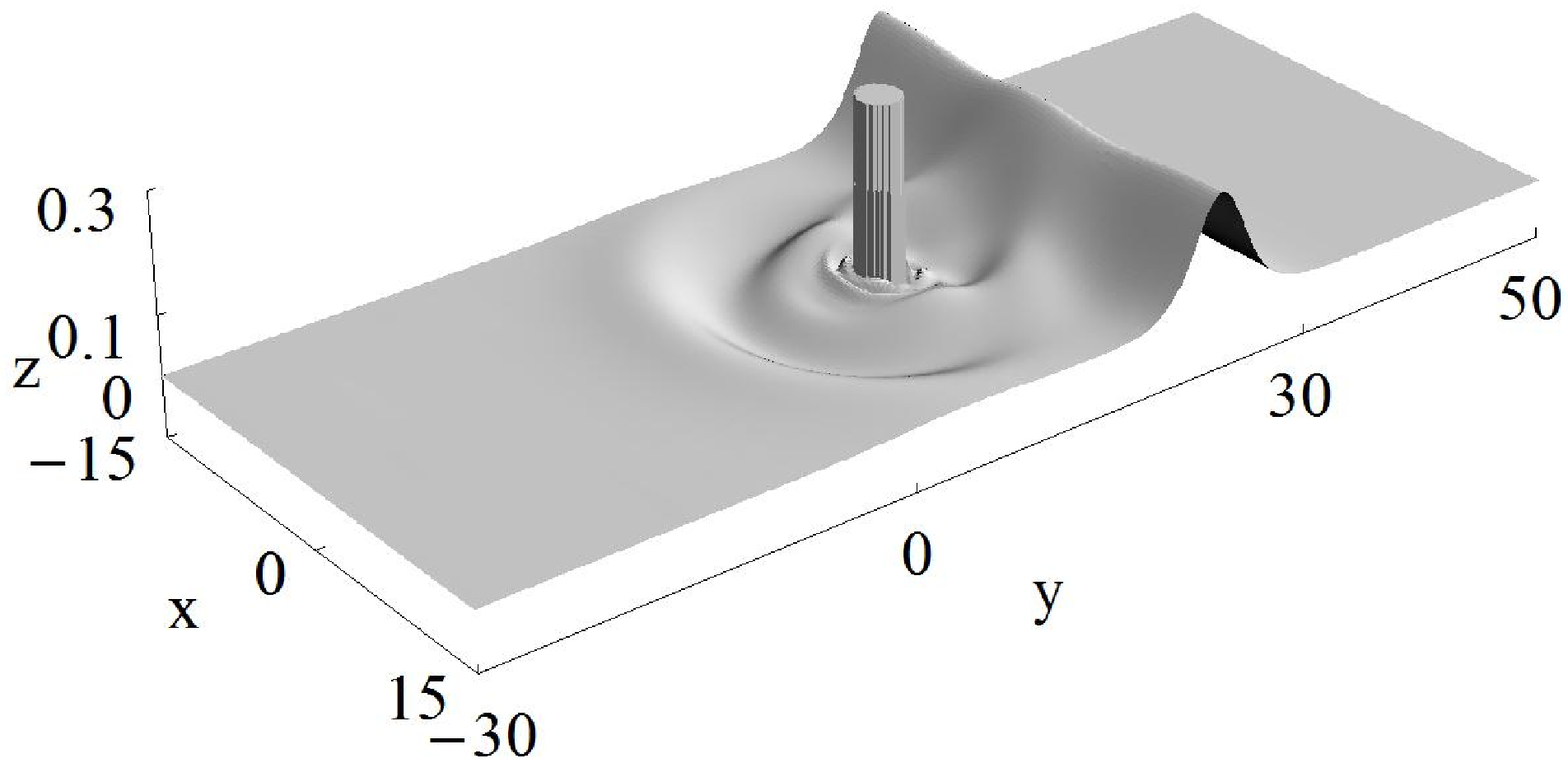} &\includegraphics[width=2.8in]{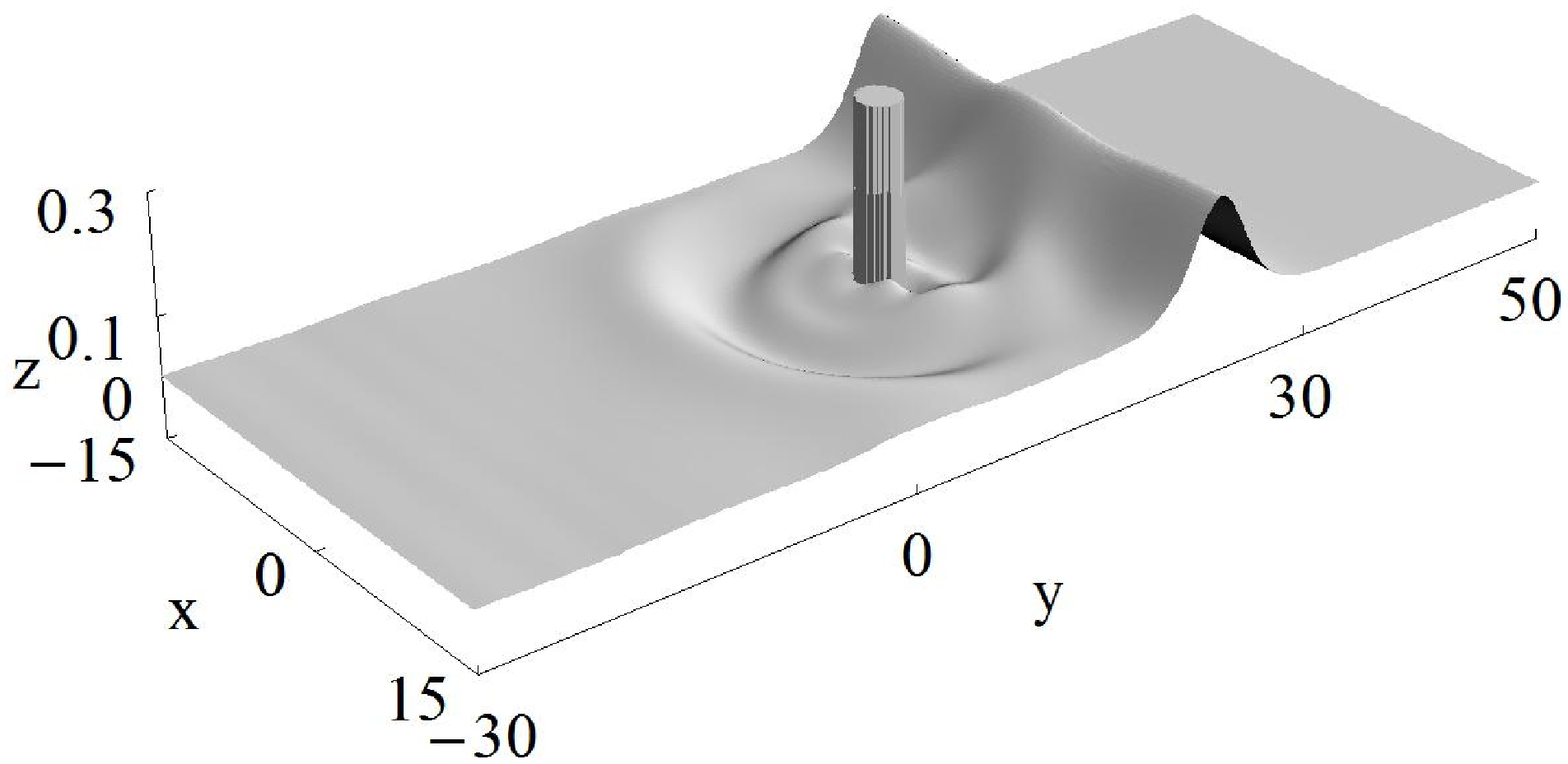}\\
BBM-BBM $t=30$ & Bona-Smith $t=30$\\
\includegraphics[width=2.8in]{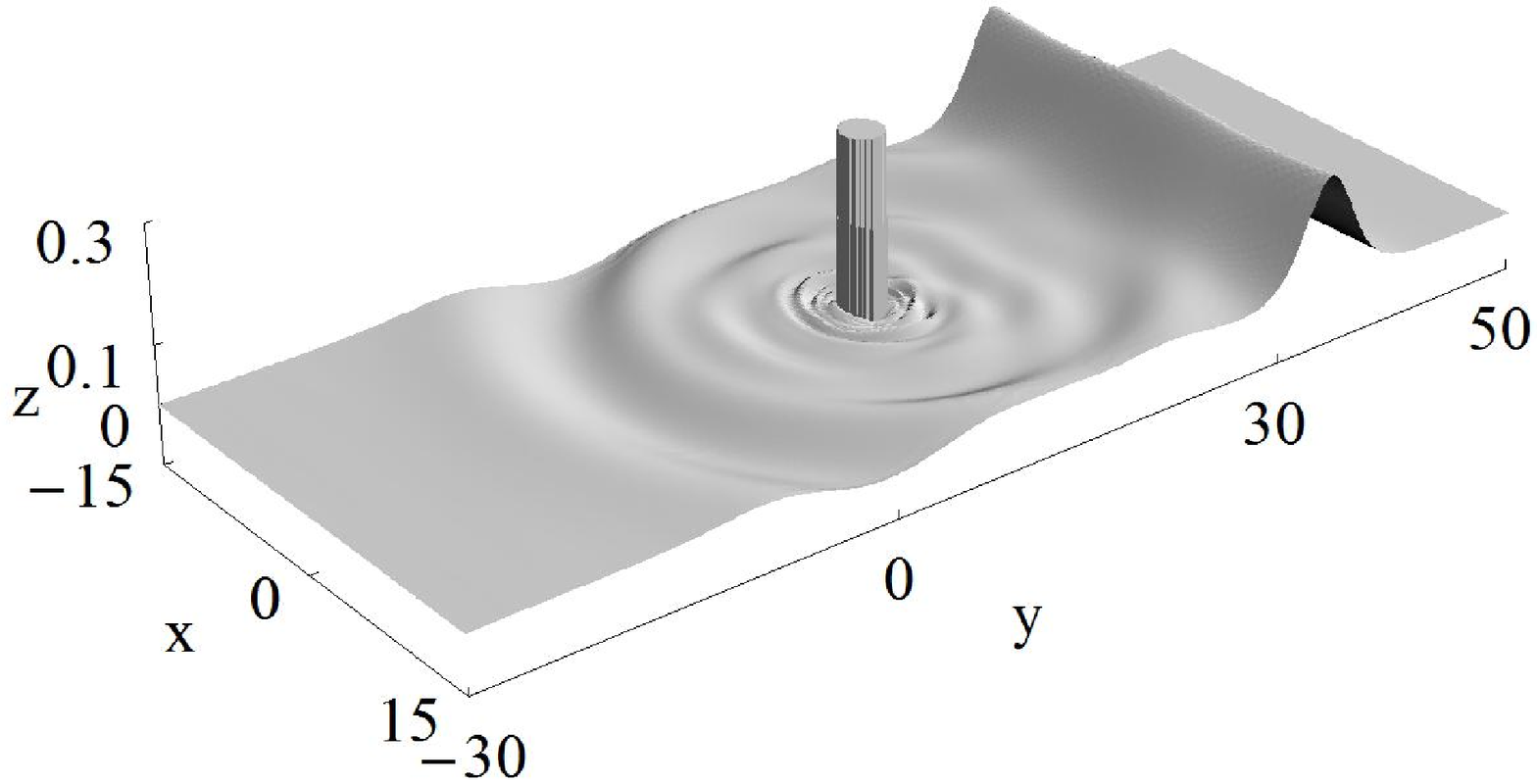} &\includegraphics[width=2.8in]{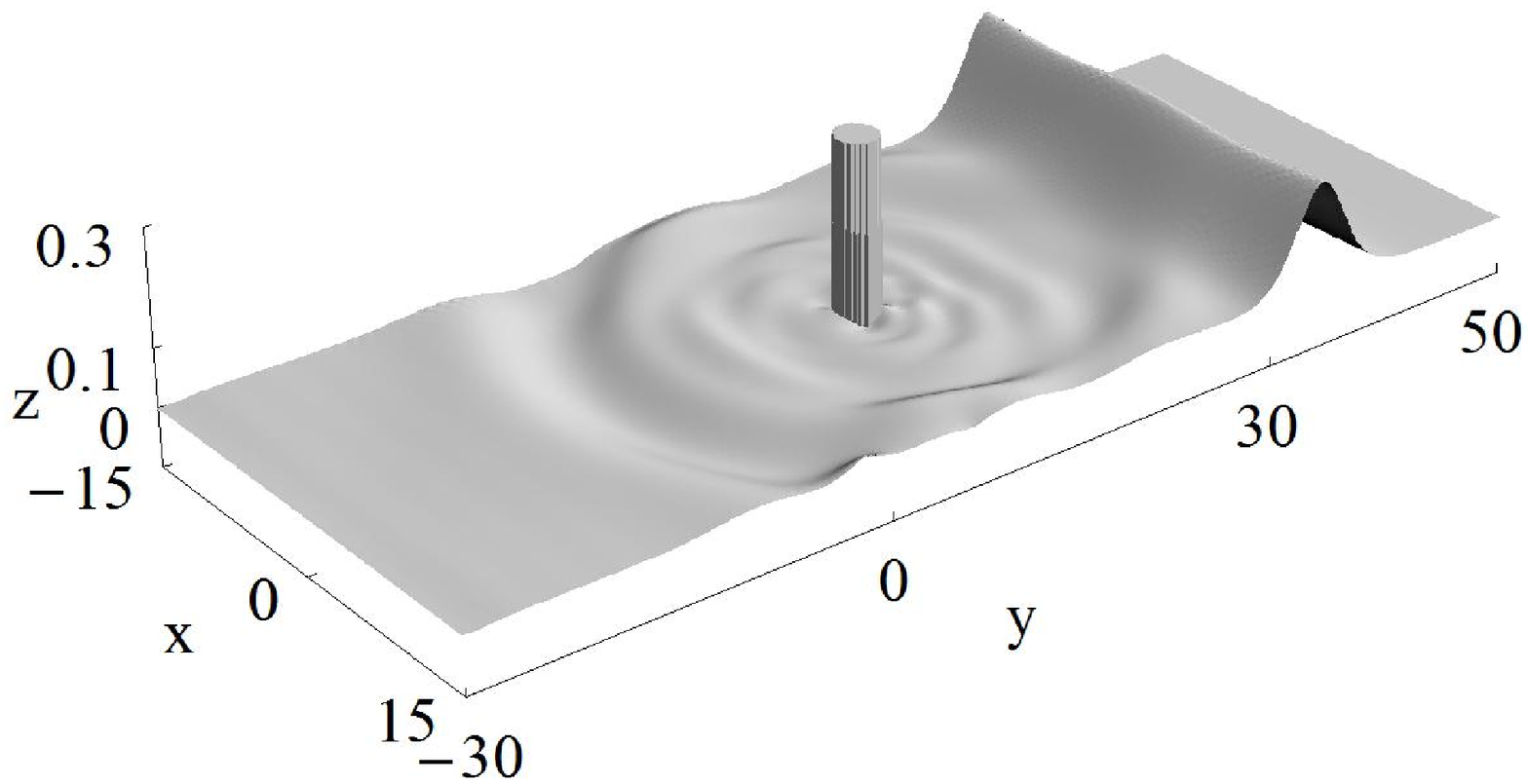}\\
BBM-BBM $t=40$ & Bona-Smith $t=40$
\end{tabular}
  \caption{Experiment 4.2 Free surface elevation at four time instances. BBM-BBM vs. Bona-Smith ($\theta^2=9/11$) systems.}
  \end{center}
\end{figure}\label{F4.1}

\begin{figure*}[p]
\begin{center}
\begin{tabular}{cc}
\includegraphics[width=3in]{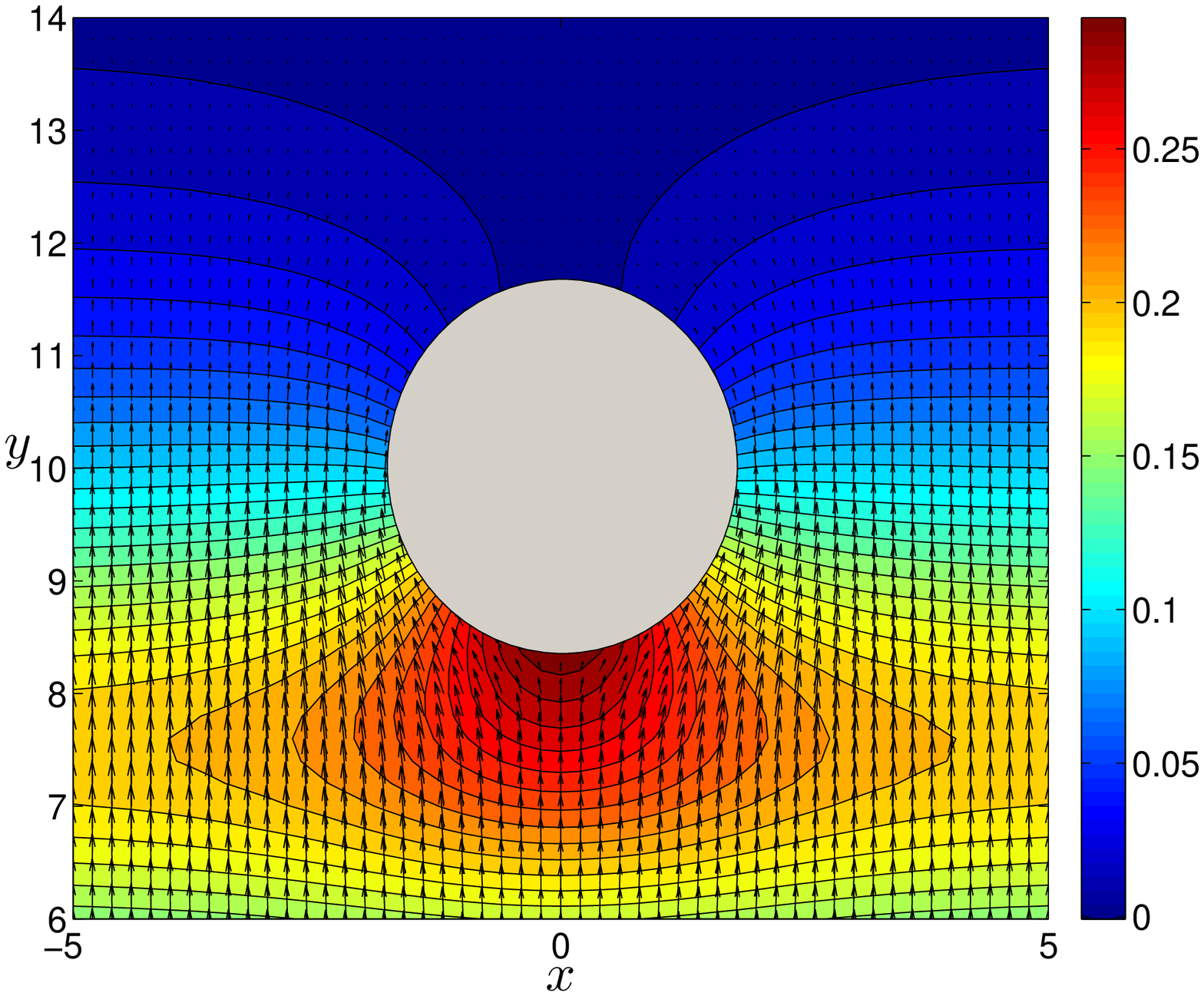} &\includegraphics[width=3in]{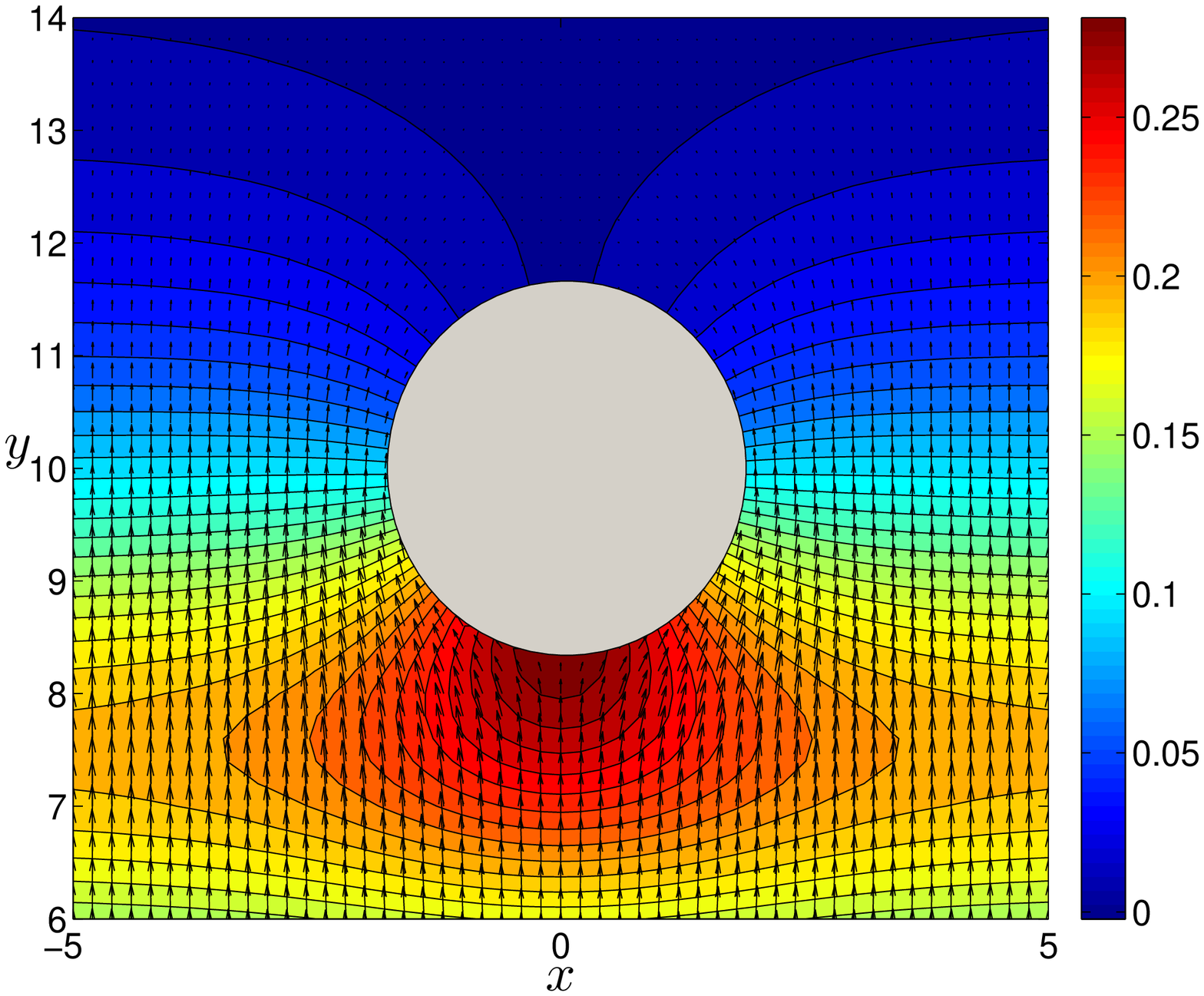}\\
BBM-BBM $t=16$ & Bona-Smith $t=16$\\
\includegraphics[width=3in]{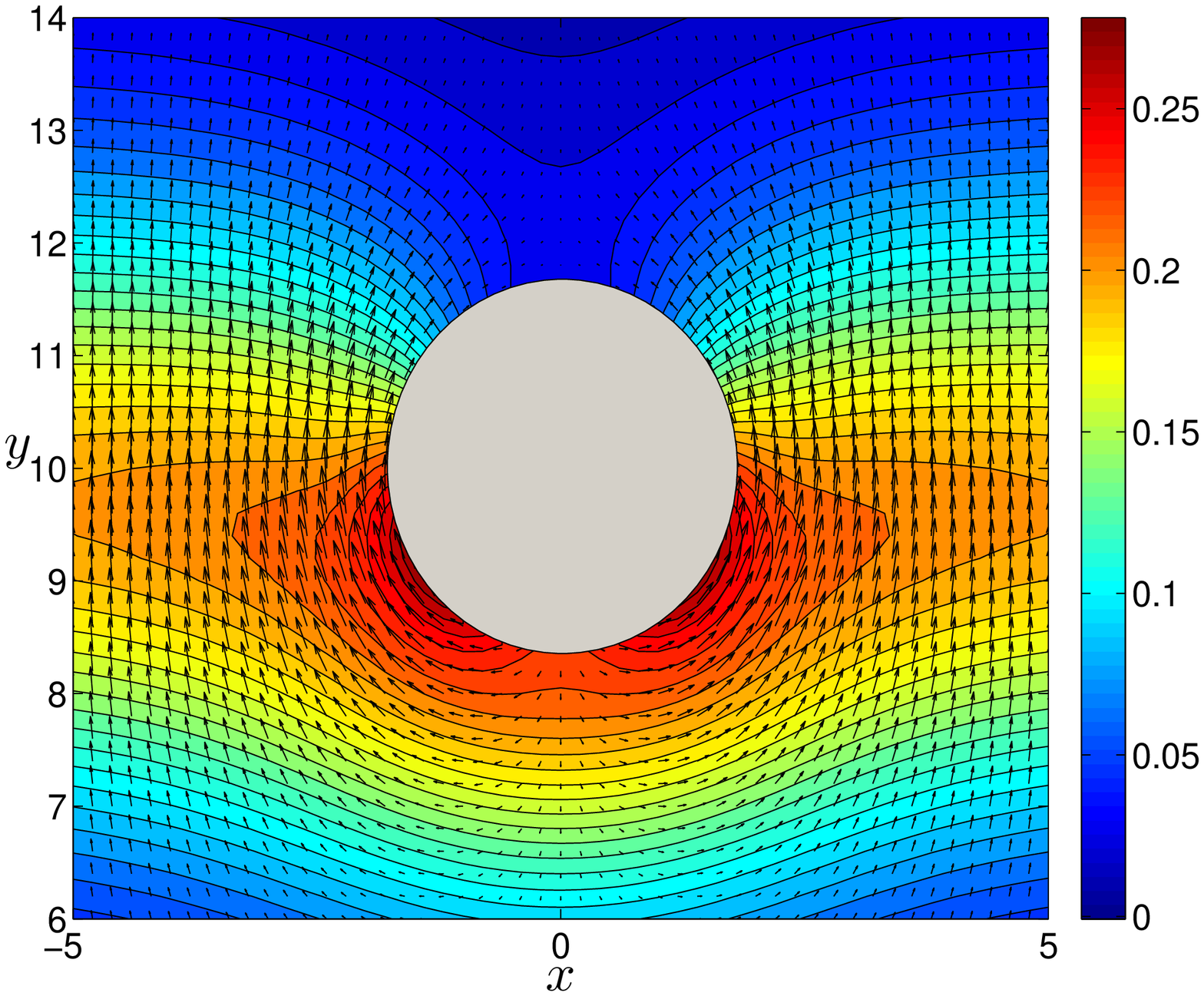} &\includegraphics[width=3in]{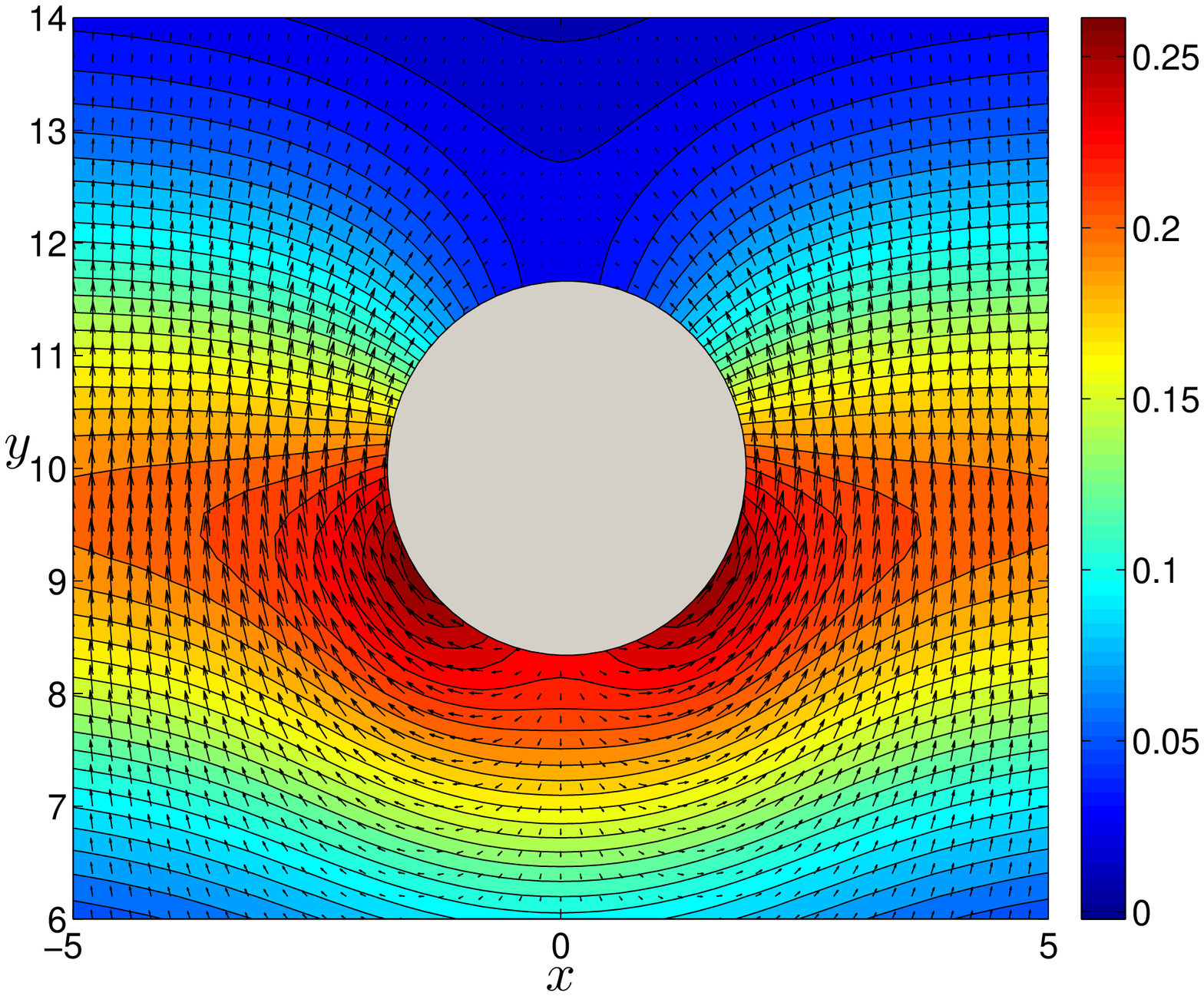}\\
BBM-BBM $t=18$ & Bona-Smith $t=18$\\
\end{tabular}
  \end{center}
\end{figure*}

\begin{figure}[p]
\begin{center}
\begin{tabular}{cc}
\includegraphics[width=3in]{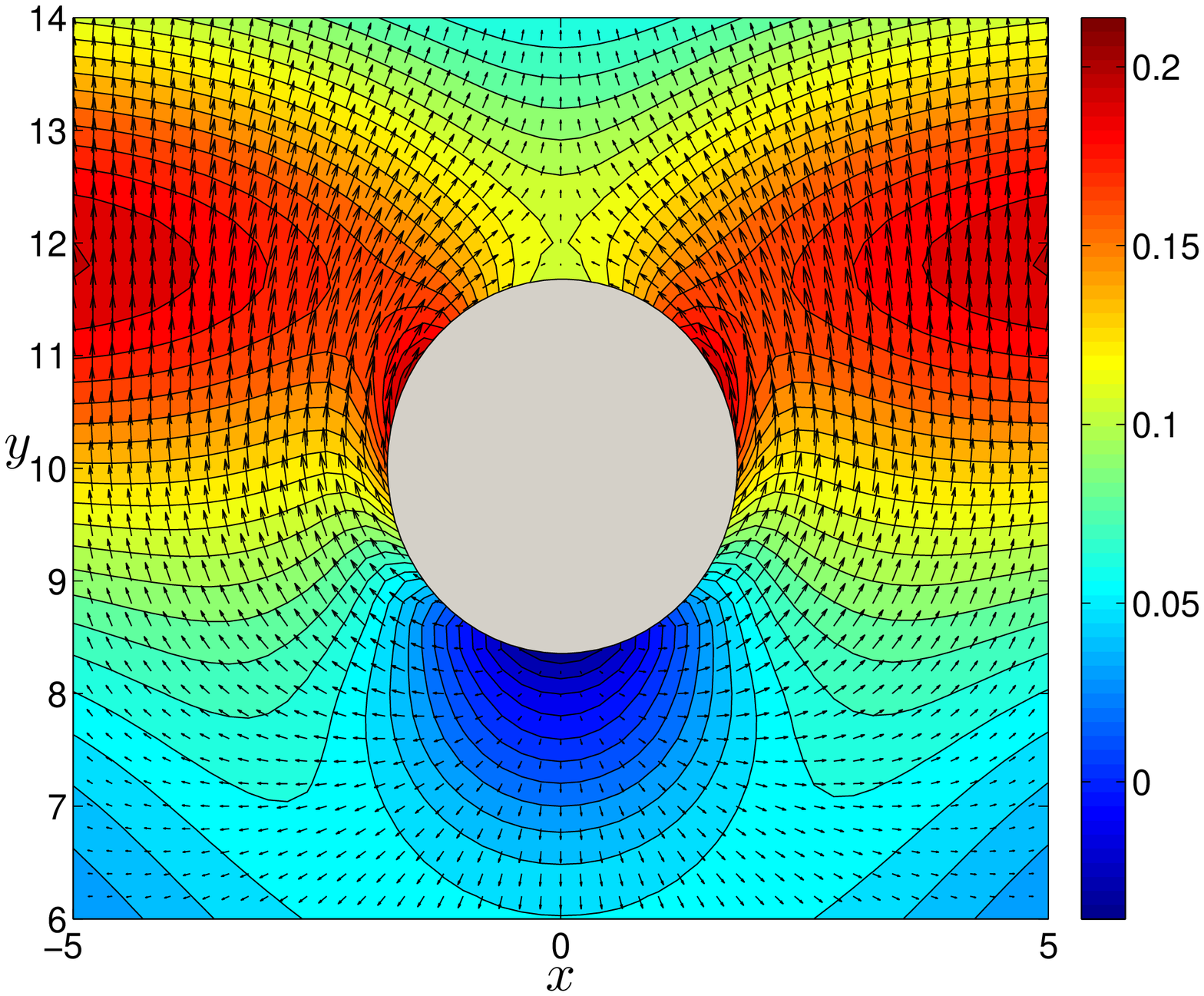} &\includegraphics[width=3in]{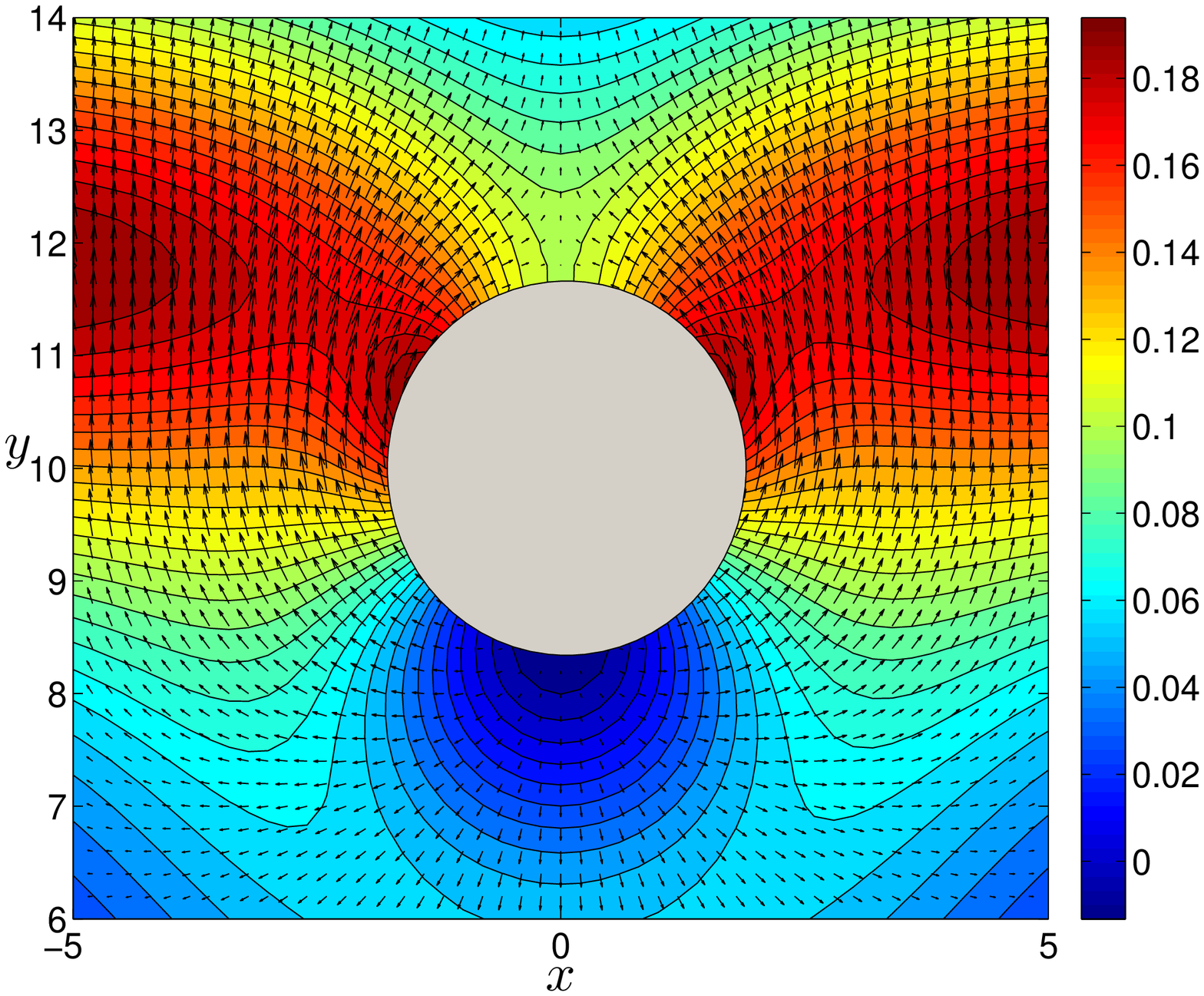}\\
BBM-BBM $t=20$ & Bona-Smith $t=20$\\
\includegraphics[width=3in]{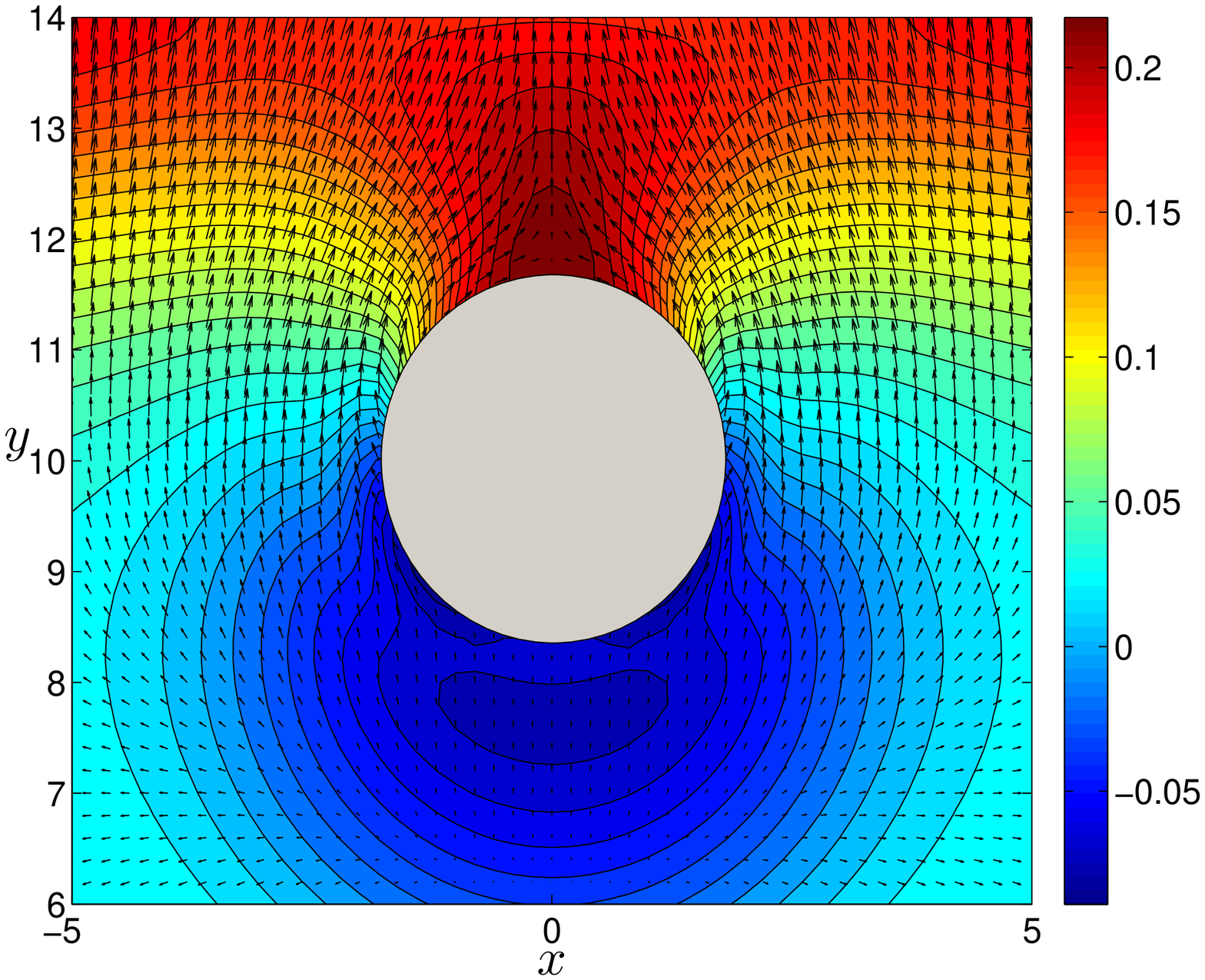} &\includegraphics[width=3in]{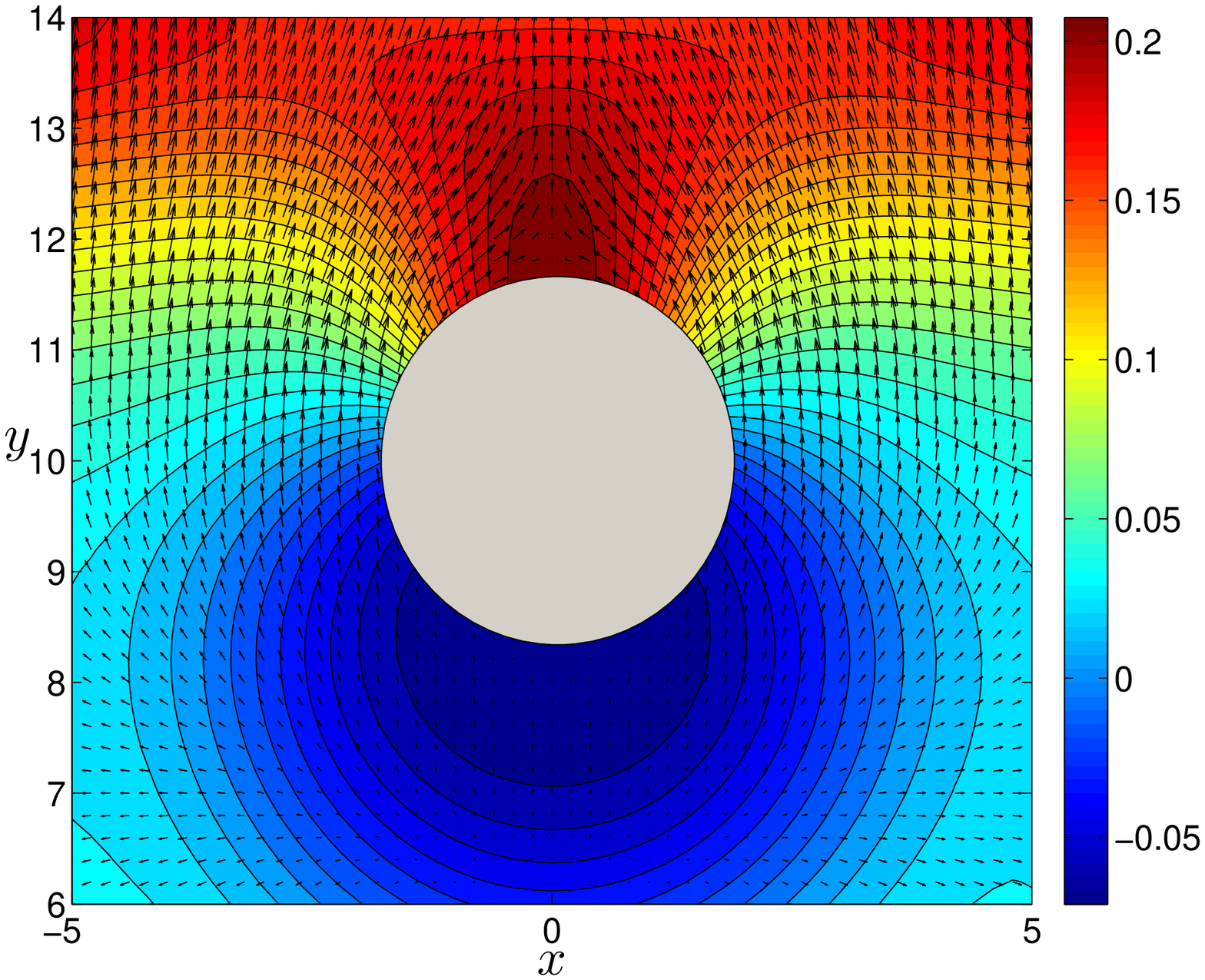}\\
BBM-BBM $t=22$ & Bona-Smith $t=22$
\end{tabular}
  \caption{Experiment 4.2. $\eta$-contour-velocity vector plots. BBM-BBM vs. Bona-Smith ($\theta^2=9/11$) systems.}
  \end{center}
\end{figure}\label{F4.2}

\begin{figure*}[p]
\begin{center}
\begin{tabular}{cc}
\includegraphics[width=3in]{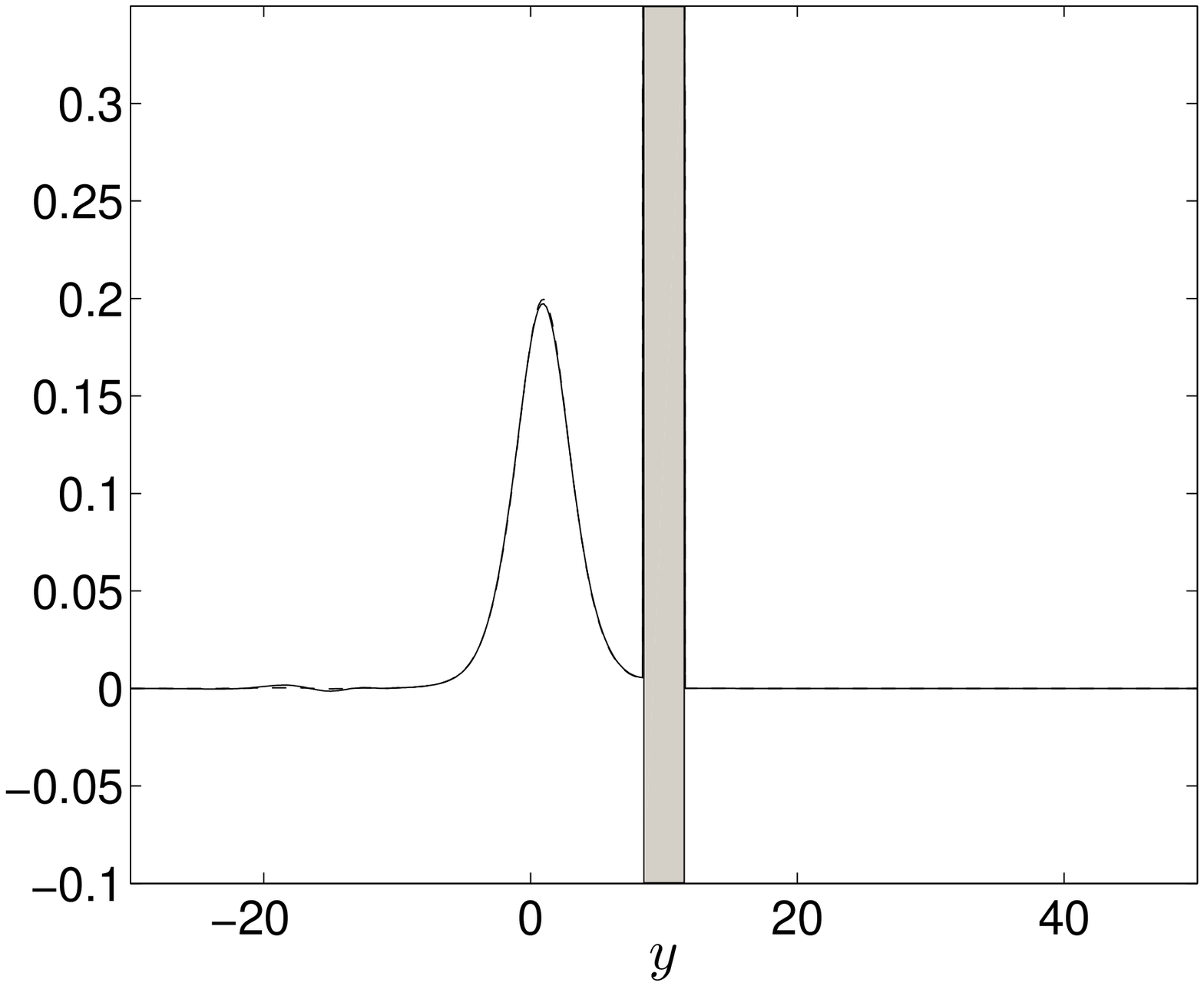} & \includegraphics[width=3in]{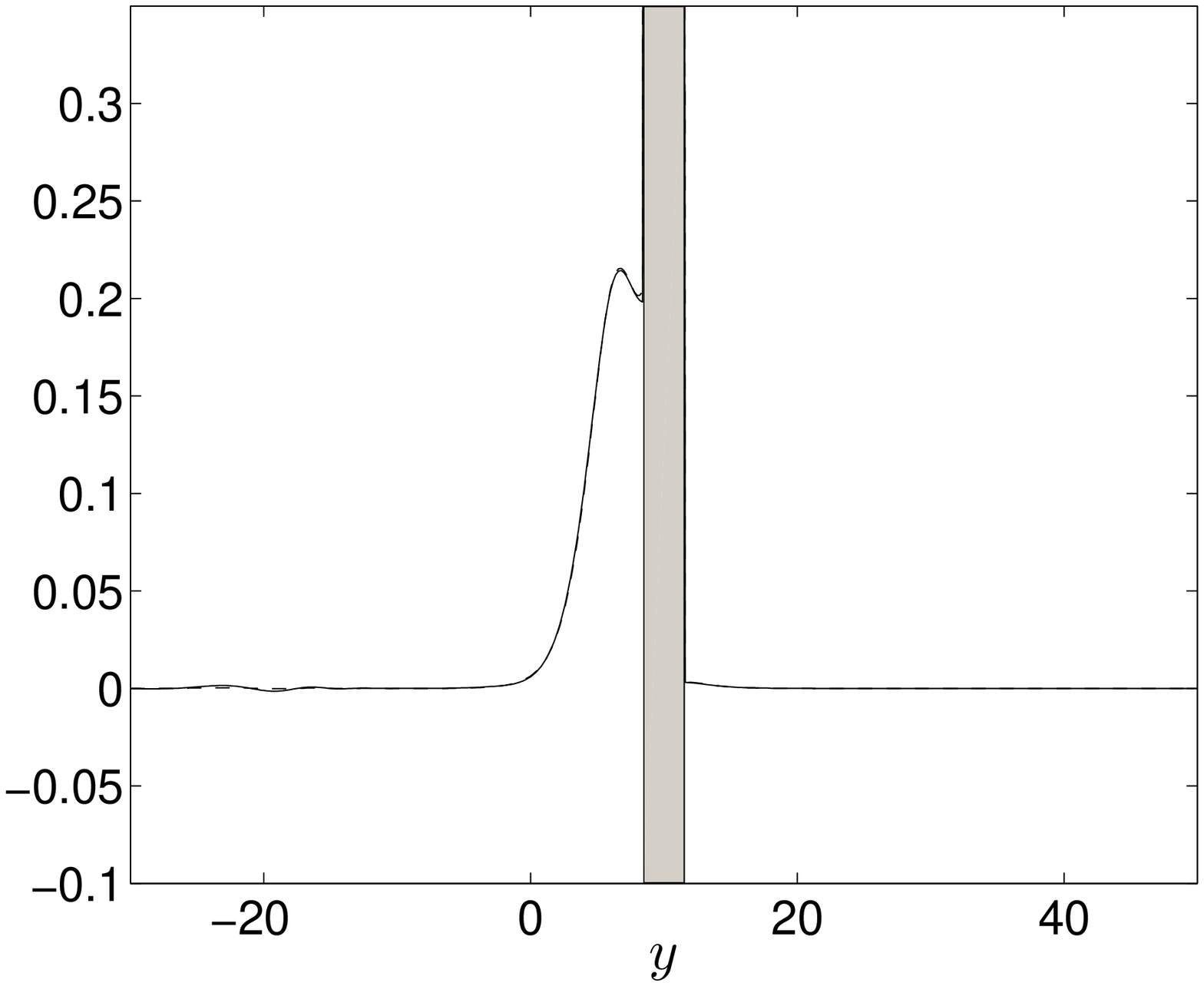}\\
$t=10$ & $t=15$ \\
\includegraphics[width=3in]{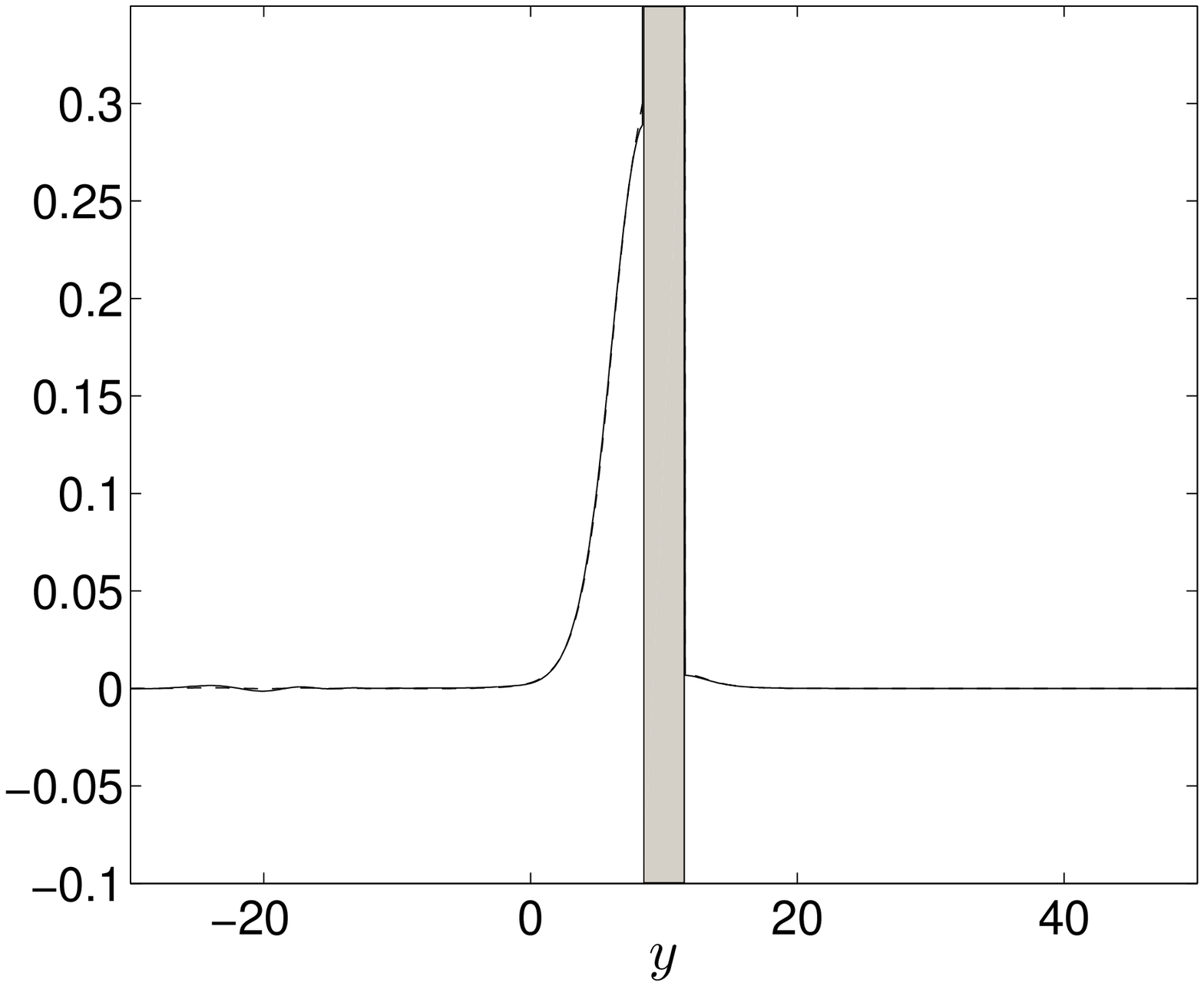} & \includegraphics[width=3in]{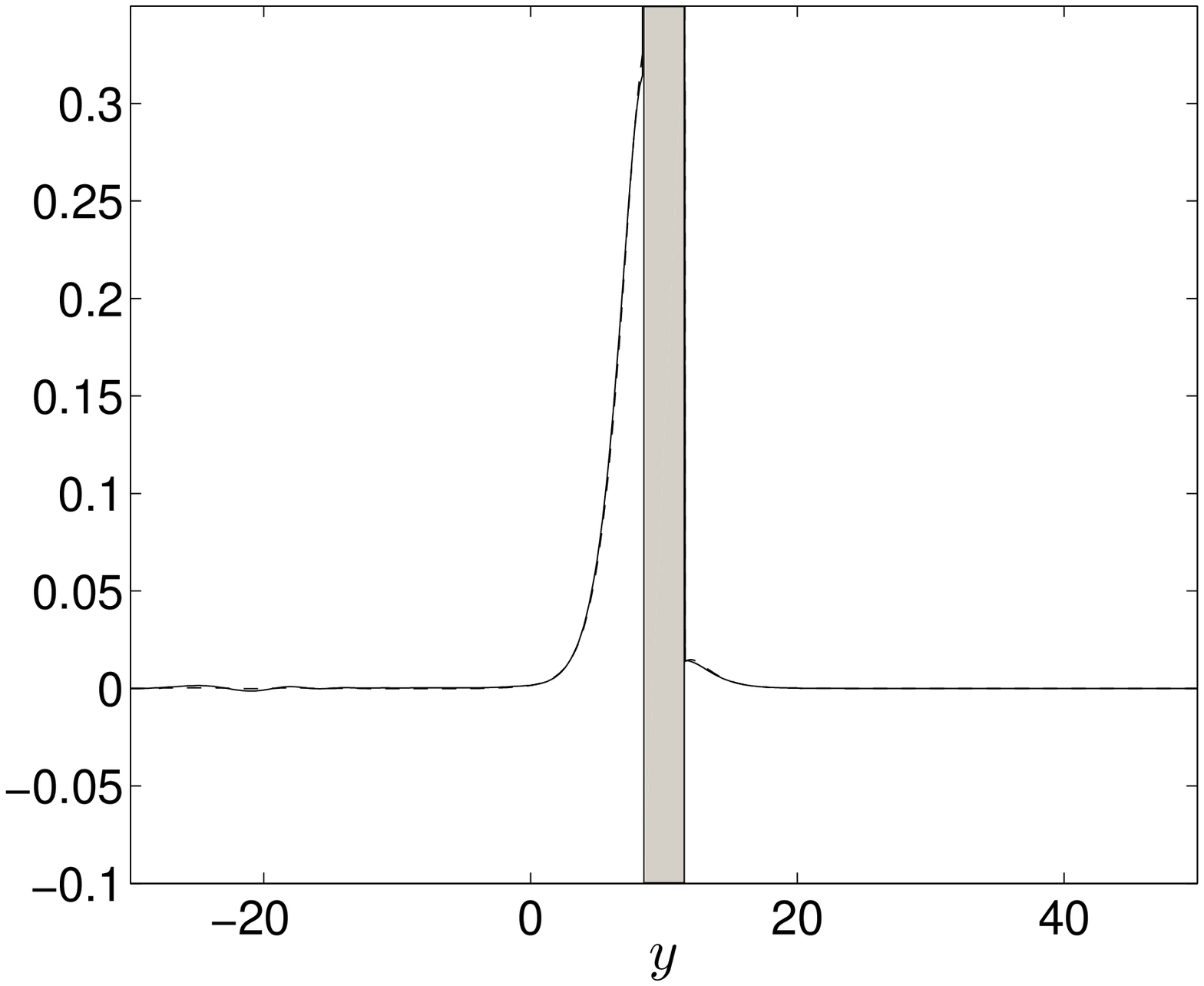}\\
$t=16$ & $t=17$ \\
\includegraphics[width=3in]{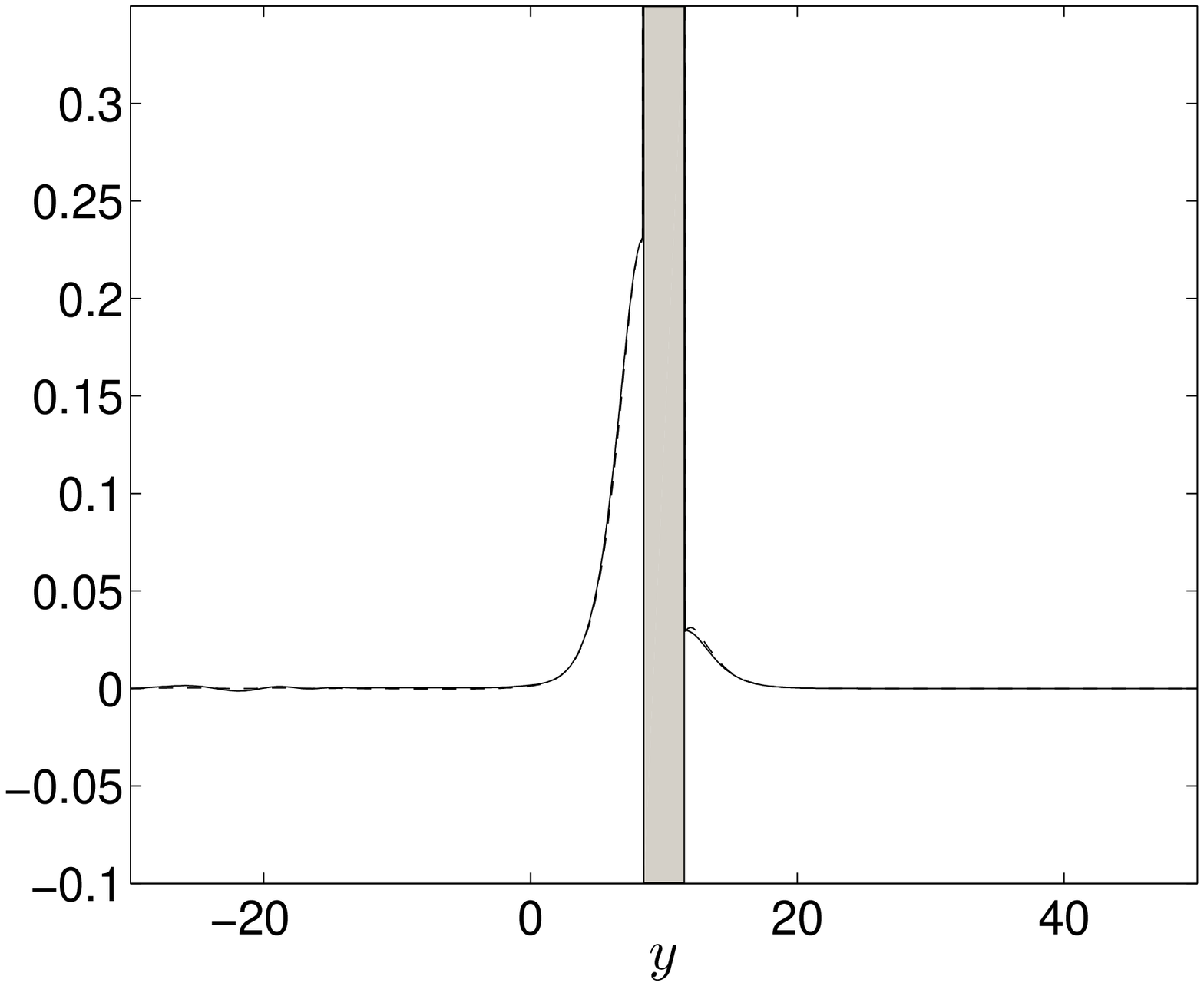} & \includegraphics[width=3in]{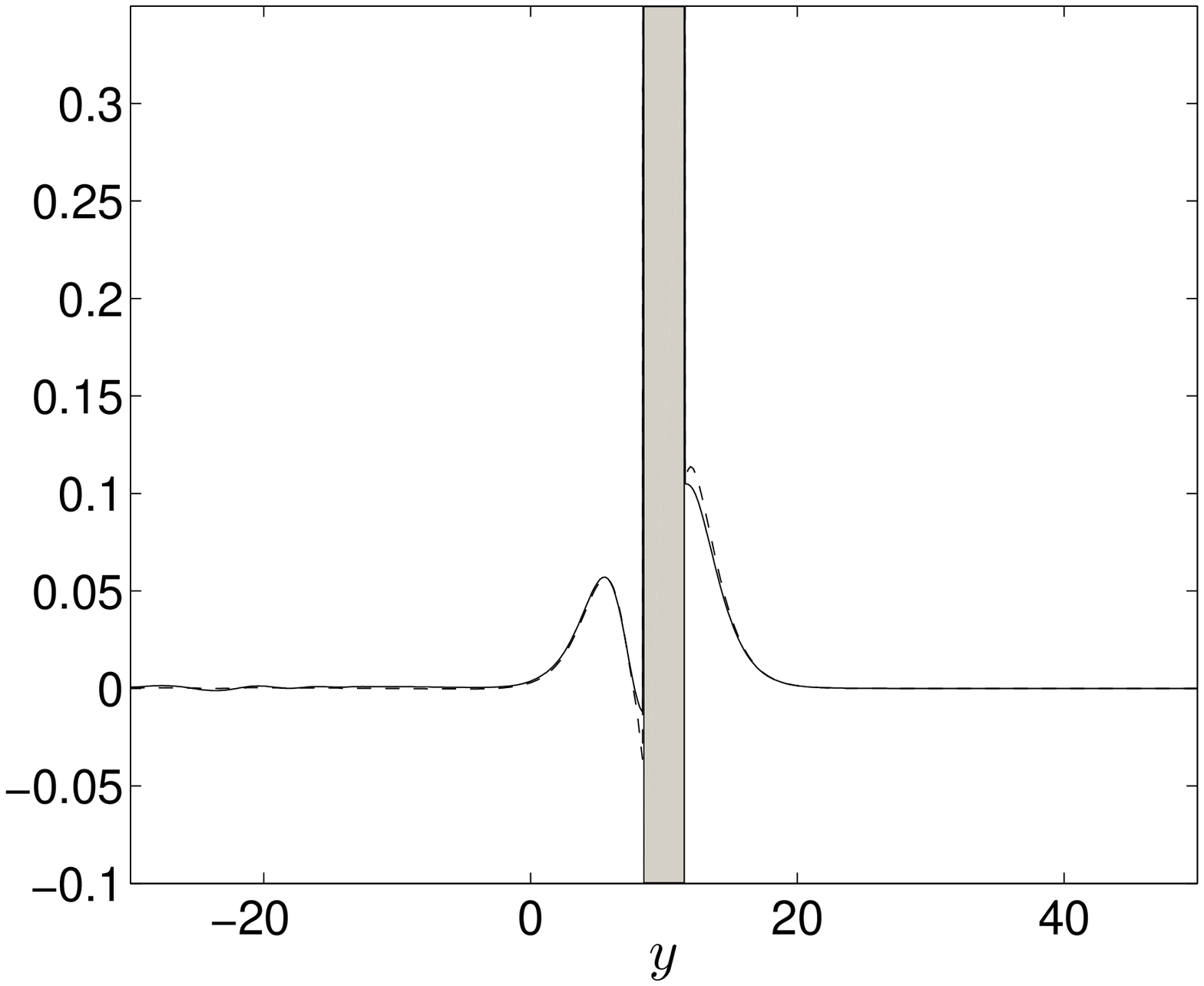}\\
$t=18$ & $t=20$ \\
\end{tabular}
\end{center}
\end{figure*}

\begin{figure}[p]
\begin{center}
\begin{tabular}{cc}
\includegraphics[width=3in]{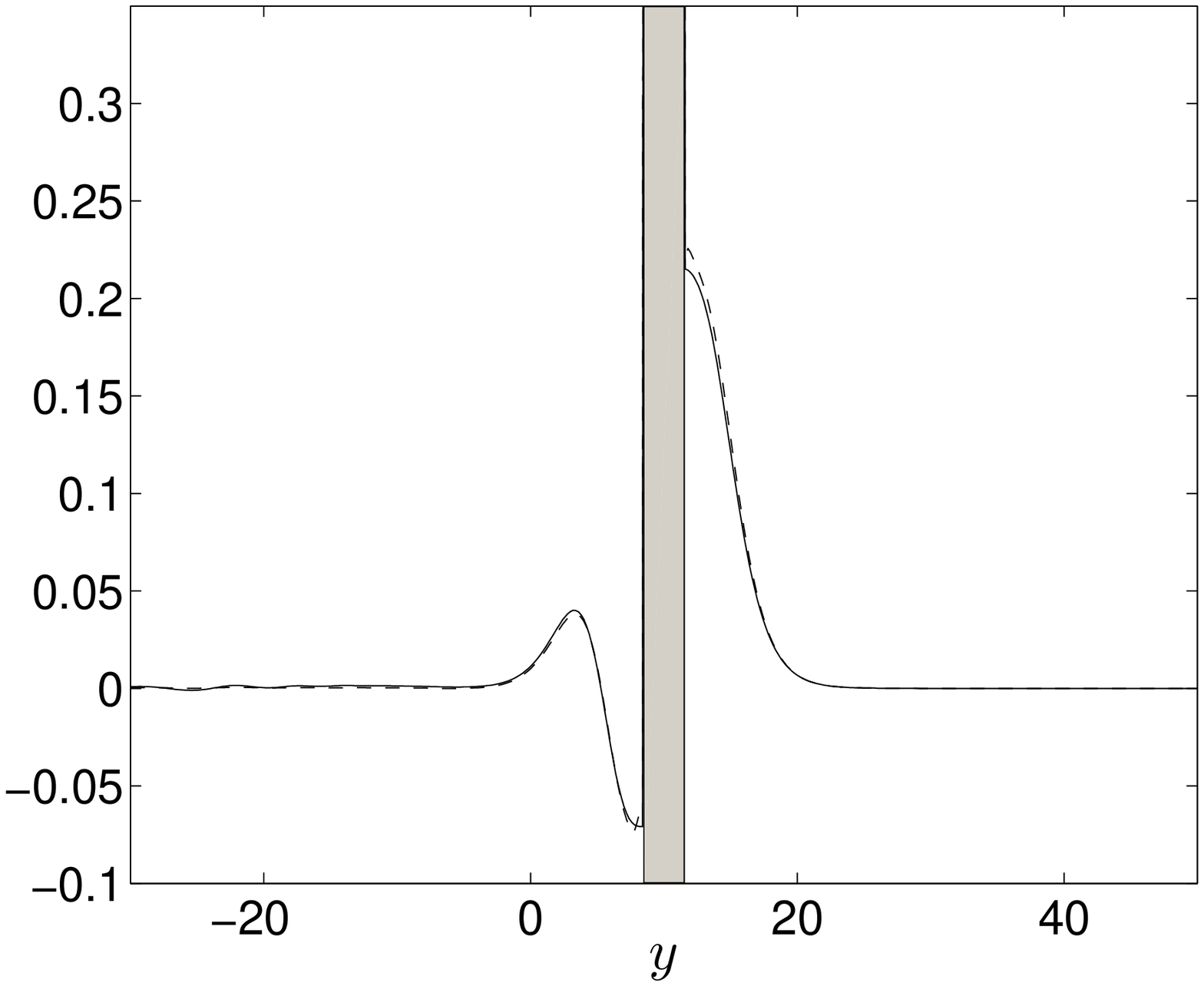} & \includegraphics[width=3in]{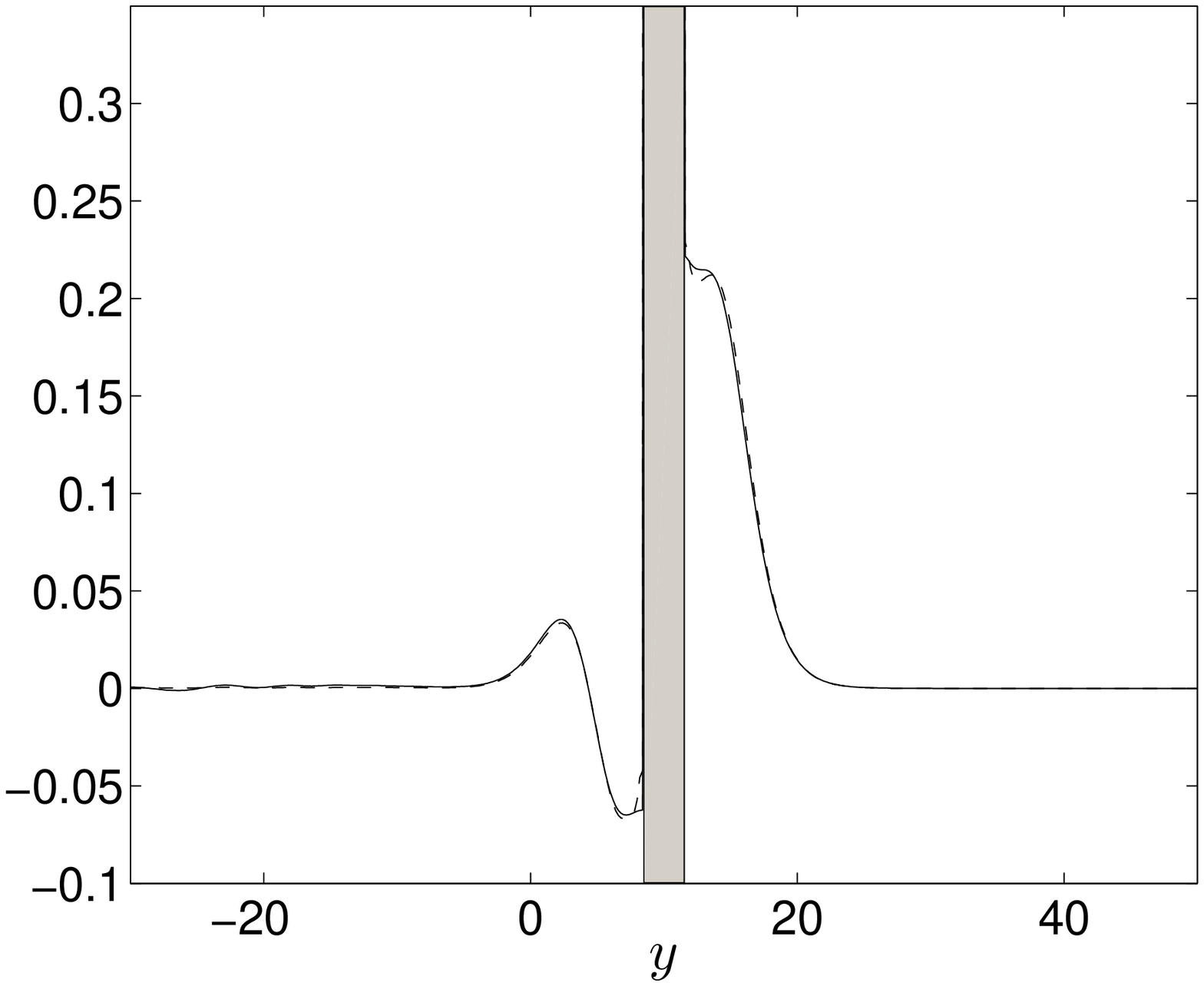}\\
$t=22$ & $t=23$ \\
\includegraphics[width=3in]{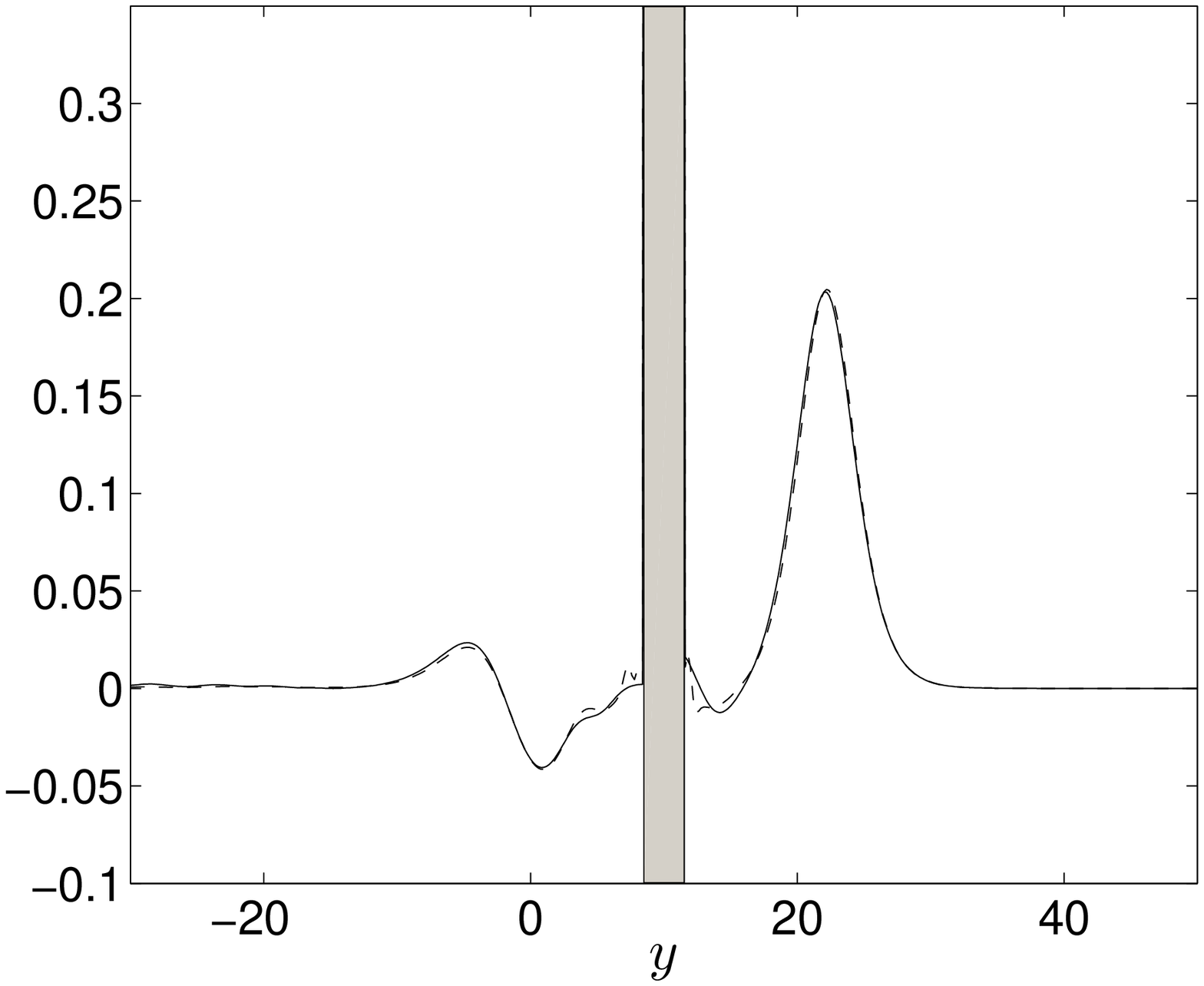} & \includegraphics[width=3in]{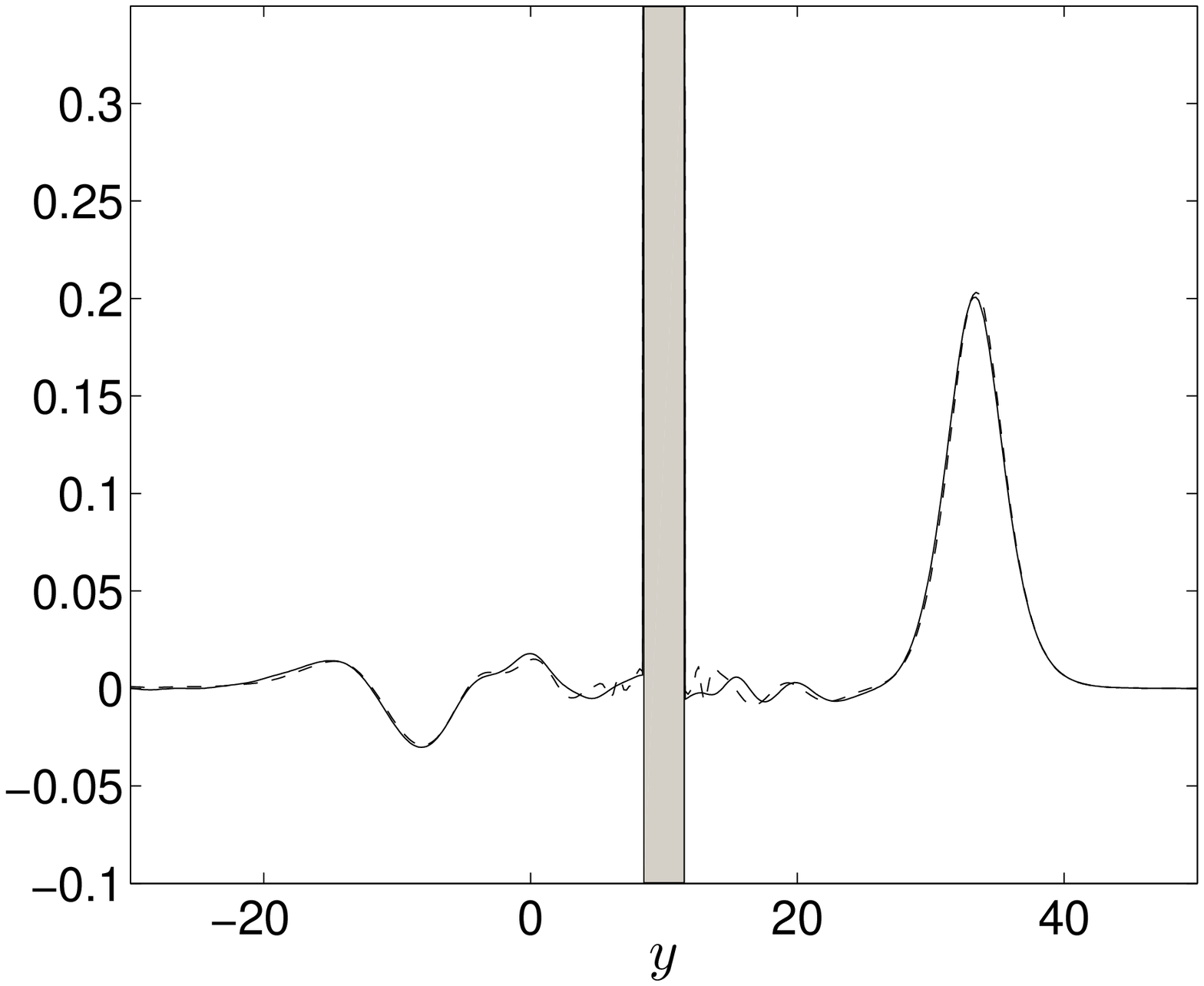}\\
$t=30$ & $t=40$ \\
\end{tabular}
  \caption{Experiment 4.2. Free surface elevation plots as functions of $y$ at $x=0$ at five time instances. BBM-BBM $- -$, Bona-Smith
 ---. }
  \end{center}\label{F4.3}
\end{figure}

\begin{figure}[p]
\begin{center}
\begin{tabular}{c}
\includegraphics[width=4in]{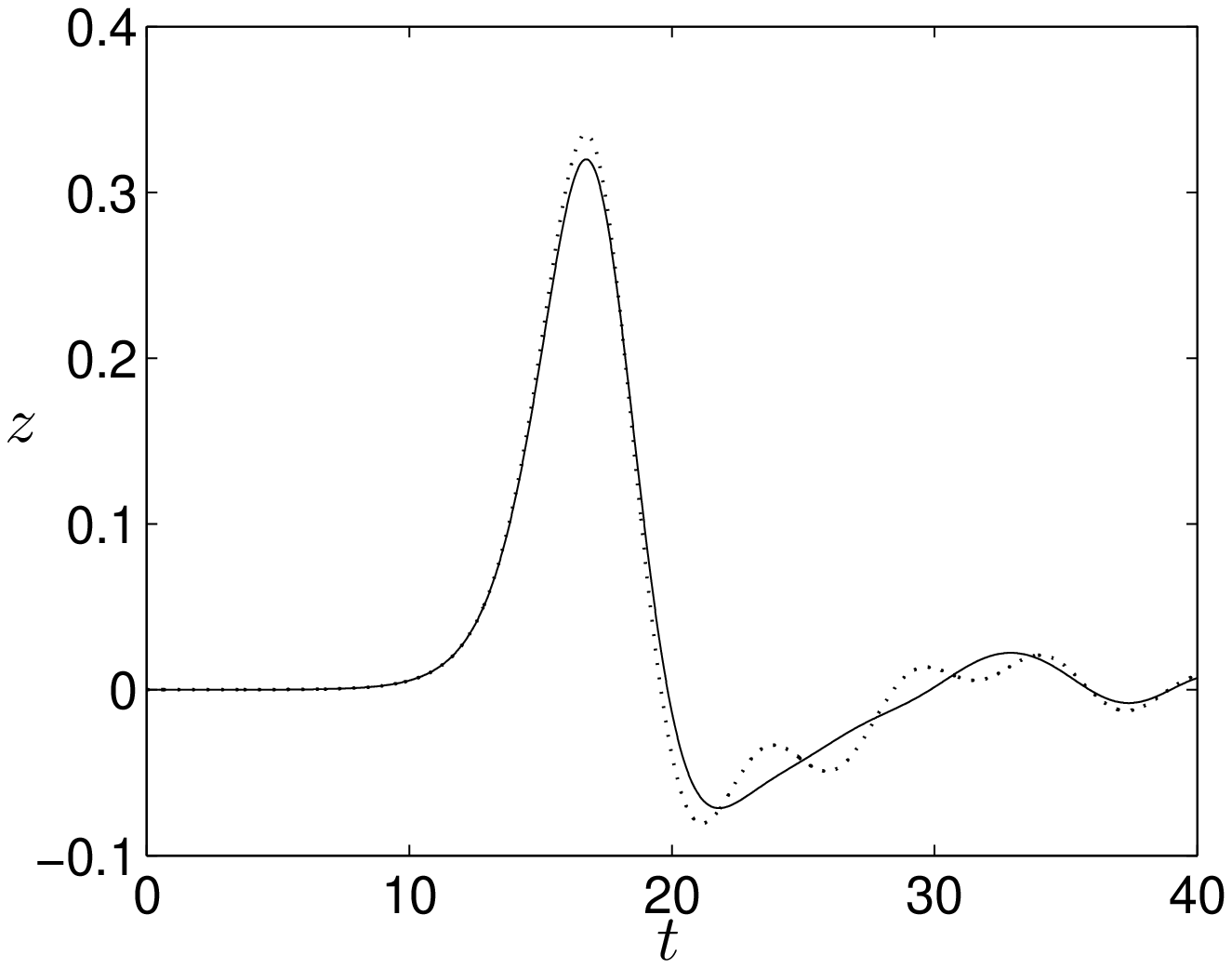}\\
\includegraphics[width=4in]{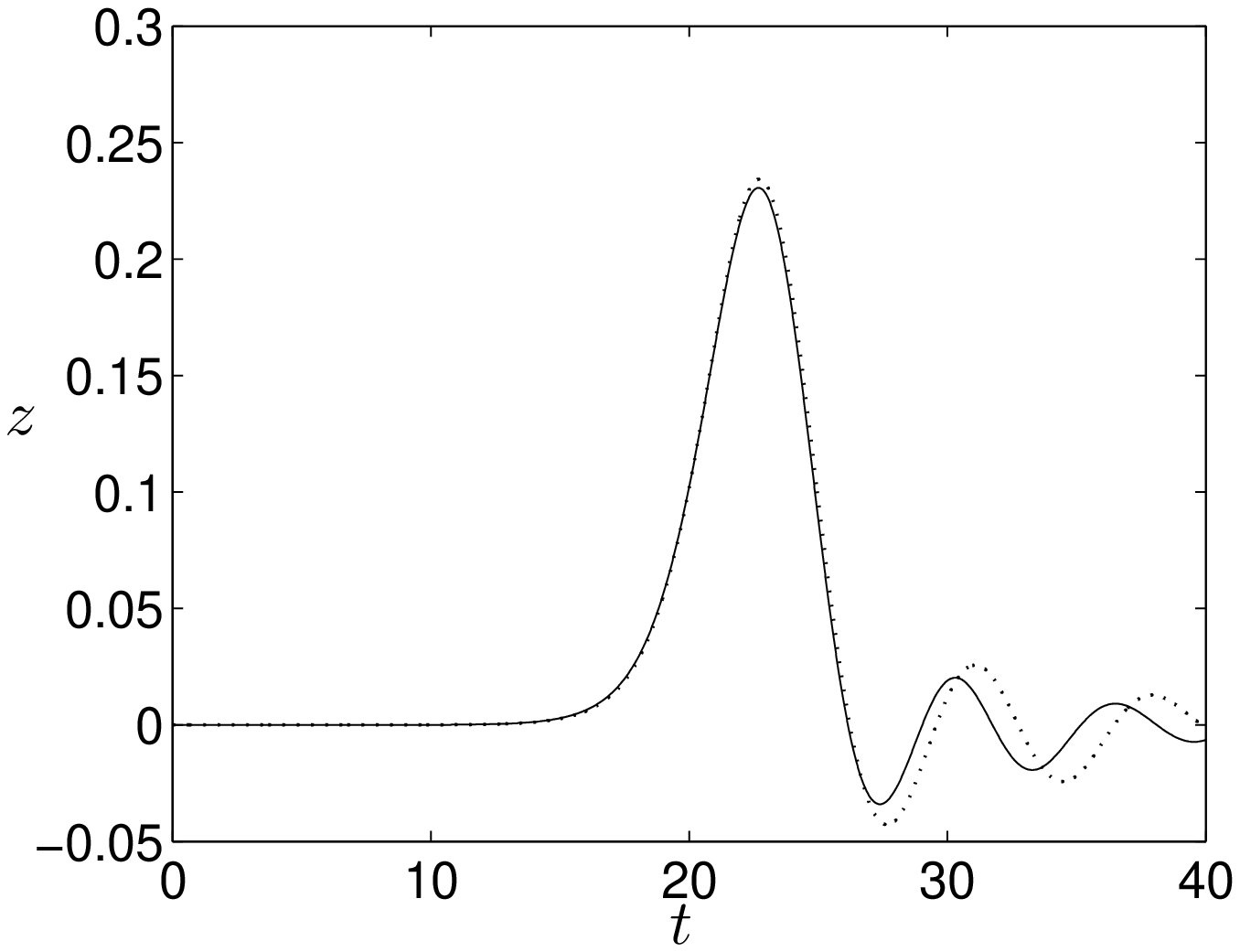}\\
\end{tabular}
\caption{Experiment 4.2. $\eta$ as a function of $t$ at $(x,y)=(0,8.5)$ and $(x,y)=(0,11.5)$. --: Bona-Smith system, $\cdots$: BBM-BBM system. }
  \end{center}
\end{figure}\label{F4.4}

\subsection{Evolution and reflection at the boundaries of a `heap' of water}

The sequence of plots of Figure 4.6 shows the temporal evolution of the free surface elevation of an initial Gaussian `heap' of water with
$$\eta_0(x,y)=2e^{-((x+40)^2+(y+40)^2)/5},\quad {\bf v}_0(x,y)=0,$$
as it collapses, forms radial outgoing riples and interacts with the reflective boundaries near a corner of the square $[-80,80]\times [-80,80]$. Again, no major differences were observed between the solutions of the two systems (BBM-BBM and Bona-Smith, $\theta^2=9/11$). The computation was effected with $N=84992$ triangular elements and $\Delta t=0.1$. Figure 4.7 shows the corresponding one-dimensional plots (along $x=-40$) of the free surface elevation for both systems at four temporal instanses.

\begin{figure*}[p]
\begin{center}
\begin{tabular}{cc}
\includegraphics[width=2.5in]{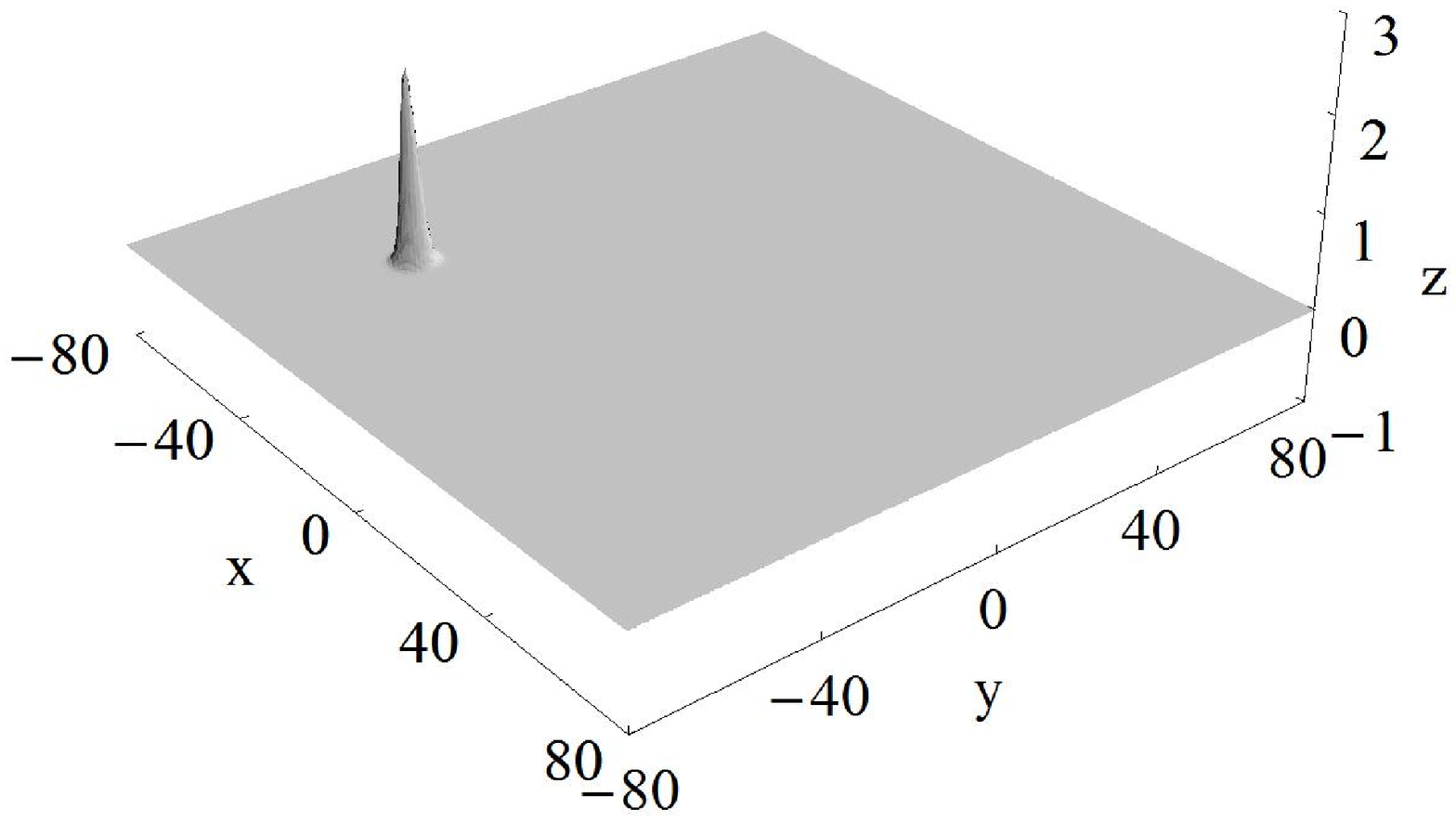} &
\includegraphics[width=2.5in]{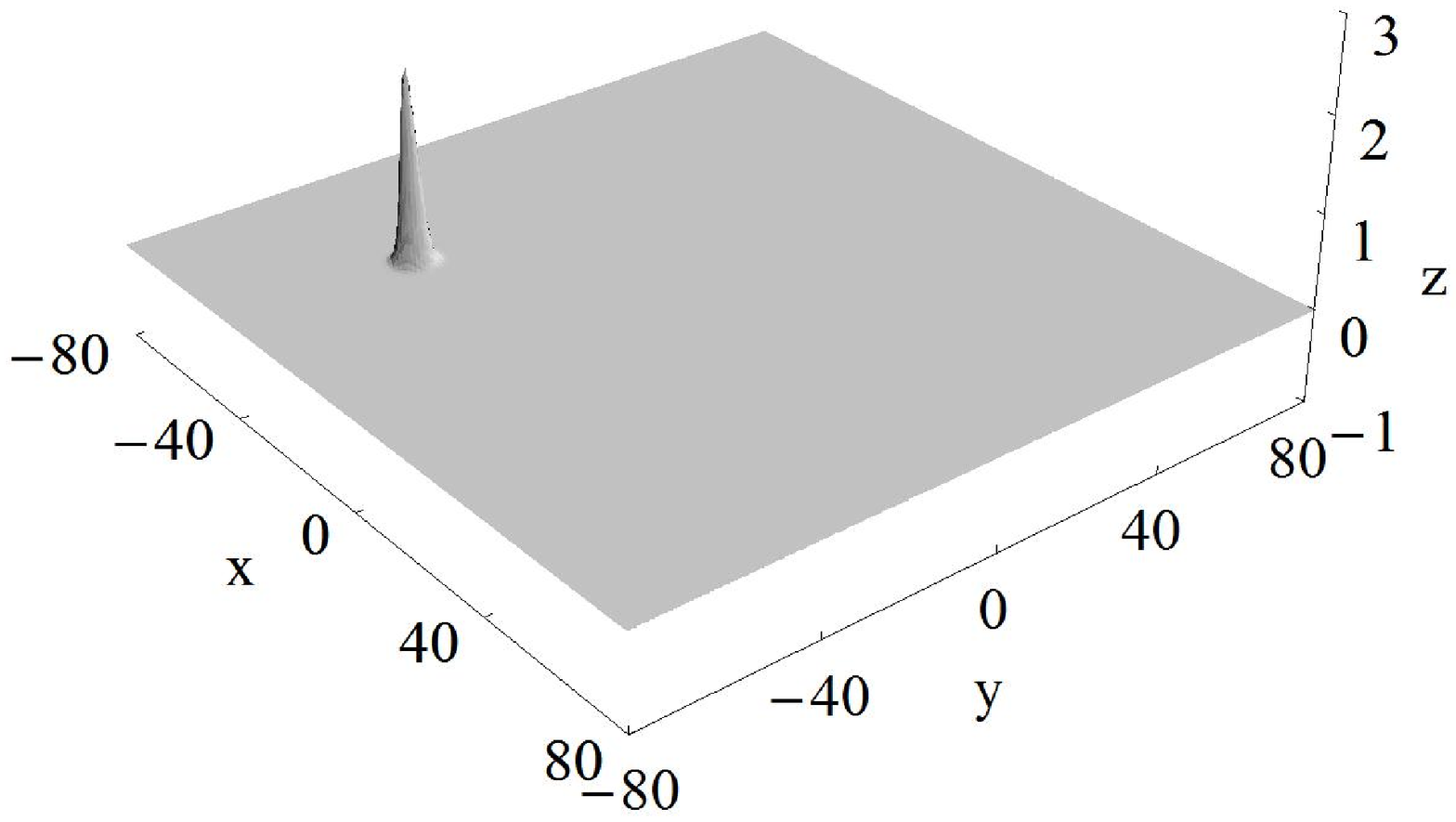}\\
BBM-BBM $t=0$ & Bona-Smith $t=0$\\
\includegraphics[width=2.5in]{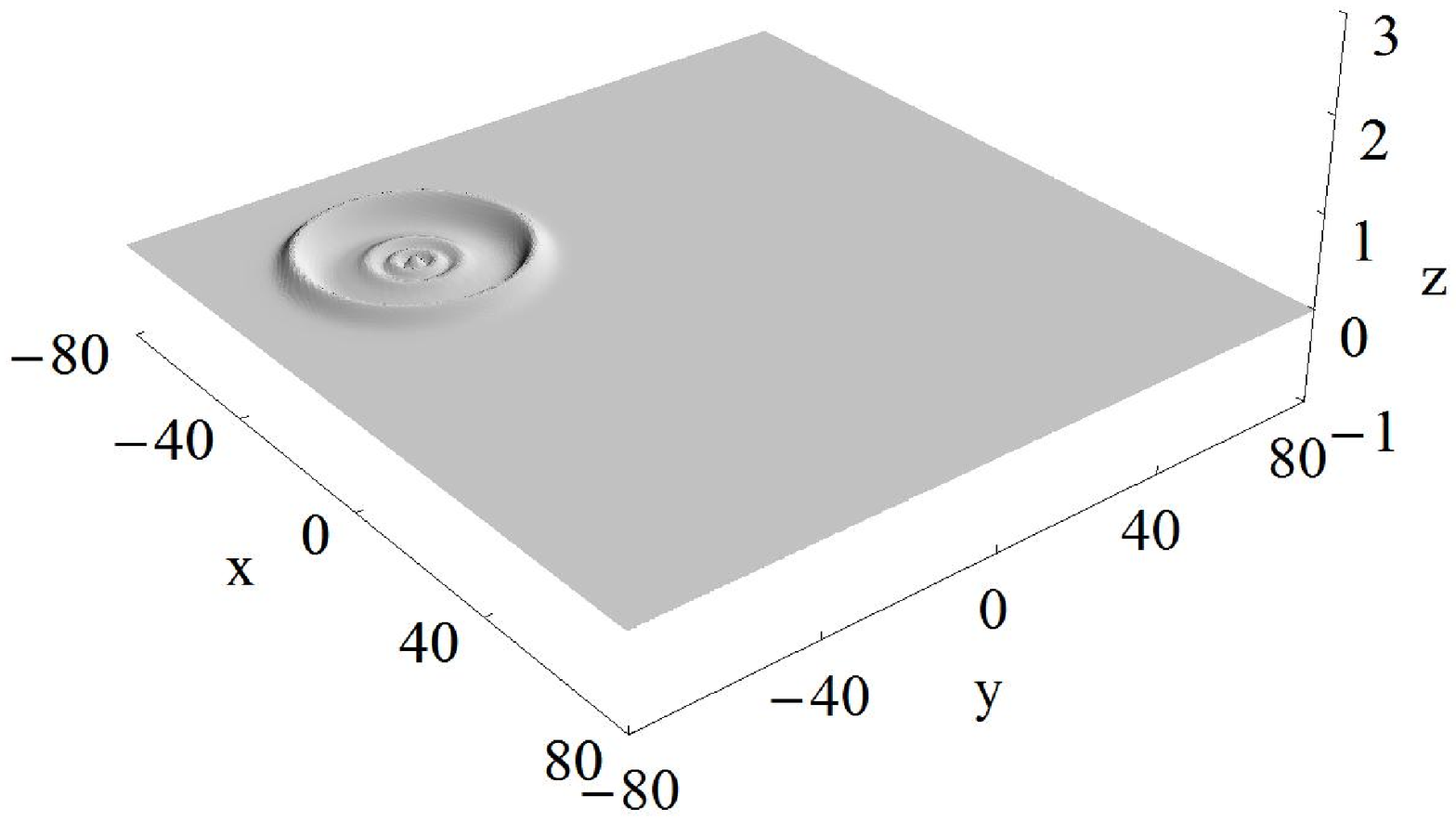} &\includegraphics[width=2.5in]{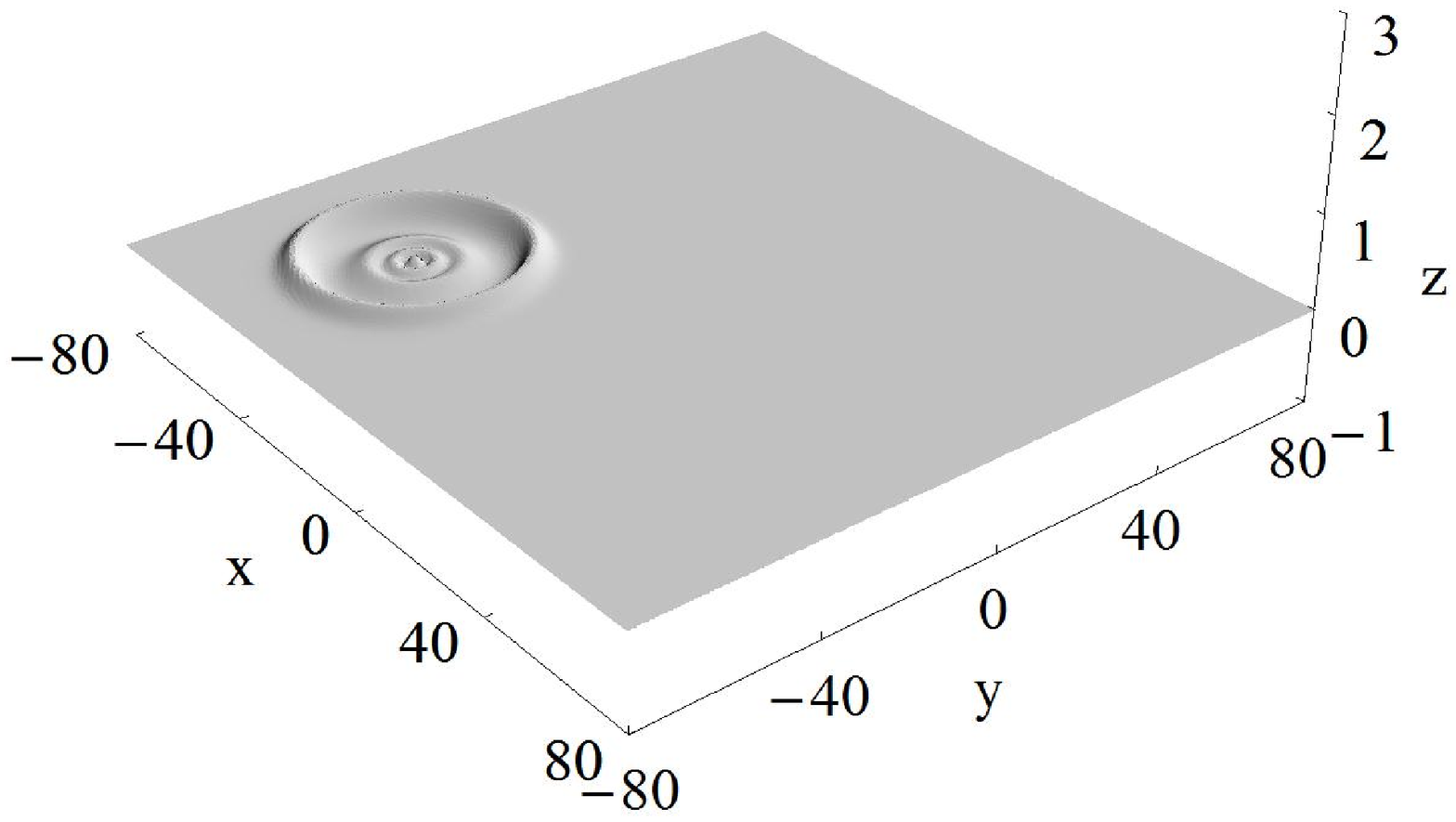}\\
BBM-BBM $t=20$ & Bona-Smith $t=20$\\
\includegraphics[width=2.5in]{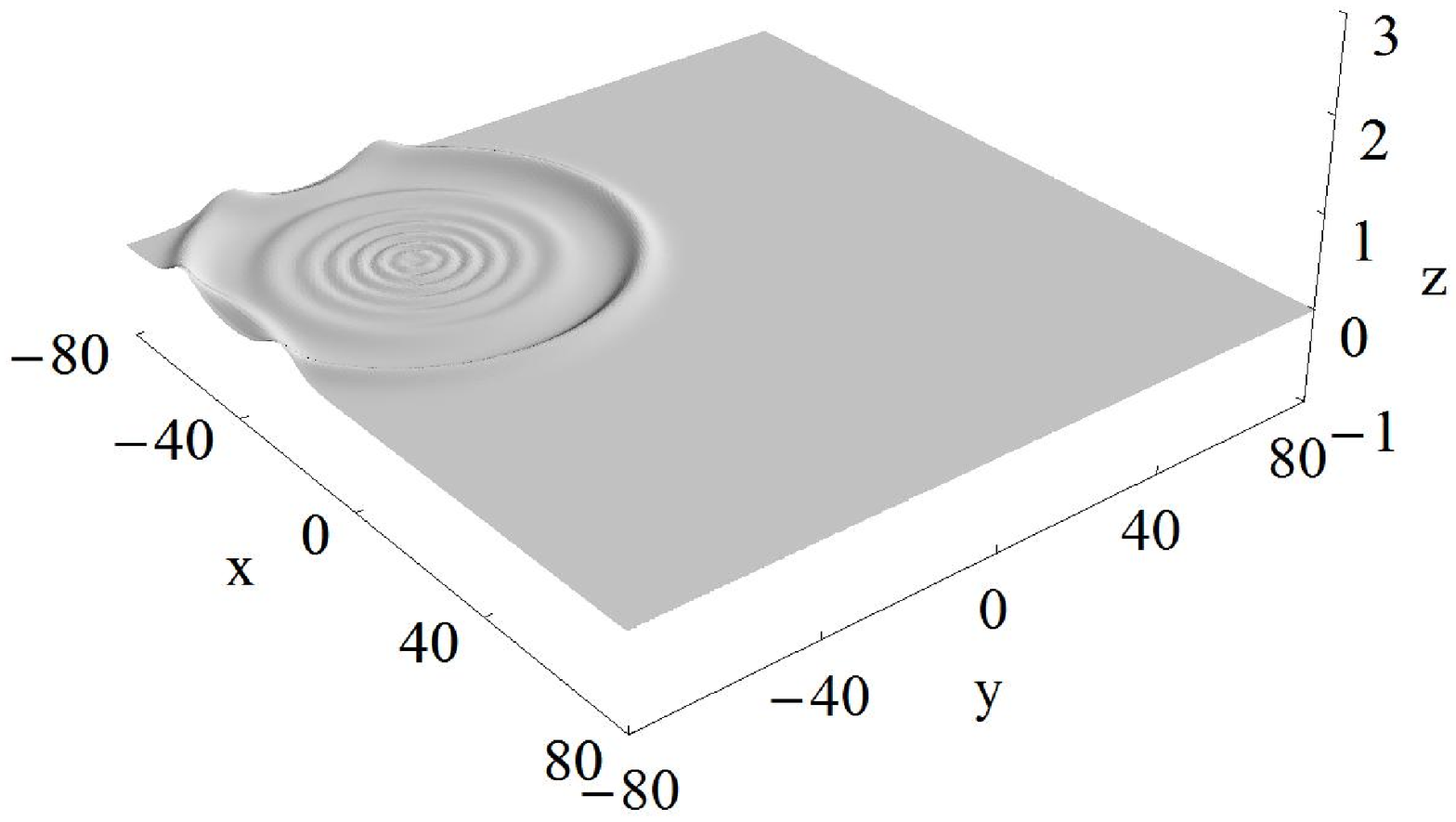} &\includegraphics[width=2.5in]{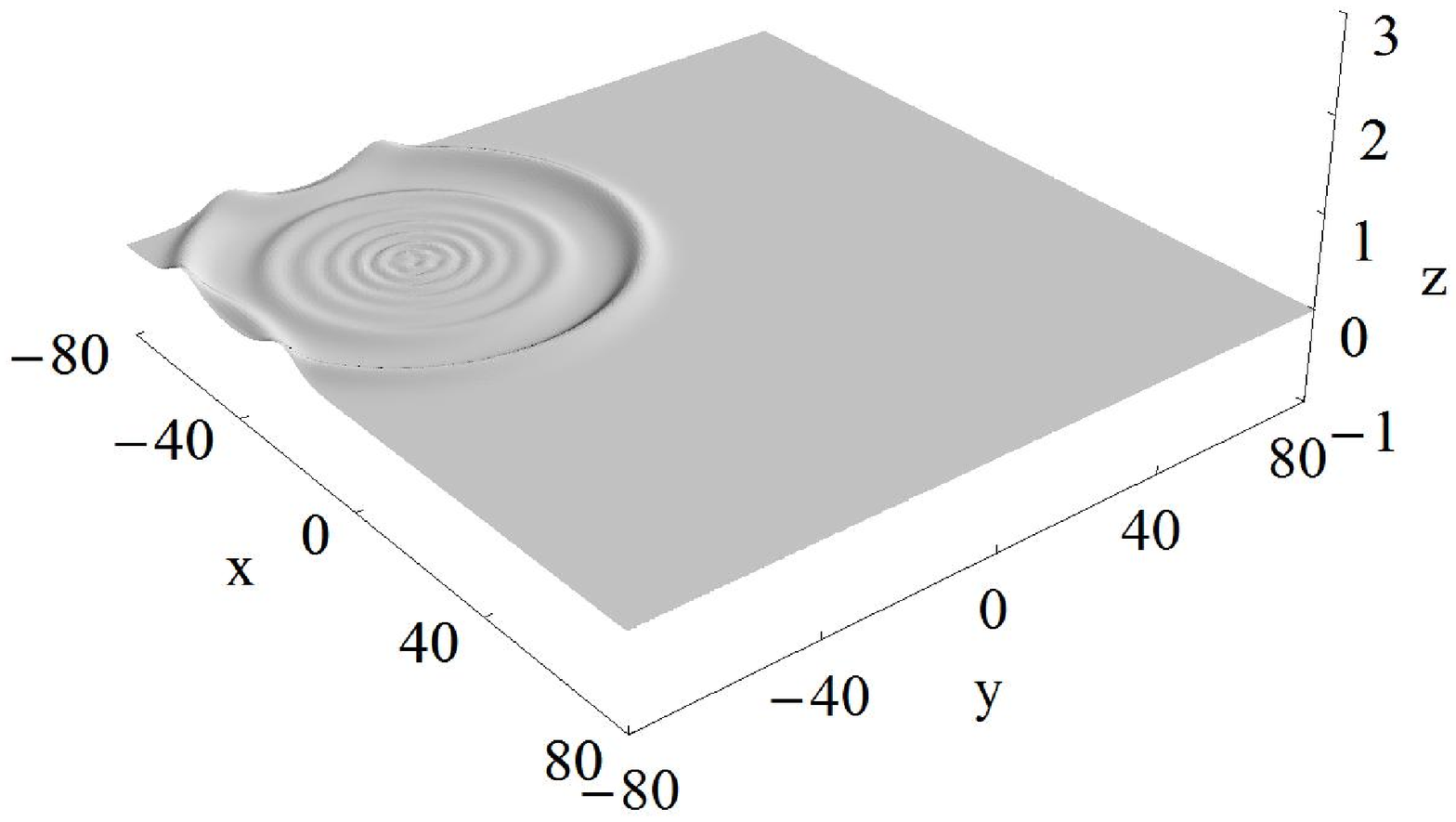}\\
BBM-BBM $t=40$ & Bona-Smith $t=40$
\end{tabular}
\end{center}
\end{figure*}

\begin{figure}[p]
\begin{center}
\begin{tabular}{cc}
\includegraphics[width=2.5in]{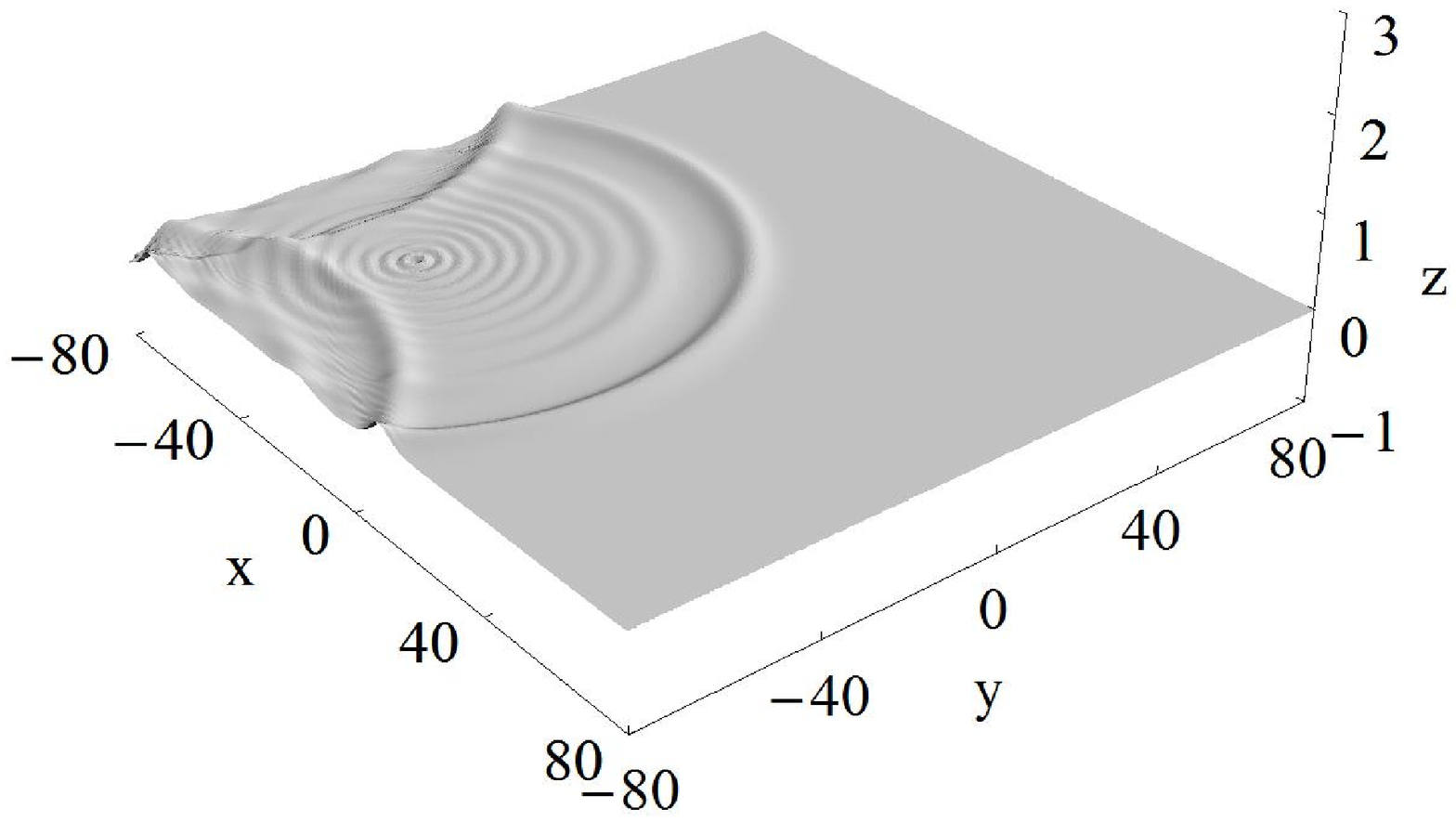} &\includegraphics[width=2.5in]{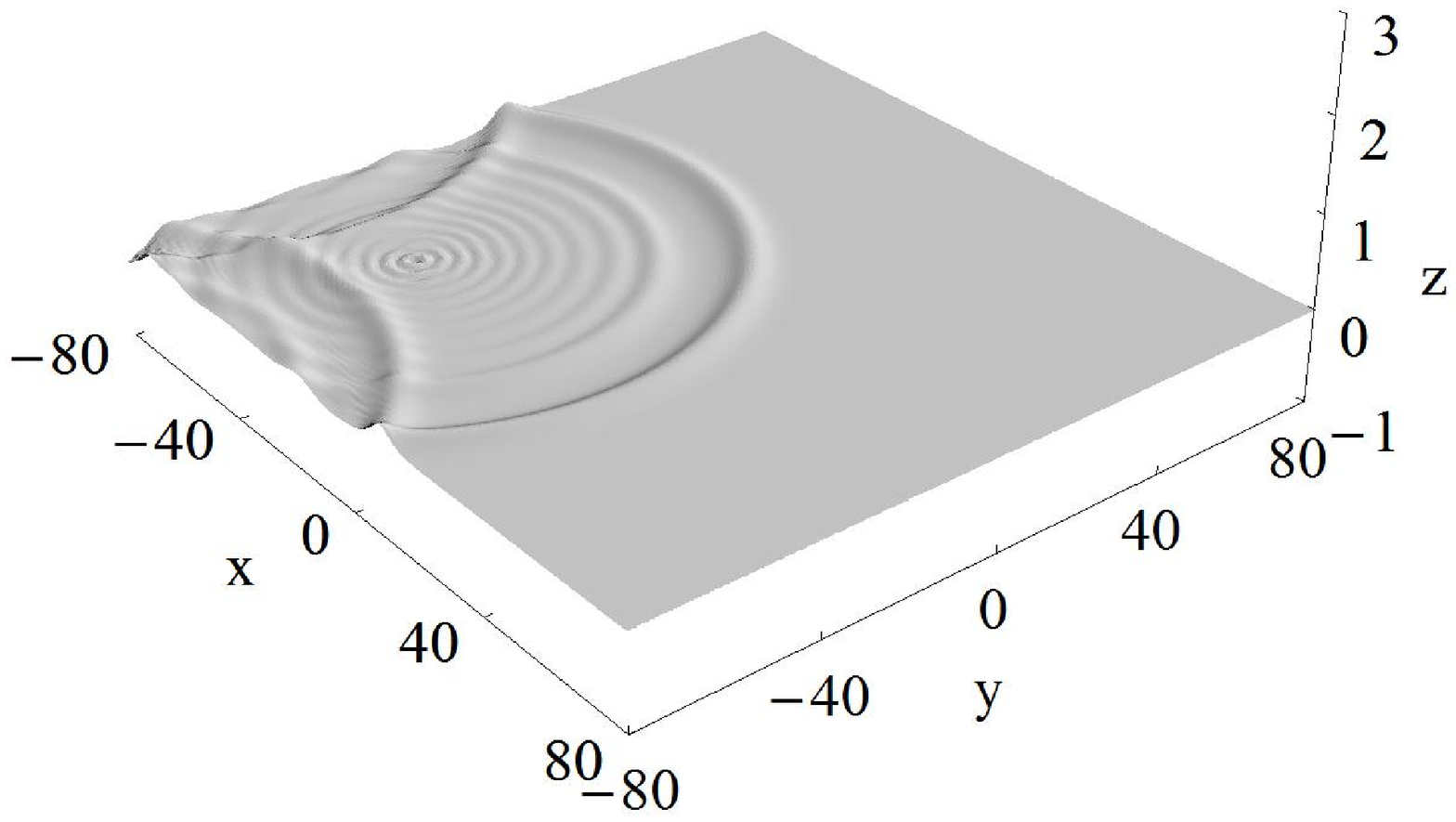}\\
BBM-BBM $t=60$ & Bona-Smith $t=60$\\
\includegraphics[width=2.5in]{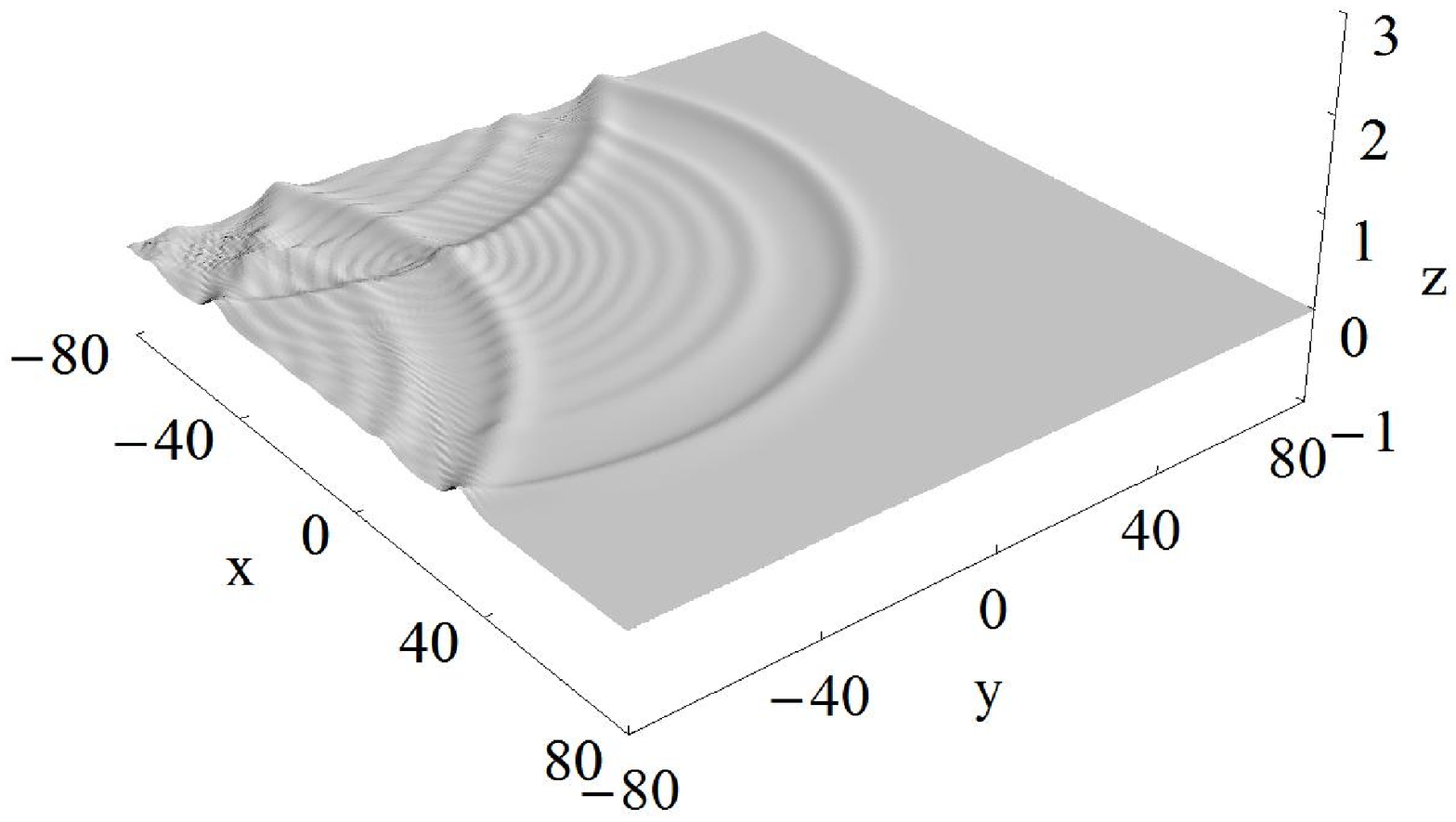} &\includegraphics[width=2.5in]{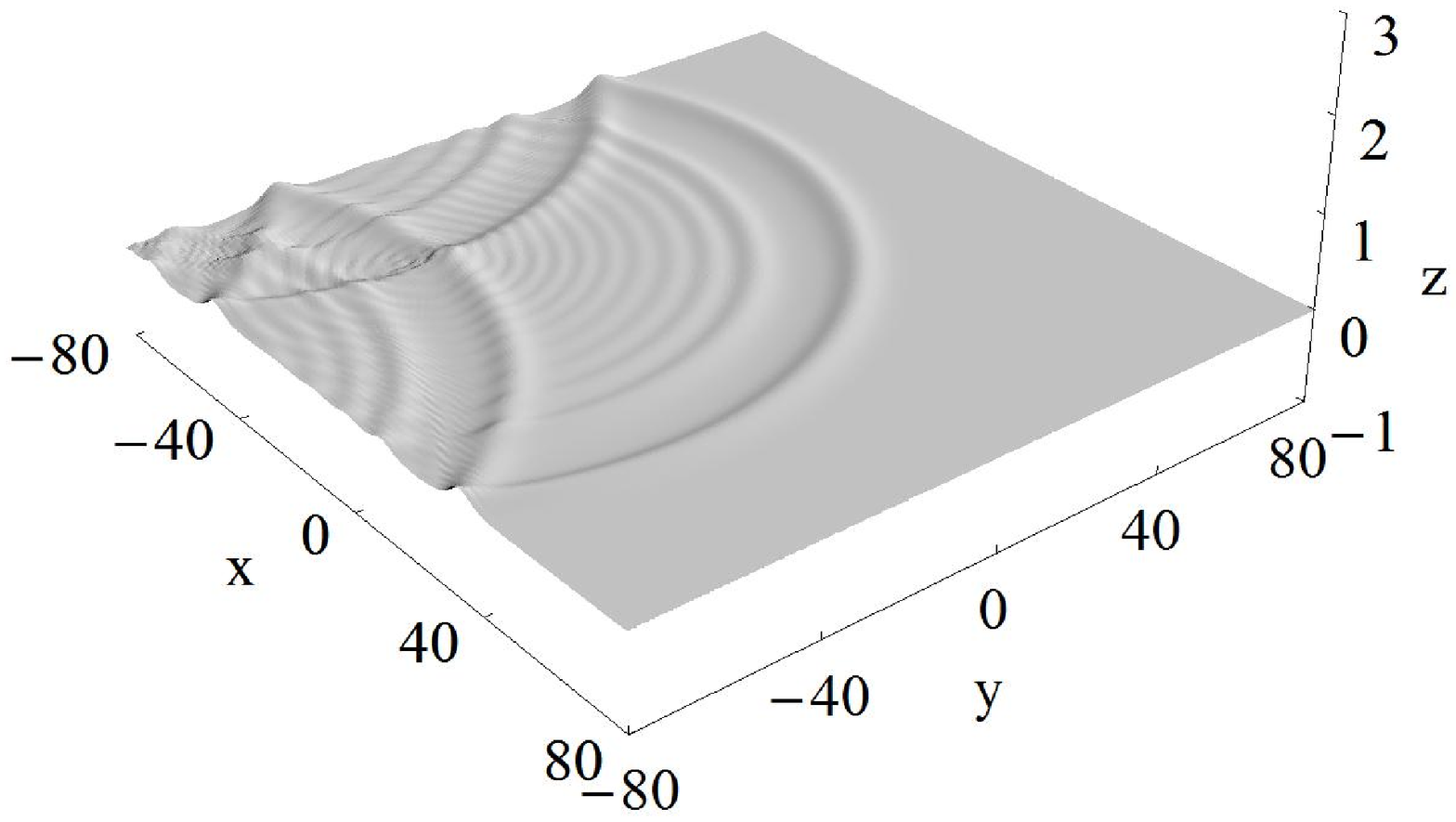}\\
BBM-BBM $t=80$ & Bona-Smith $t=80$
\end{tabular}
  \caption{Experiment 4.3. Free surface elevation at five time instances.}
  \end{center}
\end{figure}

\begin{figure}[p]
\begin{center}
\begin{tabular}{cc}
\includegraphics[width=3in]{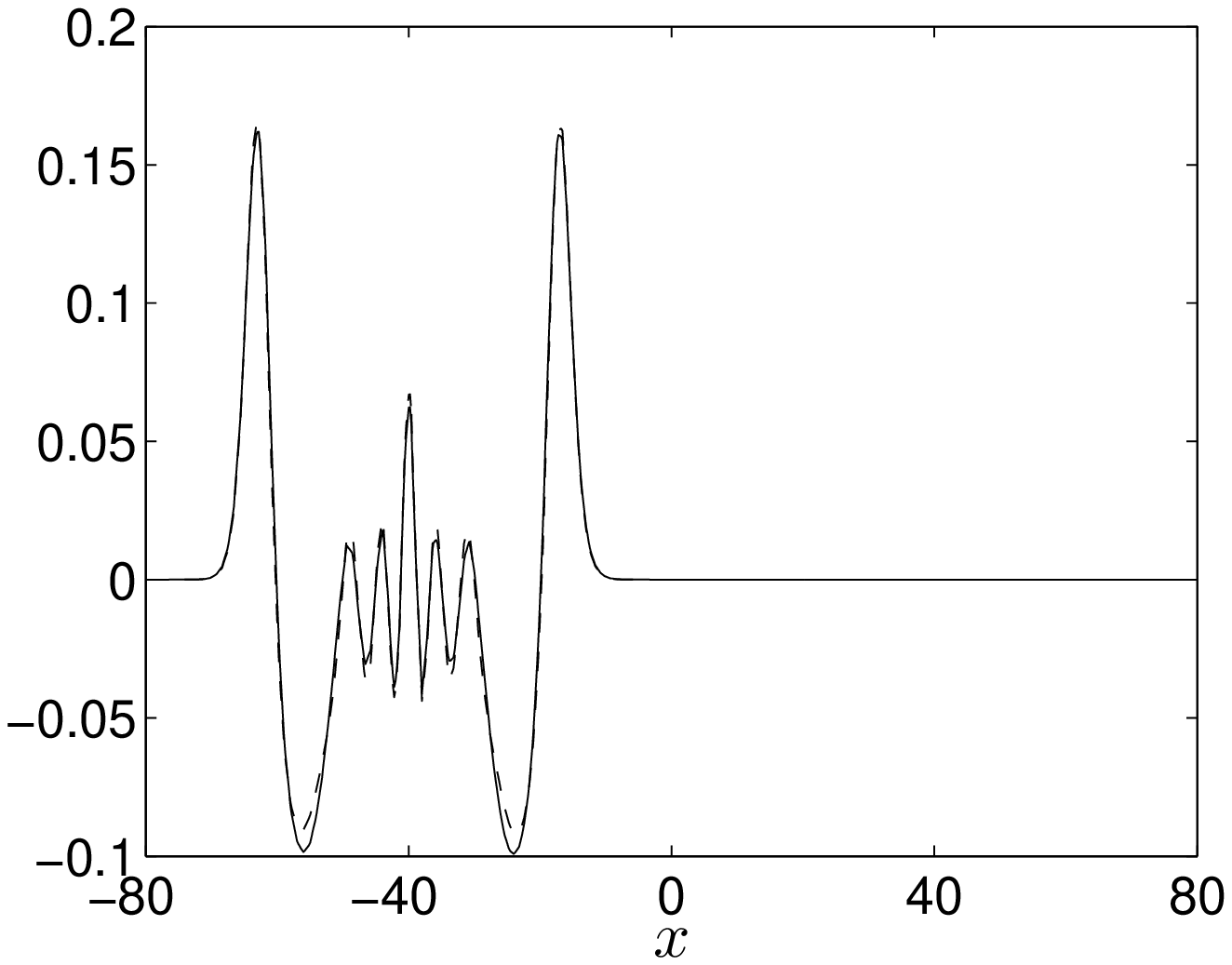} & \includegraphics[width=3in]{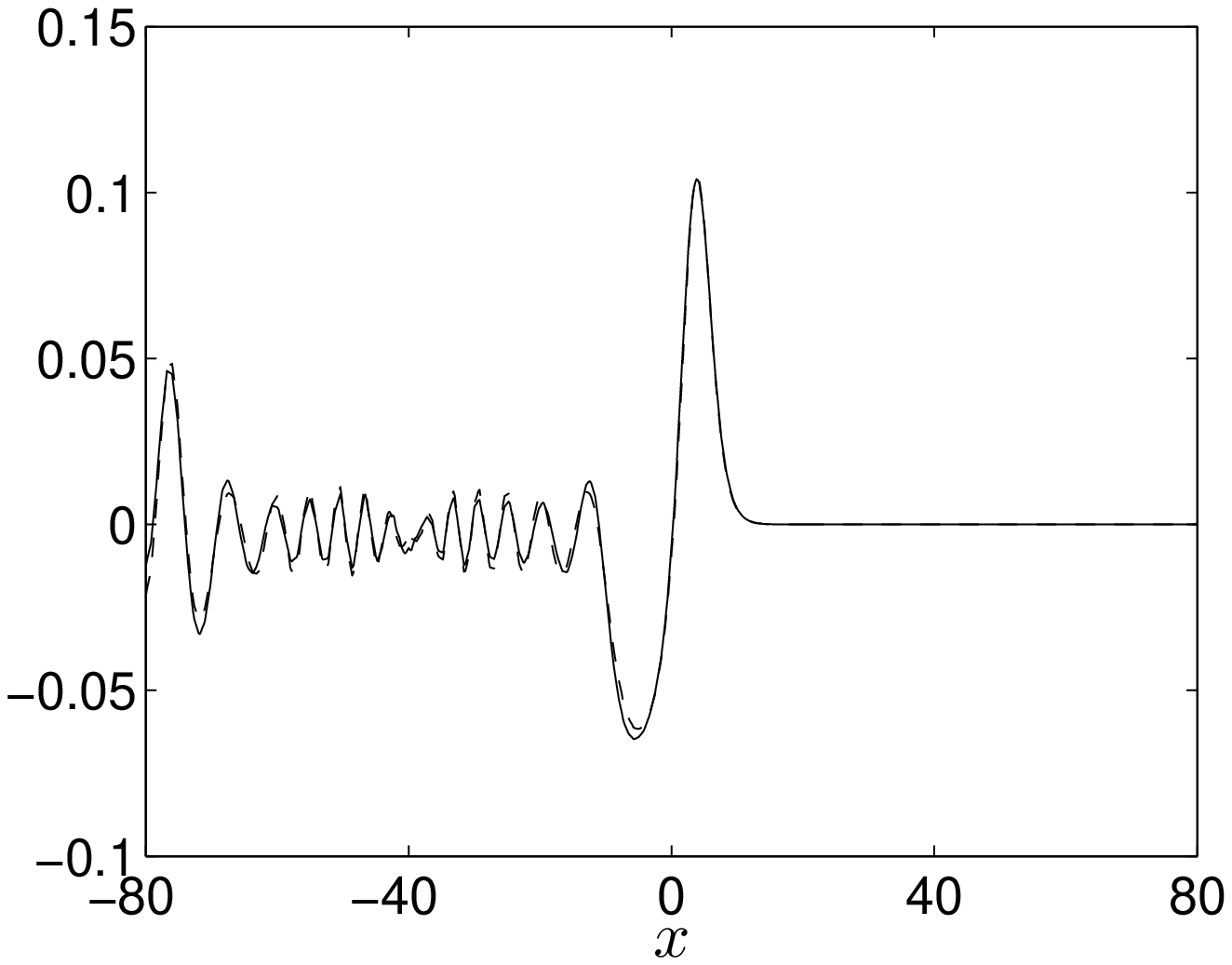}\\
$t=20$ &
$t=40$\\
\includegraphics[width=3in]{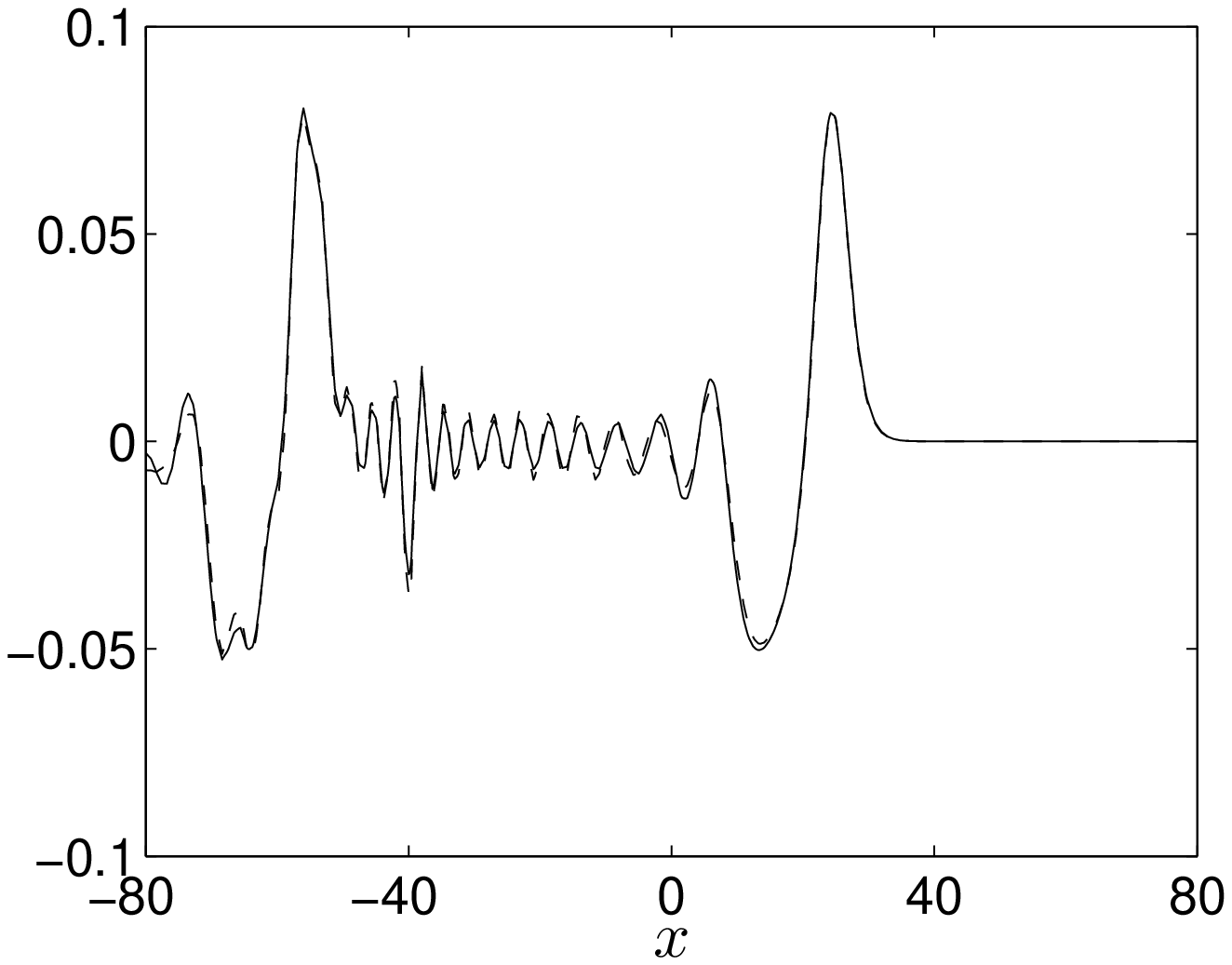}&
\includegraphics[width=3in]{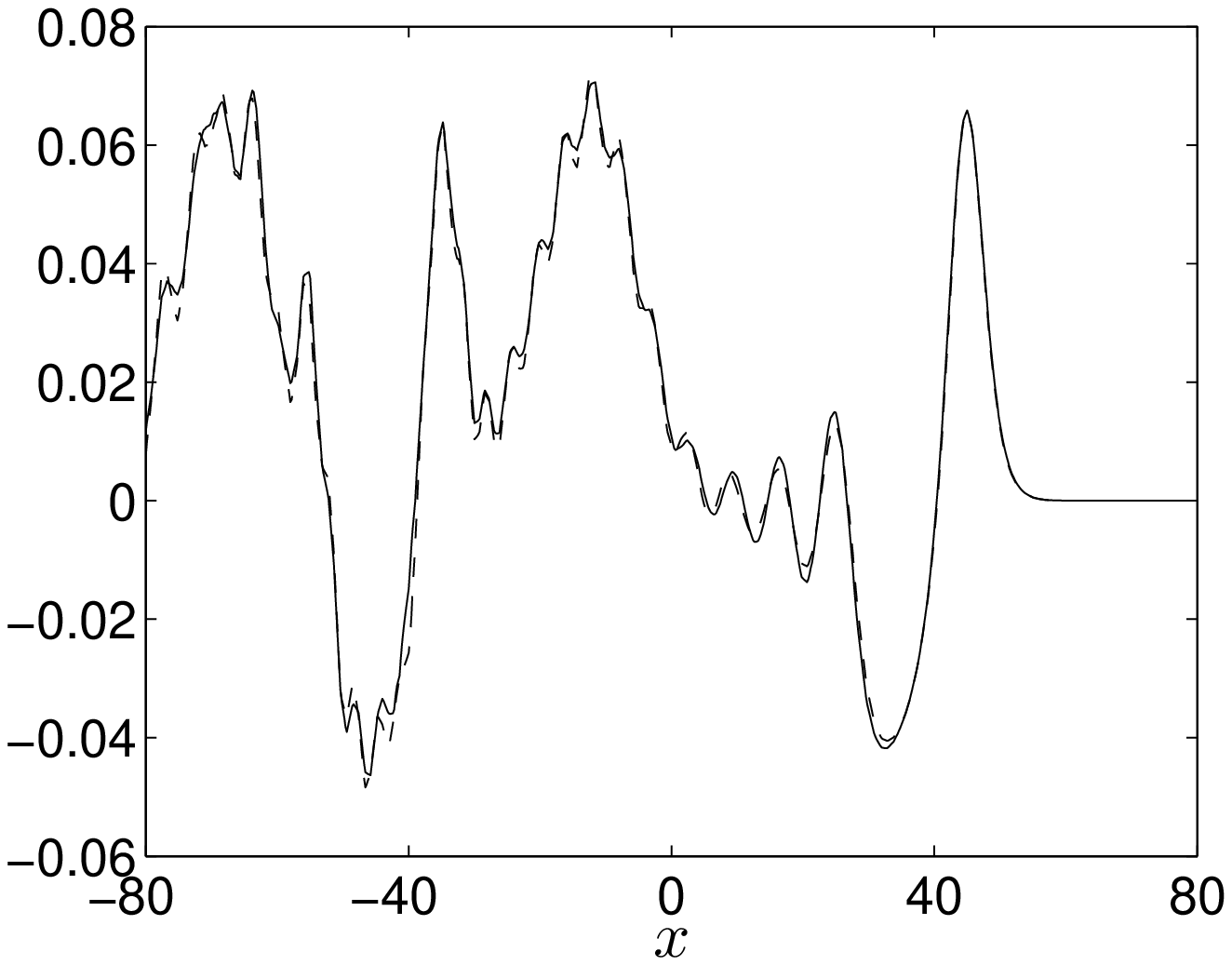}\\
$t=60$&
$t=80$
\end{tabular}
  \caption{Experiment 4.3. Free surface elevation at four time instances along $y=-40m$. BBM-BBM $- -$, Bona-Smith ($\theta^2=9/11$)
 ---. }
  \end{center}
\end{figure}

\subsection{Line wave impinging on a `port' structure}

In this experiment we integrated the BBM-BBM and the Bona-Smith ($\theta^2=9/11$) systems in dimensional variables. Our domain represents part of a `port' of depth $h_0=50m$ consisting of the rectangle $[-250,250]\times [0,2000]$ minus the rectangular `pier' $[0,100]\times [0,700]$. (All distances in meters). Normal reflective boundary conditions are assumed to hold along the boundary of the pier and the intervals $[-250,0]$ and $[100,250]$ of the $x$-axis, while homogeneous Neumann boundary conditions have been imposed on $\eta$ and ${\bf v}$ on the remaining parts of the boundary. An initial wave with
$\eta_0(x,y)=A e^{-\frac{y-1500}{6000}}$, $A=1\, m$, and $u_0(x,y)=0\, m/sec$, $v_0(x,y)=-\frac{1}{2}(\eta_0(x,y)-\frac{1}{4}\eta_0^2(x,y))\, m/sec$ travels mainly towards the negative $y$ direction shedding a dispersive tail behind. (Both systems were integrated with $N=77632$ triangles and $\Delta t=0.01\, sec$). For the impinging wave at $t=30\, sec$ we estimated that $A/h_0\cong 0.014$, $\lambda/h_0\cong 16.6$, so that the Stokes number is approximately equal to $3.9$, within the range of validity of the Boussinesq systems. The incoming wave hits the pier and the port boundary at $x=0$, is reflected backwards and interacts with the remaining boundary. Figure 4.8 shows contour plots of the free surface elevation for both systems at several temporal instances in the whole domain. In Figure 4.9 we plot the computed free surface elevation as a function of $y$ along the $x=40\, m$ line at several temporal instances as the wave hits the pier front ($y=700$) and reflects backwards. Most of the differences in the solution of the two systems are of the order of $10\, cm$ and are observed in the reflection phase. Figure 4.10 shows the temporal history of the free surface elevation for both systems at the point $(x,y)=(43.75,700)$ at the front of the pier. The maximum run-up observed at that point was equal to $z=0.976\, m$ (at $t=38.6\, sec$) for the BBM-BBM system and to $z=0.988\, m$ (at $t=38.7$) for the Bona-Smith system.

\begin{figure}[p]
\begin{center}
\begin{tabular}{cc}
\includegraphics[width=2.5in]{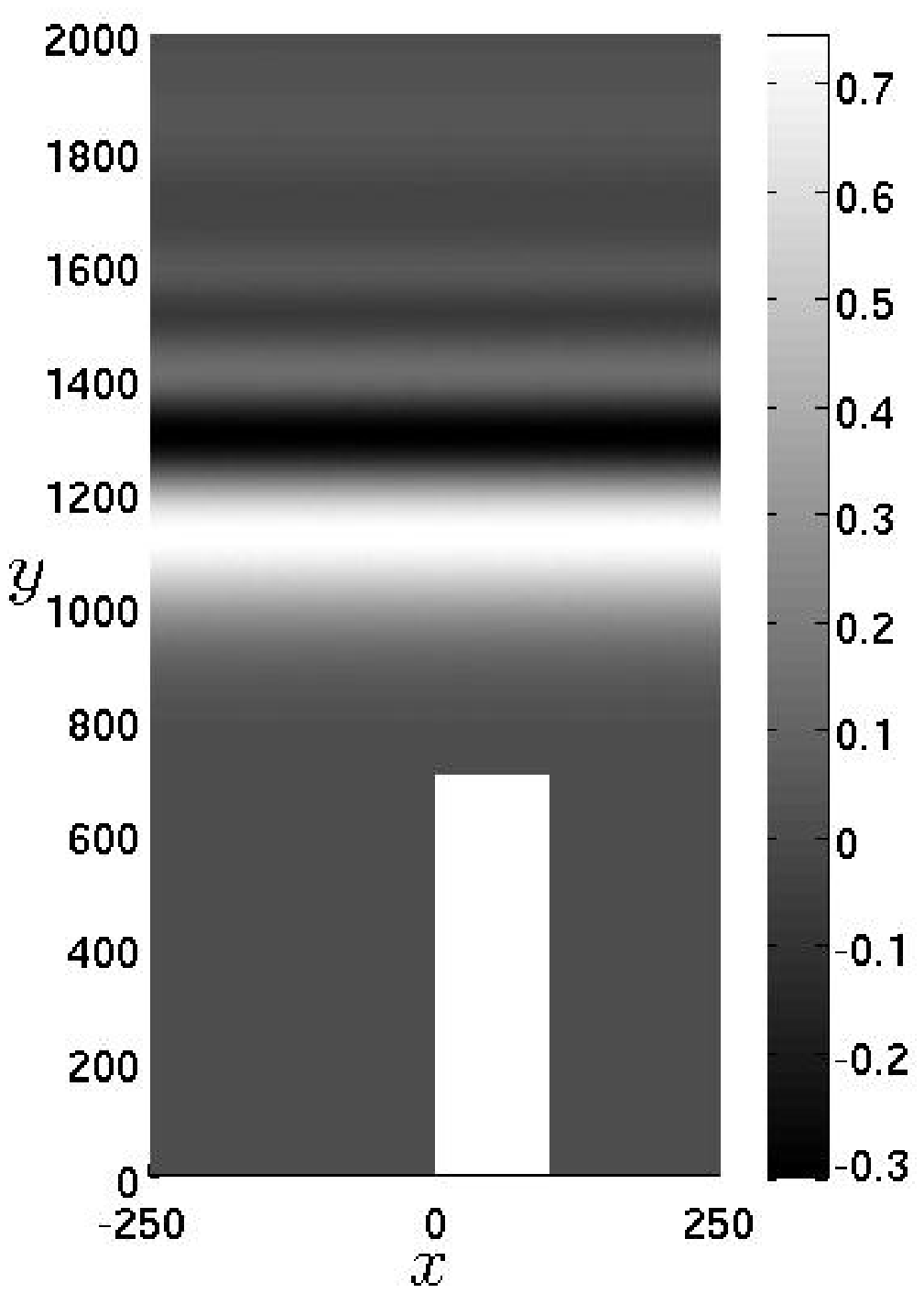} &\includegraphics[width=2.5in]{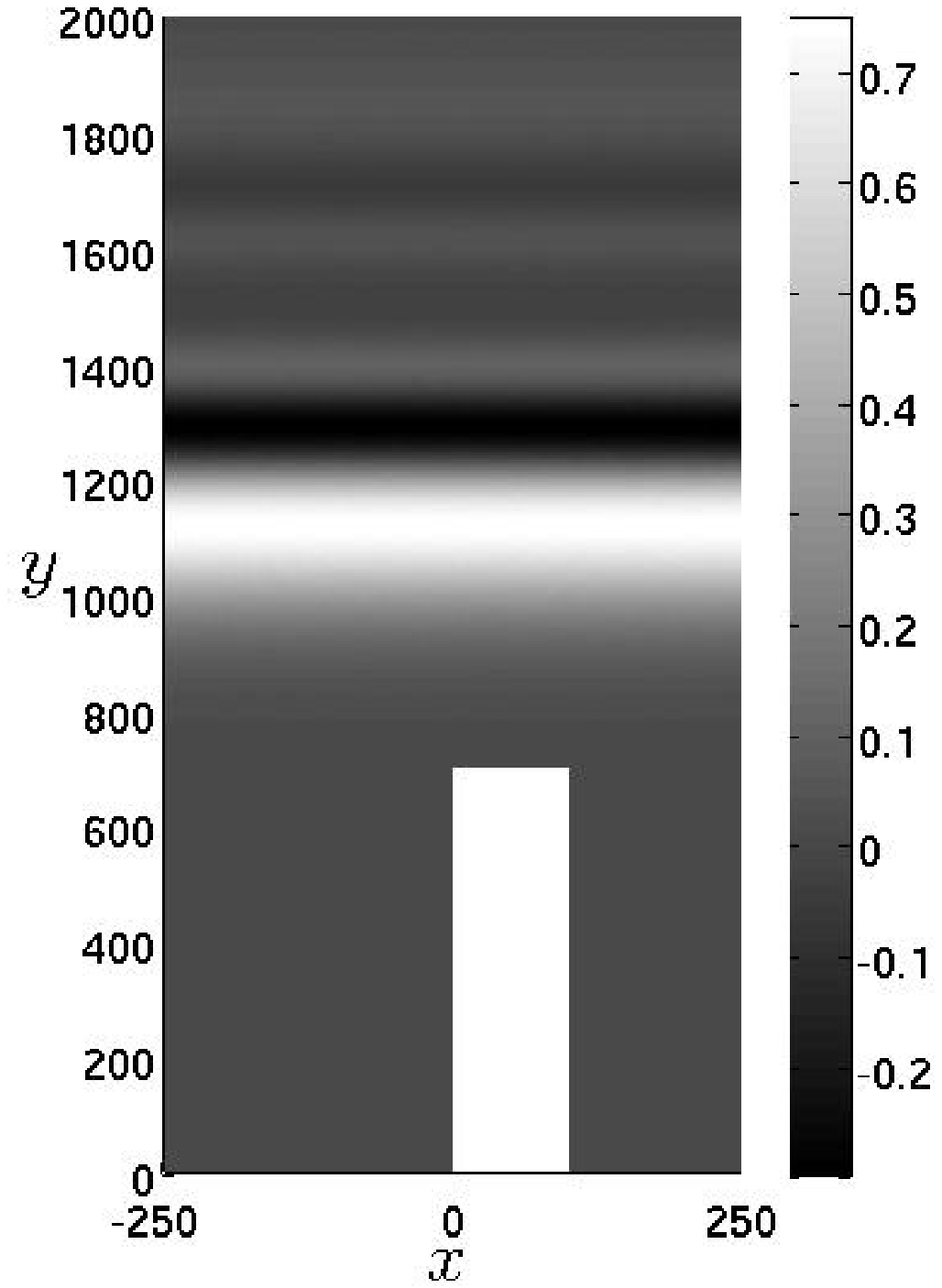}\\
BBM-BBM $t=20$ & Bona-Smith $t=20$\\
\includegraphics[width=2.5in]{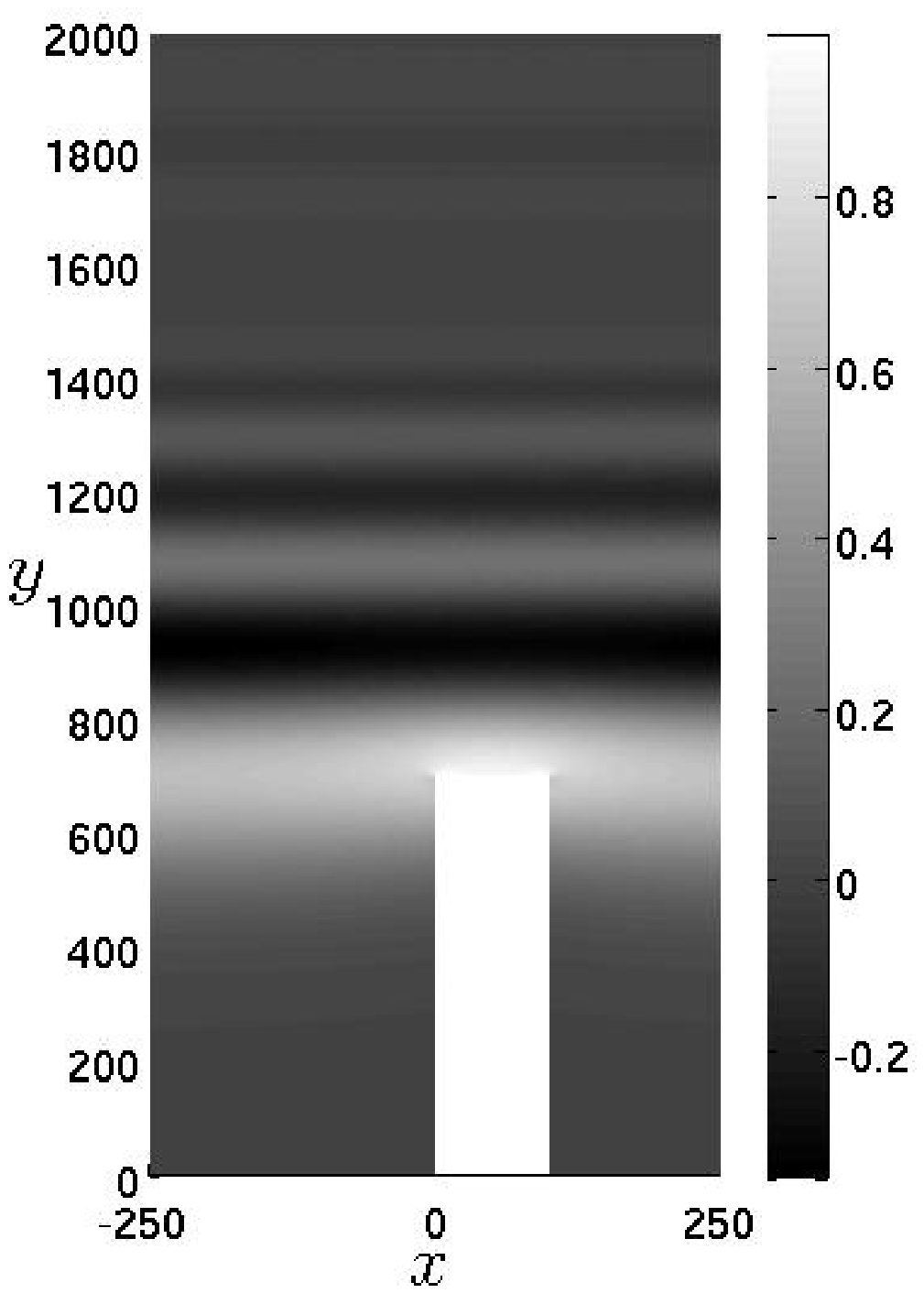} &\includegraphics[width=2.5in]{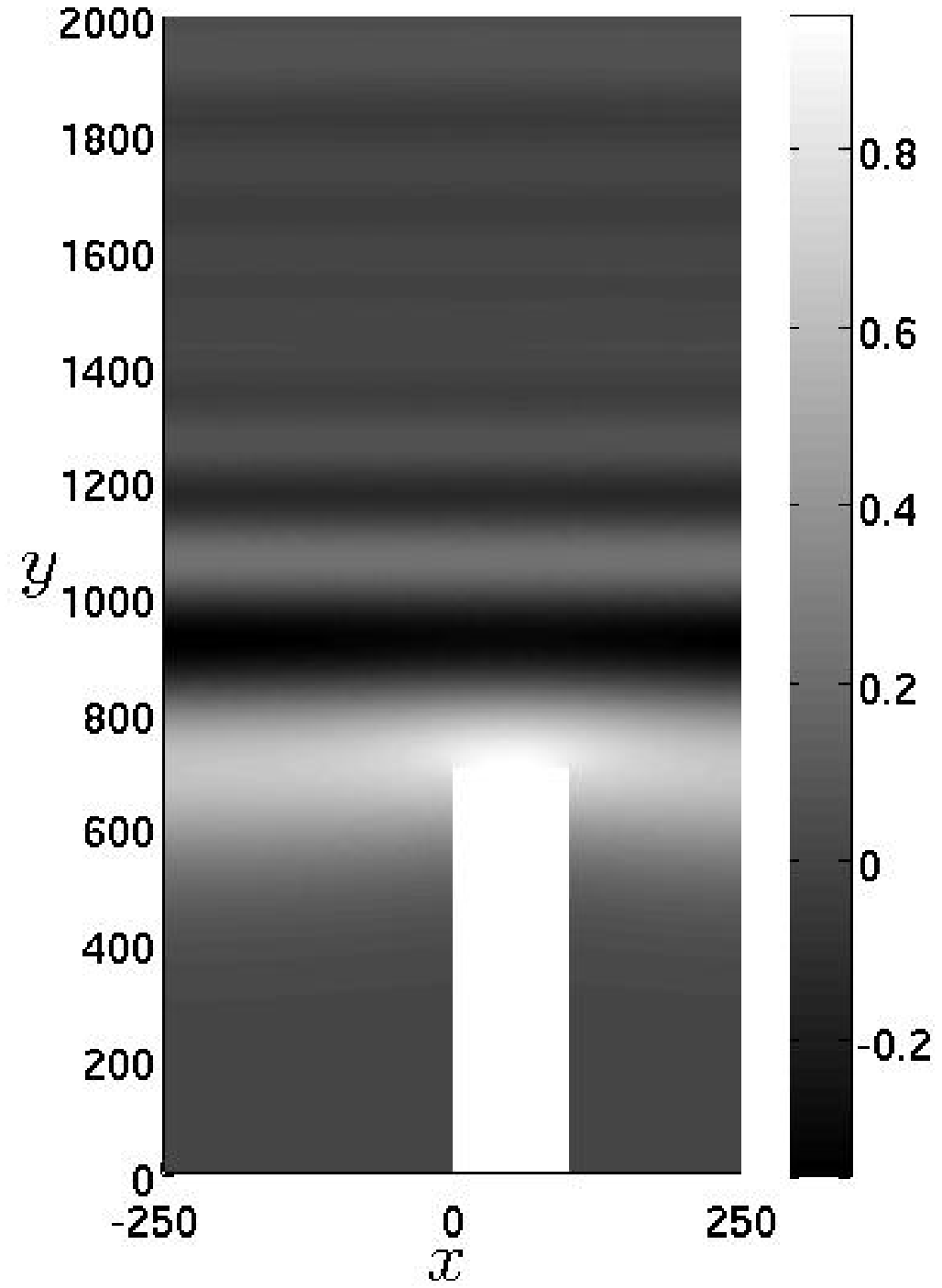}\\
BBM-BBM $t=40$ & Bona-Smith $t=40$\\
\includegraphics[width=2.5in]{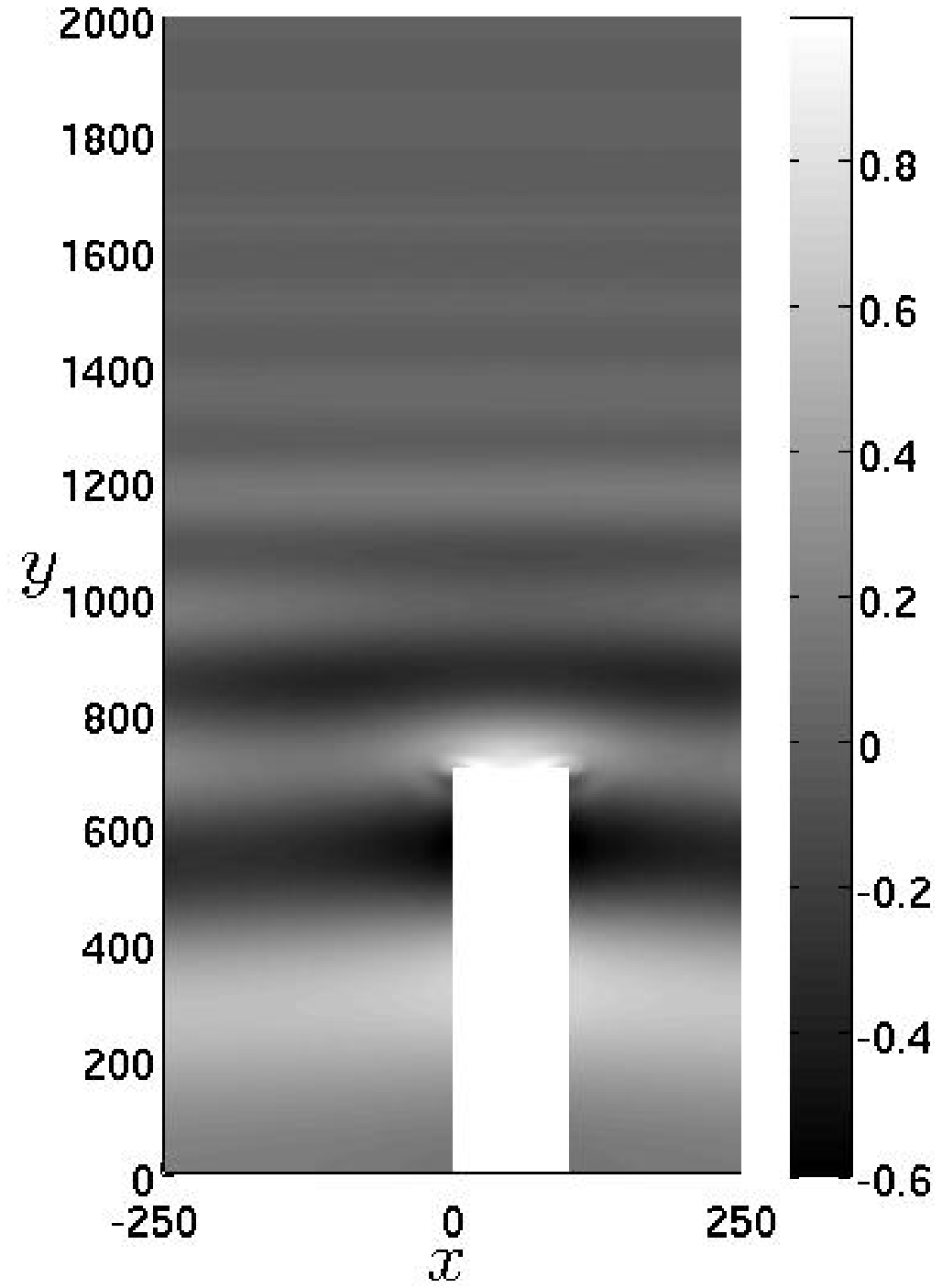} &\includegraphics[width=2.5in]{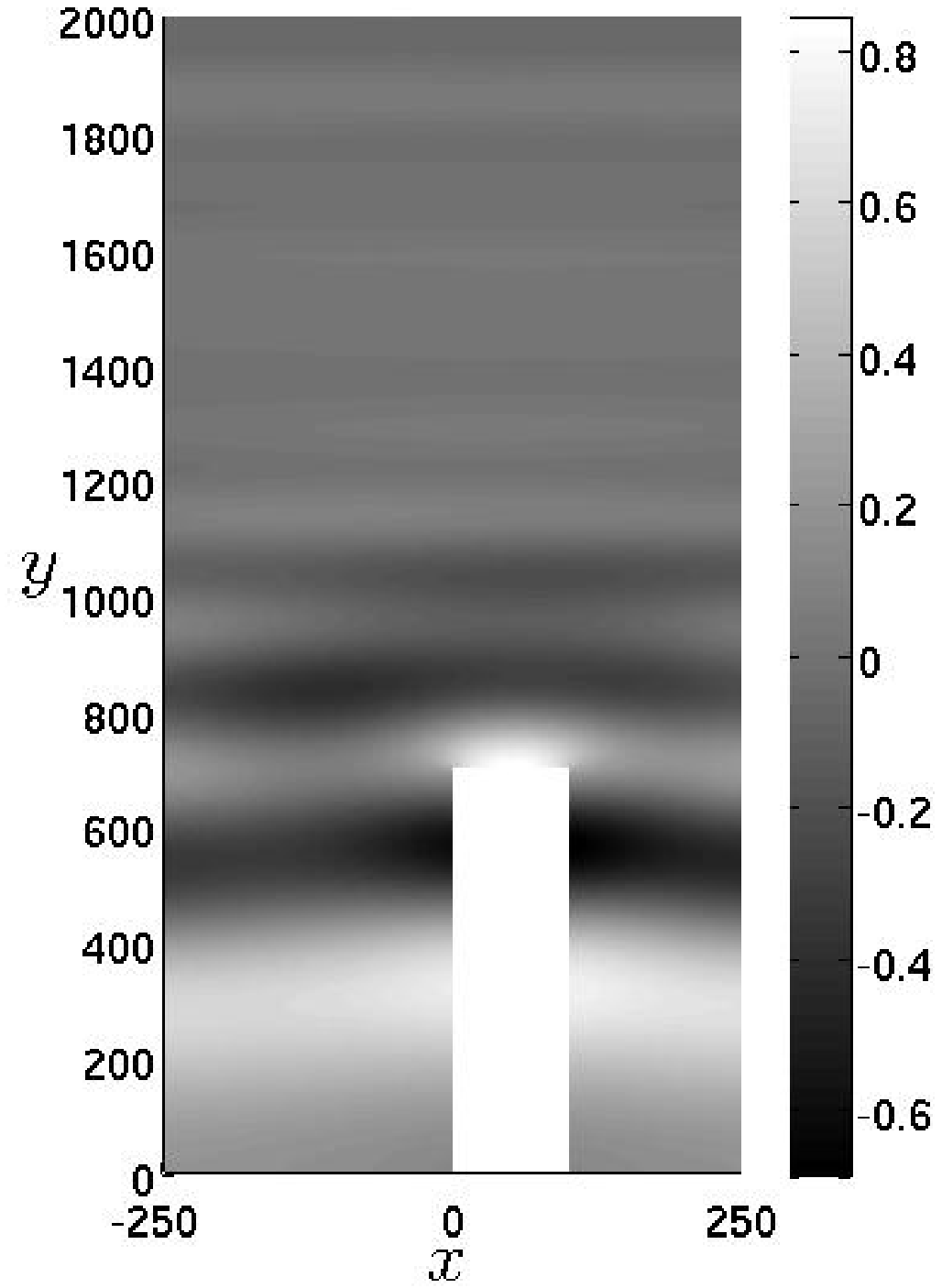}\\
BBM-BBM $t=60$ & Bona-Smith $t=60$
\end{tabular}
\end{center}
\end{figure}

\begin{figure}[p]
\begin{center}
\begin{tabular}{cc}
\includegraphics[width=2.5in]{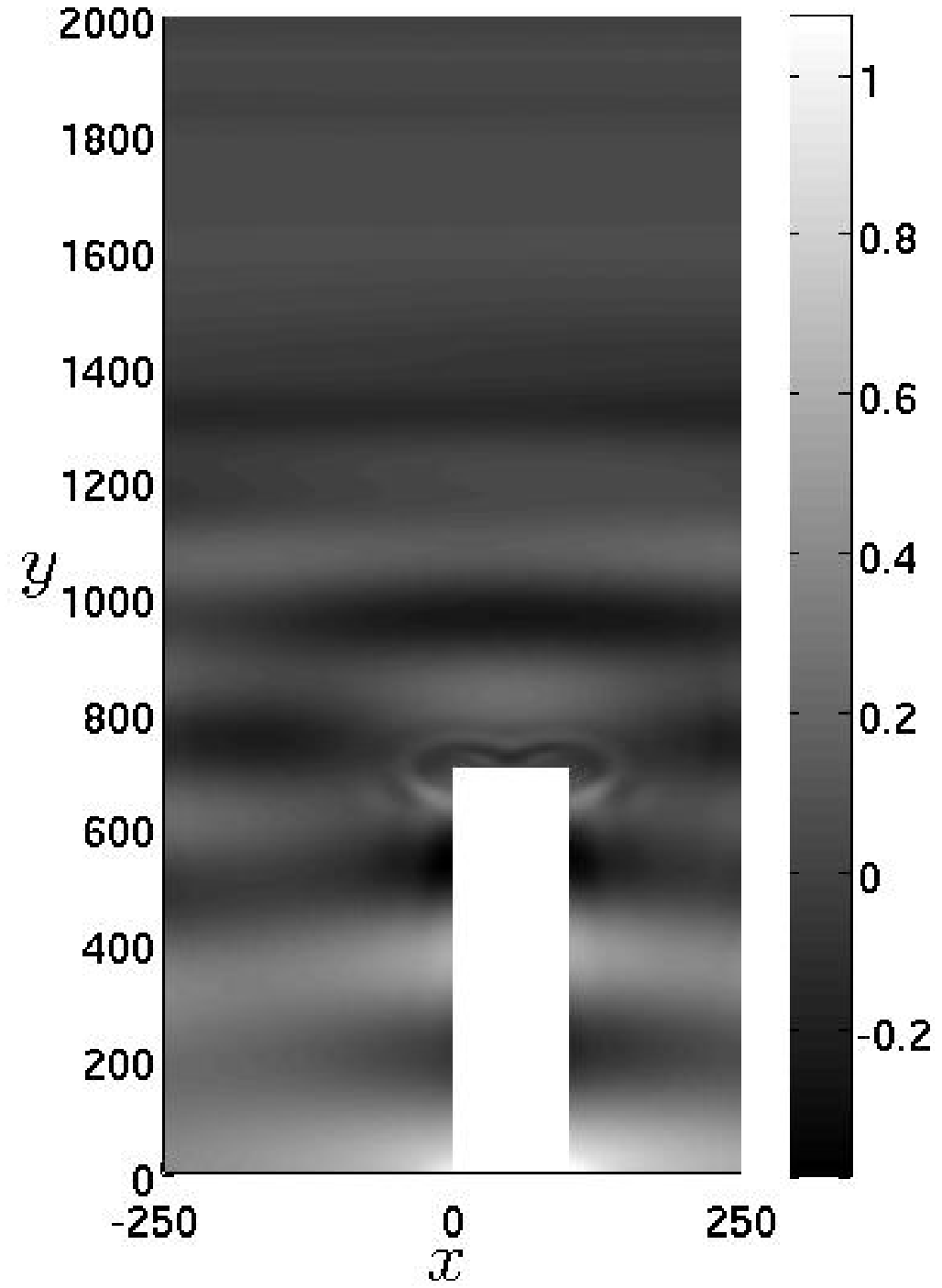} &\includegraphics[width=2.6in]{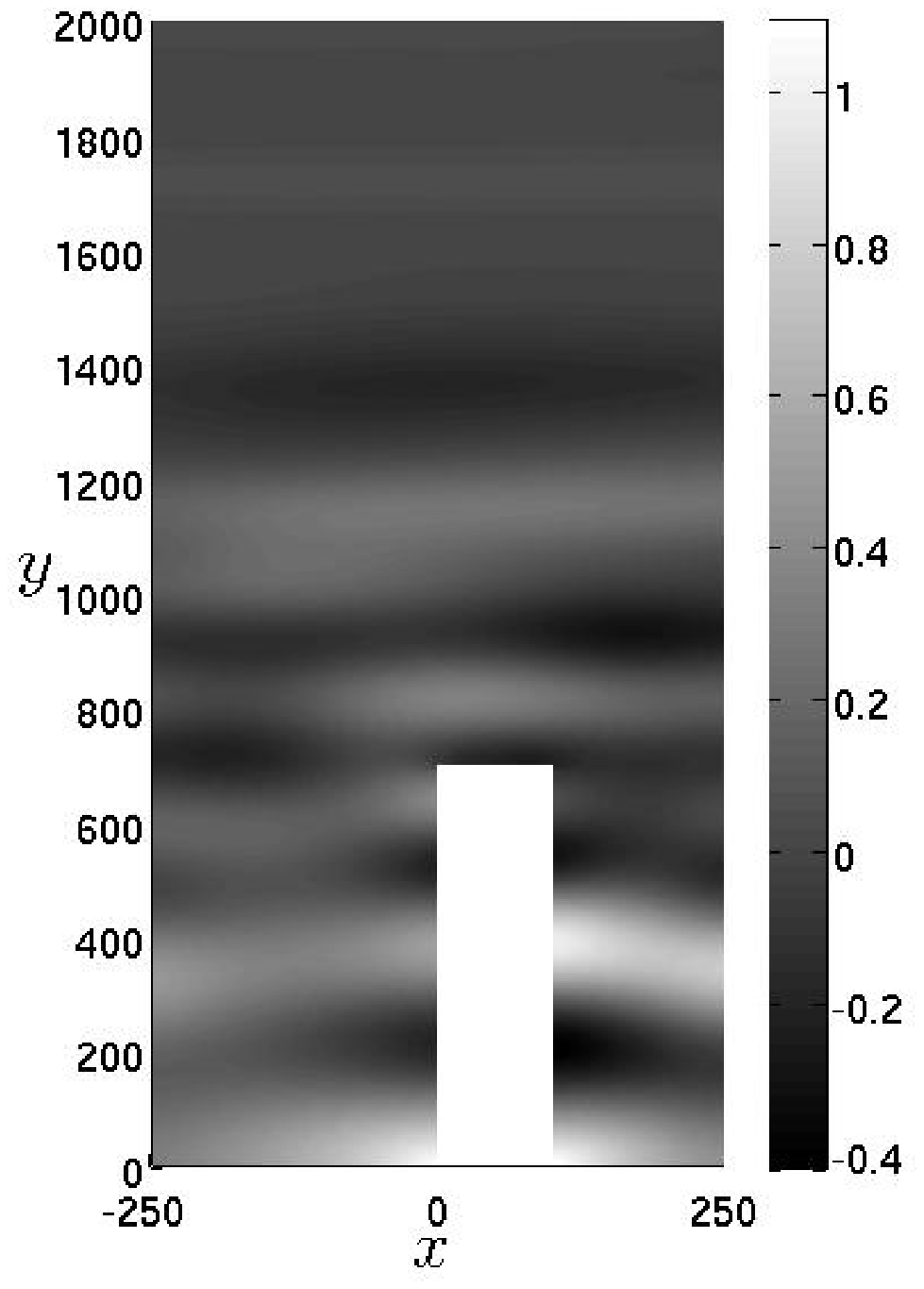}\\
BBM-BBM $t=80$ & Bona-Smith $t=80$
\end{tabular}
  \caption{Experiment 4.4. Free surface elevation at four time instances. BBM-BBM and Bona-Smith ($\theta^2=9/11$) systems. 
(elevation and $x,y$ in meters).}
  \end{center}
\end{figure}

\begin{figure}[p]
\begin{center}
\begin{tabular}{cc}
\includegraphics[width=3in]{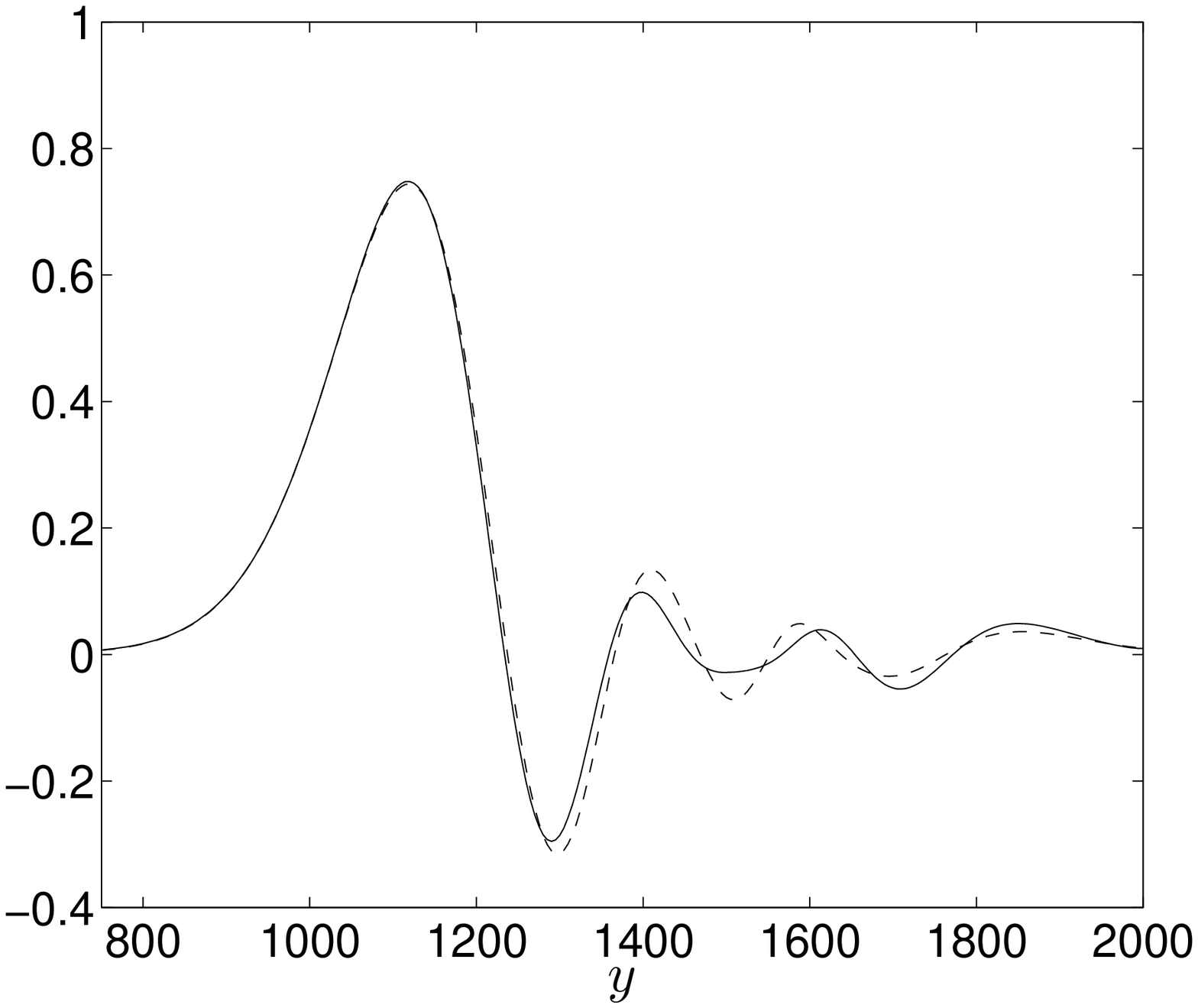} & \includegraphics[width=3in]{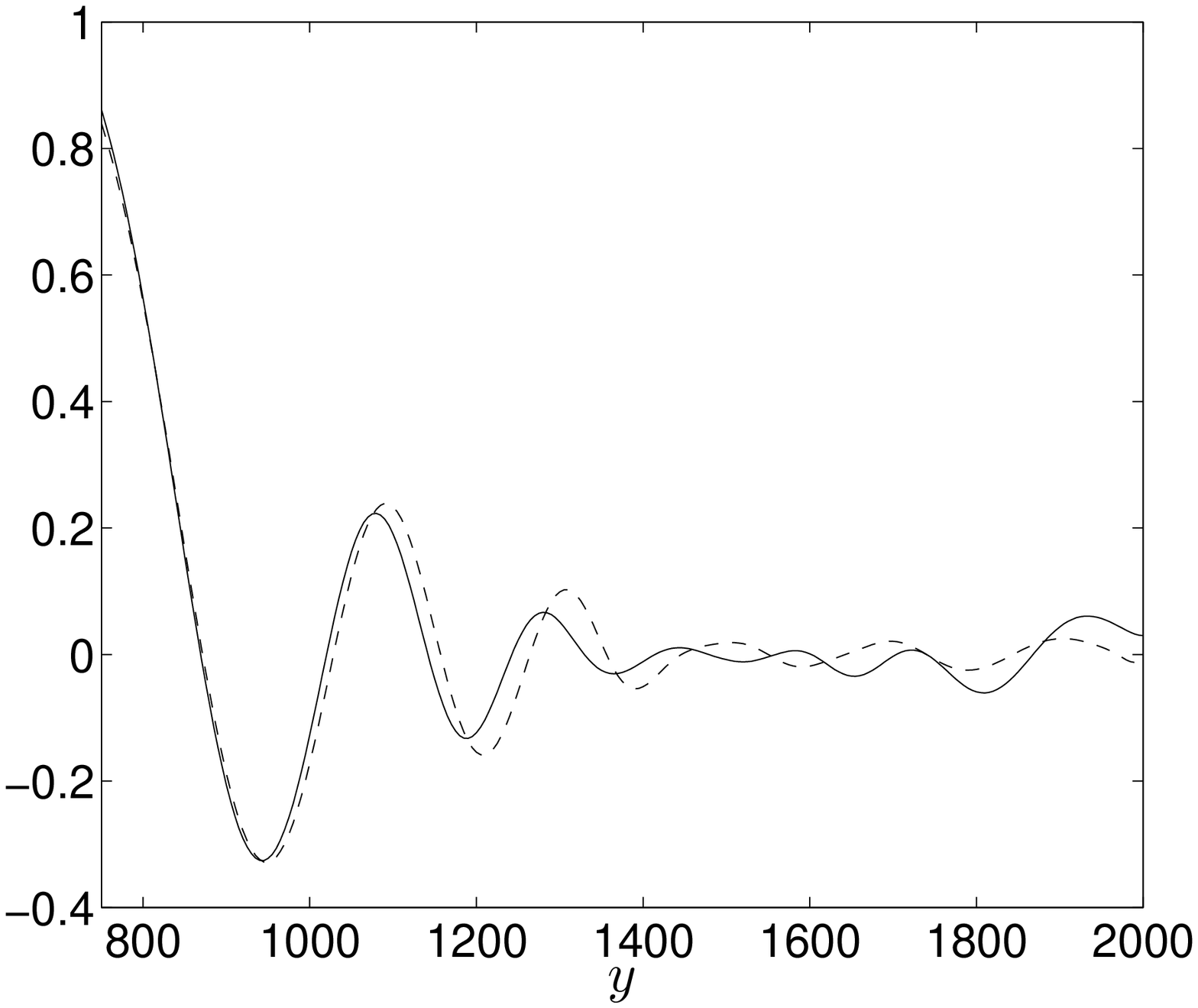}\\
$t=20s$ &
$t=39s$\\
\includegraphics[width=3in]{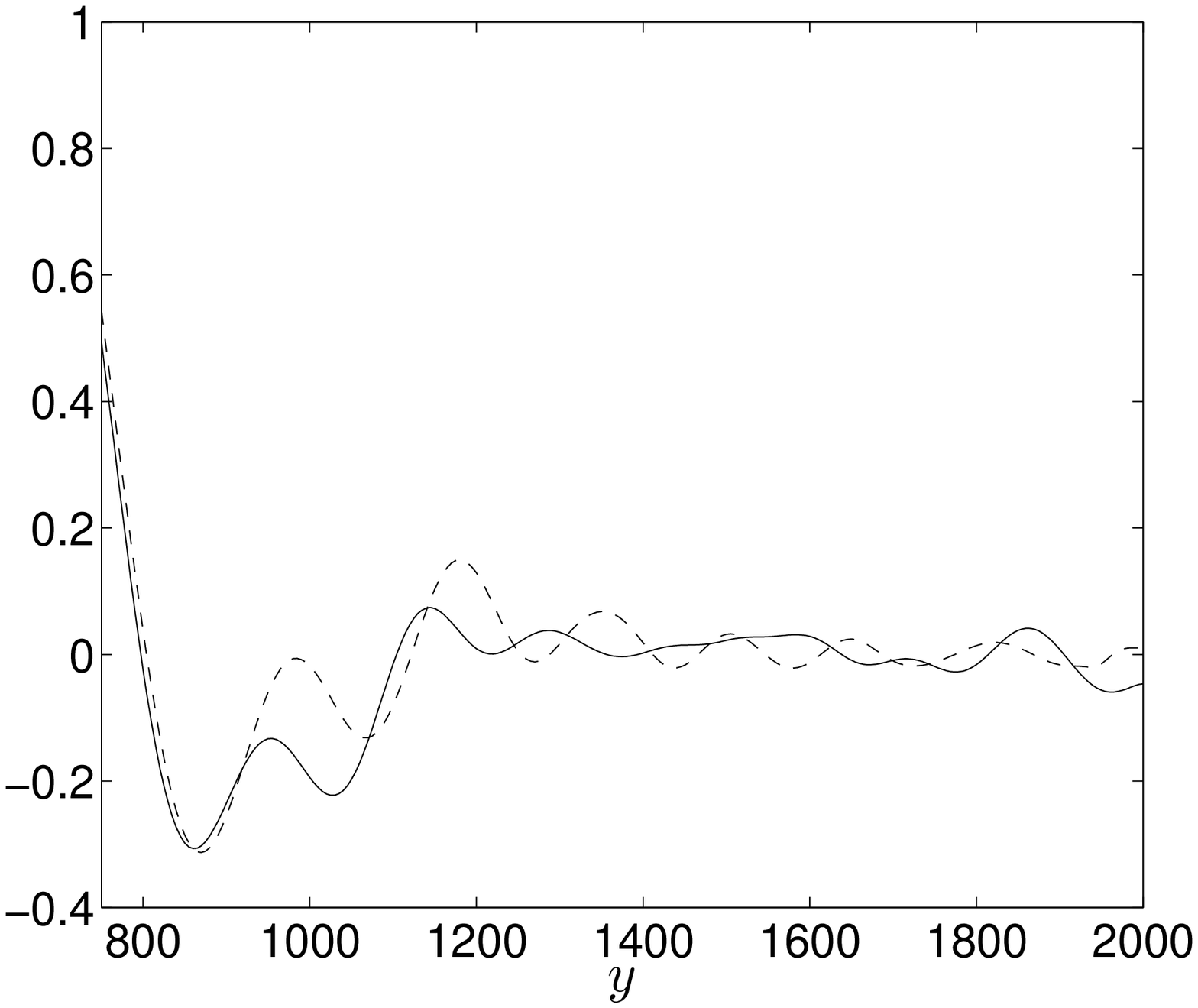}&
\includegraphics[width=3in]{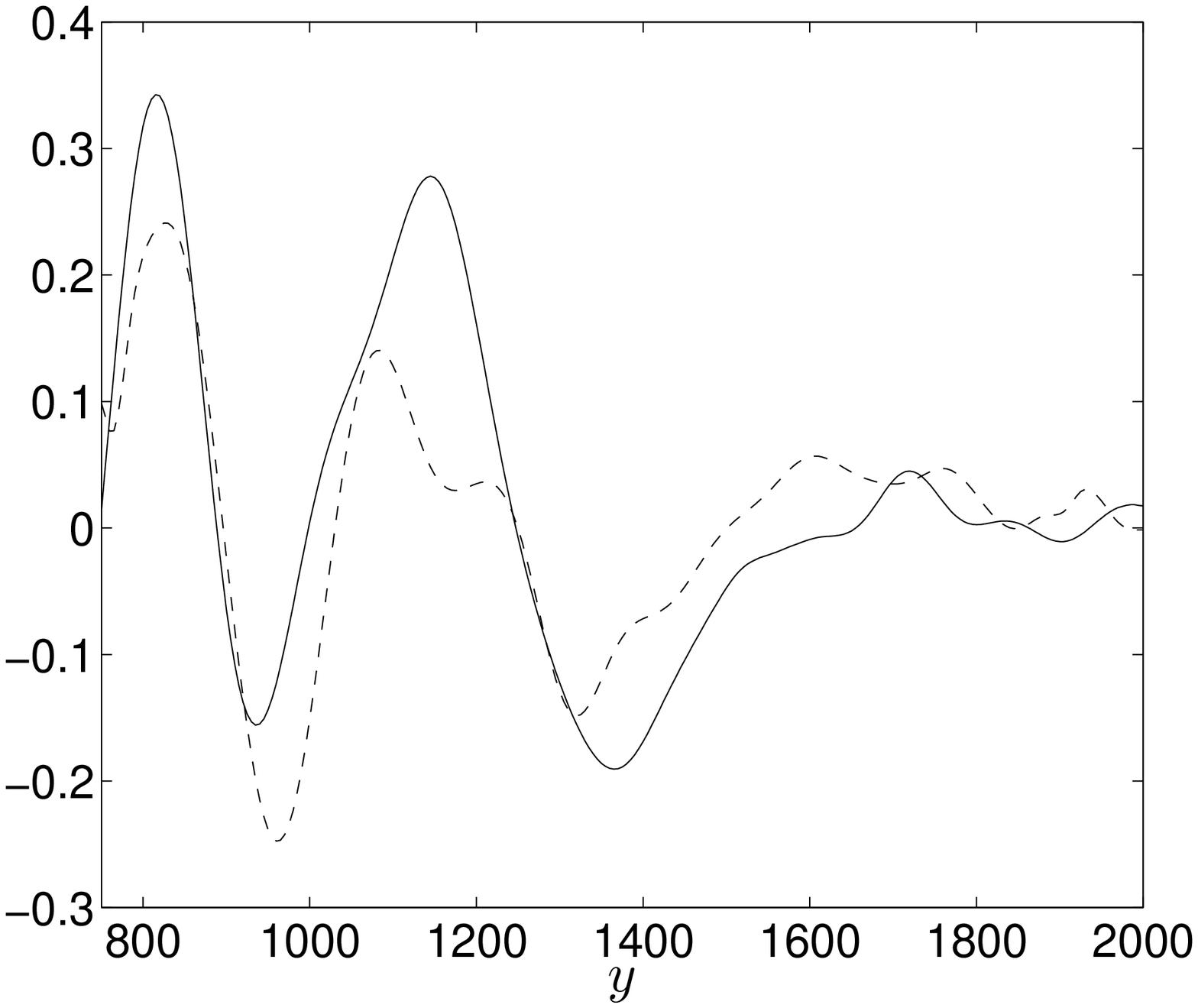}\\
$t=60s$&
$t=80s$
\end{tabular}
\caption{Experiment 4.4. Free surface elevation (in meters) as function of $y$ at four time instances along $x=40m$. BBM-BBM $- -$, Bona-Smith ($\theta^2=9/11$) ---. }
  \end{center}
\end{figure}

\begin{figure}[p]
\begin{center}
\includegraphics[width=3.5in]{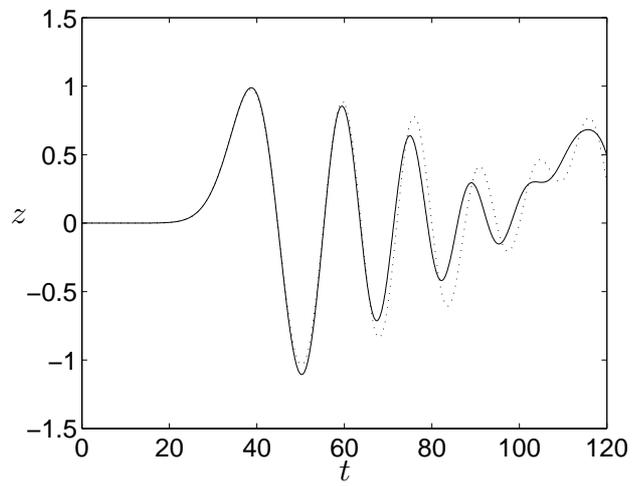} 
\caption{Experiment 4.4. Free surface elevation (in meters) as a function of $t$ at $(x,y)=(43.75,700)$. --: Bona-Smith ($\theta^2=9/11$) system, $\cdots$: BBM-BBM system.}
  \end{center}
\end{figure}
\clearpage

\end{document}